\documentclass[pdflatex,sn-mathphys-num]{sn-jnl}

\usepackage{graphicx}%
\usepackage{multirow}%
\usepackage{amsmath,amssymb,amsfonts}%
\usepackage{amsthm}%
\usepackage{mathrsfs}%
\usepackage[title]{appendix}%
\usepackage{xcolor}%
\usepackage{textcomp}%
\usepackage{manyfoot}%
\usepackage{booktabs}%
\usepackage{algorithm}%
\usepackage{algorithmicx}%
\usepackage{algpseudocode}%
\usepackage{listings}%
\usepackage{url}
\usepackage{caption}
\usepackage{pdflscape}
\usepackage{lmodern}

\newcommand{\lpt}{ASKAP~J1832--0911}
\newcommand{\mJy}{mJy\,beam$^{-1}$}
\newcommand{\pccm}{pc\,cm$^{-3}$}
\newcommand{\hi}{H\,{\sc i}}

\begin{document}

\title[J1832-0911 X-ray Radio]{Detection of X-ray Emission from a Bright Long-Period Radio Transient}


\author*[1]{\fnm{Ziteng} \sur{Wang}}\email{ziteng.wang@curtin.edu.au}
\author[2,3]{\fnm{Nanda} \sur{Rea}}
\author[4]{\fnm{Tong} \sur{Bao}}
\author[5]{\fnm{David~L.} \sur{Kaplan}}
\author[6]{\fnm{Emil} \sur{Lenc}}
\author[7,8,9]{\fnm{Zorawar} \sur{Wadiasingh}}
\author[10,9,11]{\fnm{Jeremy} \sur{Hare}}
\author[6,12]{\fnm{Andrew} \sur{Zic}}
\author[5]{\fnm{Akash} \sur{Anumarlapudi}}
\author[1]{\fnm{Apurba} \sur{Bera}}
\author[13,14,15]{\fnm{Paz} \sur{Beniamini}}
\author[16]{\fnm{A.~J.} \sur{Cooper}}
\author[17]{\fnm{Tracy~E.} \sur{Clarke}}
\author[18,12]{\fnm{Adam~T.} \sur{Deller}}
\author[19,6]{\fnm{J.~R.} \sur{Dawson}}
\author[20,1,21]{\fnm{Marcin} \sur{Glowacki}}
\author[1]{\fnm{Natasha} \sur{Hurley-Walker}}
\author[1]{\fnm{S.~J.} \sur{McSweeney}}
\author[17]{\fnm{Emil~J.} \sur{Polisensky}}
\author[17]{\fnm{Wendy~M.} \sur{Peters}}
\author[10,22]{\fnm{George} \sur{Younes}}
\author[6,23]{\fnm{Keith~W.} \sur{Bannister}}
\author[23,12]{\fnm{Manisha} \sur{Caleb}}
\author[1]{\fnm{Kristen~C.} \sur{Dage}}
\author[1]{\fnm{Clancy~W.} \sur{James}}
\author[24]{\fnm{Mansi~M.} \sur{Kasliwal}}
\author[24]{\fnm{Viraj} \sur{Karambelkar}}
\author[18,6]{\fnm{Marcus~E.} \sur{Lower}}
\author[25]{\fnm{Kaya} \sur{Mori}}
\author[24,26]{\fnm{Stella~Koch} \sur{Ocker}}
\author[27,28]{\fnm{Miguel} \sur{P\'erez-Torres}}
\author[29]{\fnm{Hao} \sur{Qiu}}
\author[23,6]{\fnm{Kovi} \sur{Rose}}
\author[18]{\fnm{Ryan~M.} \sur{Shannon}}
\author[30]{\fnm{Rhianna} \sur{Taub}}
\author[31]{\fnm{Fayin} \sur{Wang}}
\author[18,12]{\fnm{Yuanming} \sur{Wang}}
\author[31]{\fnm{Zhenyin} \sur{Zhao}}
\author[1]{\fnm{N.~D.~R.} \sur{Bhat}}
\author[23,12]{\fnm{Dougal} \sur{Dobie}}
\author[23]{\fnm{Laura~N.} \sur{Driessen}}
\author[23,12]{\fnm{Tara} \sur{Murphy}}
\author[18]{\fnm{Akhil} \sur{Jaini}}
\author[6]{\fnm{Xinping} \sur{Deng}}
\author[18]{\fnm{Joscha~N.} \sur{Jahns-Schindler}}
\author[23,12,6]{\fnm{Y.~W.~J.} \sur{Lee}}
\author[6]{\fnm{Joshua} \sur{Pritchard}}
\author[6]{\fnm{John} \sur{Tuthill}}
\author[32]{\fnm{Nithyanandan} \sur{Thyagarajan}}

\affil[1]{International Centre for Radio Astronomy Research, Curtin University, Kent Street, Bentley WA, 6102, Australia}

\affil[2]{Institute of Space Sciences (ICE), CSIC, Campus UAB, Carrer de Can Magrans s/n, E-08193, Barcelona, Spain}

\affil[3]{Institut d’Estudis Espacials de Catalunya (IEEC), 08860 Castelldefels (Barcelona), Spain}

\affil[4]{INAF – Osservatorio Astronomico di Brera, Via E. Bianchi 46, 23807 Merate (LC), Italy}

\affil[5]{Center for Gravitation, Cosmology, and Astrophysics, Department of Physics, University of Wisconsin-Milwaukee, P.O. Box 413, Milwaukee, 53201, WI, USA}

\affil[6]{Australia Telescope National Facility, CSIRO Space and Astronomy, PO Box 76, Epping, NSW 1710, Australia}

\affil[7]{Department of Astronomy, University of Maryland College Park, College Park, MD 20742, USA}

\affil[8]{Astrophysics Science Division, NASA/GSFC, Greenbelt, MD 20771, USA}

\affil[9]{Center for Research and Exploration in Space Science and Technology, NASA/GSFC, Greenbelt, MD 20771, USA}

\affil[10]{NASA Goddard Space Flight Center, Greenbelt, Maryland, 20771, USA}

\affil[11]{The Catholic University of America, 620 Michigan Ave., N.E. Washington, DC 20064, USA}

\affil[12]{ARC Centre of Excellence for Gravitational Wave Discovery (OzGrav), Hawthorn, VIC 3122, Australia}

\affil[13]{Department of Natural Sciences, The Open University of Israel, P.O Box 808, Ra'anana 4353701, Israel}

\affil[14]{Astrophysics Research Center of the Open university (ARCO), The Open University of Israel, P.O Box 808, Ra'anana 4353701, Israel}

\affil[15]{Department of Physics, The George Washington University, 725 21st Street NW, Washington, DC 20052, USA}

\affil[16]{Astrophysics, The University of Oxford, Keble Road, Oxford, OX1 3RH, UK}

\affil[17]{Naval Research Laboratory,  4555 Overlook Ave SW,  Washington,  DC 20375, USA}

\affil[18]{Centre for Astrophysics and Supercomputing, Swinburne University of Technology, Hawthorn, VIC, 3122, Australia}

\affil[19]{School of Mathematical and Physical Sciences and Astrophysics and Space Technologies Research Centre, Macquarie University, 2109, NSW, Australia.}

\affil[20]{Institute for Astronomy, University of Edinburgh, Royal Observatory, Edinburgh, EH9 3HJ, United Kingdom}

\affil[21]{Inter-University Institute for Data Intensive Astronomy, Department of Astronomy, University of Cape Town, Cape Town, South Africa}

\affil[22]{Center for Space Sciences and Technology CRESST II, UMBC, 1000 Hilltop Cir, Baltimore, MD 21250, USA}

\affil[23]{Sydney Institute for Astronomy, School of Physics, The University of Sydney, NSW 2006, Australia}

\affil[24]{Cahill Center for Astronomy and Astrophysics, California Institute of Technology, Pasadena, CA 91125, USA}

\affil[25]{Columbia Astrophysics Laboratory, Columbia University, New York, NY 10027, USA}

\affil[26]{Observatories of the Carnegie Institution for Science, Pasadena, CA 91101, USA}

\affil[27]{Instituto de Astrof\'isica de Andaluc\'ia (IAA-CSIC), Consejo Superior de Investigaciones Cient\'ificas (CSIC), Glorieta de la Astronom\'ia, s/n, Granada, E-18008, Spain}

\affil[28]{School of Sciences, European University Cyprus, Diogenes Street, Engomi, 1516 Nicosia, Cyprus}

\affil[29]{SKA Observatory,  Jodrell Bank, Cheshire SK11 9FT, UK}

\affil[30]{Dept. Physics \& Astronomy, Wayne State University,  666 W. Hancock St, Detroit, MI, 48201, USA}

\affil[31]{School of Astronomy and Space Science, Nanjing University, Nanjing 210093, People’s Republic of China}

\affil[32]{CSIRO, Space and Astronomy, P. O. Box 1130, Bentley, WA 6102, Australia}


\abstract{Recently, a class of long-period radio transients (LPTs) has been discovered, exhibiting emission on timescales thousands of times longer than radio pulsars  \citep{2022Natur.601..526H, 2023Natur.619..487H, 2024NatAs.tmp..107C, 2024arXiv240707480D, 2024arXiv240811536D}. 
Several models had been proposed implicating either a strong magnetic field neutron star \citep{2024MNRAS.533.2133C}, isolated white dwarf pulsar \cite{2022Ap&SS.367..108K}, or a white dwarf binary system with a low-mass companion \cite{2024arXiv240905978Q}. 
While several models for LPTs also predict X-ray emission  \citep{2024MNRAS.533.2133C, 2023A&A...674L...9S}, no LPTs have been detected in X-rays despite extensive searches \citep{2022Natur.601..526H, 2023Natur.619..487H, 2024NatAs.tmp..107C, 2024arXiv240707480D, 2024arXiv240811536D, 2022ApJ...940...72R}.
Here we report the discovery of an extremely bright LPT (10--20\,Jy in radio), ASKAP\,J1832$-$0911, which has coincident radio and X-ray emission, both with a 44.2-minute period. The X-ray and radio luminosities are correlated and vary by several orders of magnitude.
These properties are unique amongst known Galactic objects and require a new explanation.
We consider a $> 0.5$\,Myr old magnetar with a $\gtrsim 10^{13}$\,G crustal field, or an extremely magnetised white dwarf in a binary system with a dwarf companion, to be plausible explanations for \lpt, although both explanations pose significant challenges to formation and emission theories. The X-ray detection also establishes a new class of hour-scale periodic X-ray transients of luminosity $\sim 10^{33}$\,erg\,s$^{-1}$ associated with exceptionally bright coherent radio emission.
}




\maketitle

\section*{MAIN}

\begin{figure}
    \centering
    \includegraphics[width=0.85\linewidth]{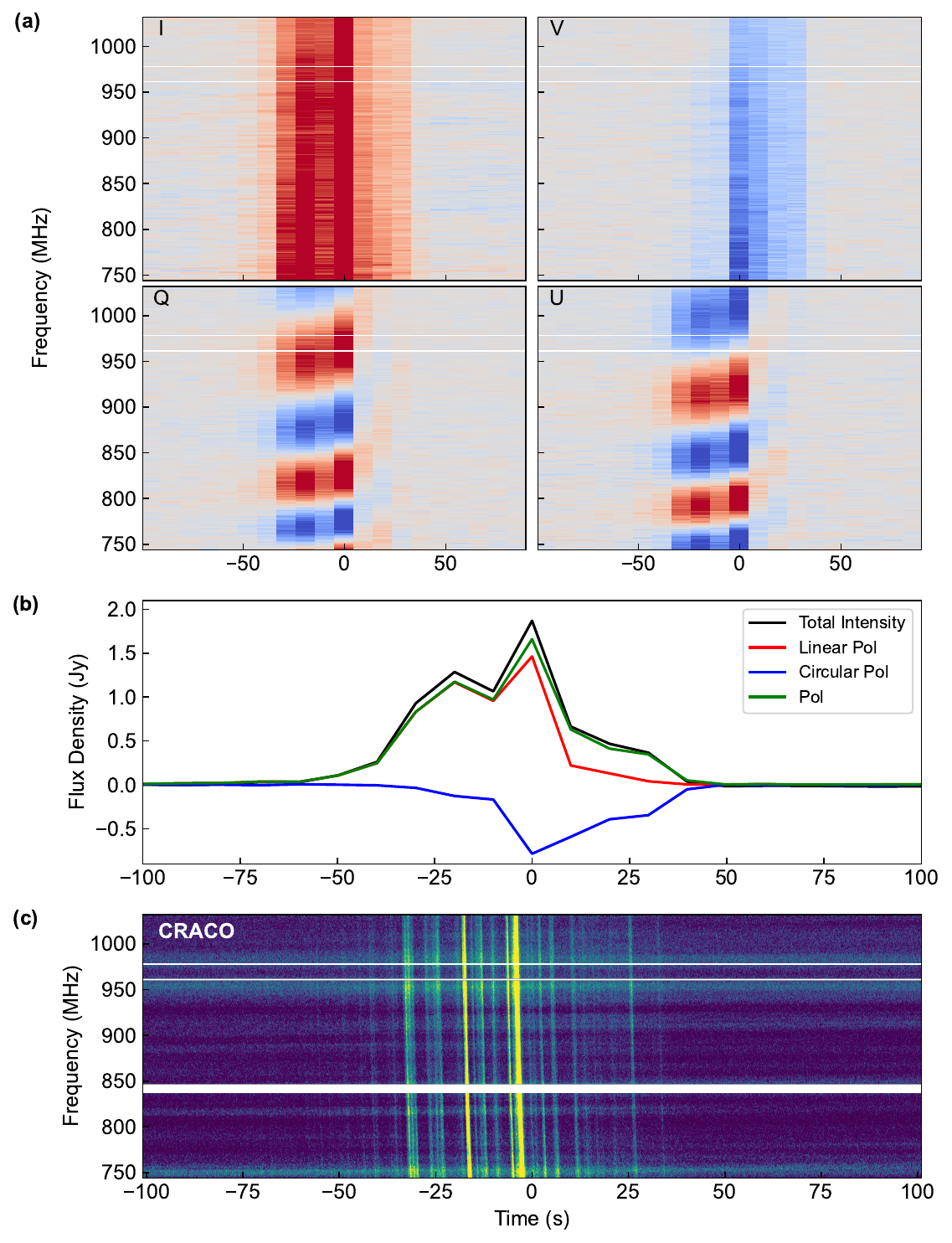}
    \caption{Dynamic spectra and light curve for the discovery observation of \lpt.  (a) From left to right, and top to bottom, the panels show the Stokes I (total intensity), V (circular polarisation), Q and U (linear polarisations) flux density as a function of frequency and time. The Stokes Q and U show a Faraday rotation of $+89.1\pm0.1\,$rad\,m$^{-2}$. (b) The total intensity and polarised emission intensity lightcurve of \lpt. We Faraday de-rotate the Stokes Q and U data with the rotation measure estimated above to obtain the linearly polarised intensity. (c) High-time resolution (13.4\,ms) dynamic spectrum of \lpt\ reveals fine structures of the pulse. The fine structures allow us to measure the dispersion measure of the pulse to be $458\pm14\,$\pccm.}
    \label{fig:J1832_discovery}
\end{figure}

\lpt\ was identified as a compact circularly polarised transient radio source using the Australian SKA Pathfinder \citep[ASKAP;][]{2021PASA...38....9H} as part of the ASKAP Variables and Slow Transients \citep[VAST;][]{2013PASA...30....6M, 2021PASA...38...54M} survey (see Methods).
The light curve at 10-second resolution (Figure~\ref{fig:J1832_discovery}) shows a signal duration of approximately two minutes, with a maximum peak flux density of 1870$\pm$6\,\mJy, and radio spectral index $S_\nu\propto\nu^\alpha$, with $\alpha=-1.5\pm0.1$.
The pulse displayed substantial linear and circular polarisation, with a total fractional polarisation of $92 \pm 3\%$. Notably, the first half of the pulse was predominantly linearly polarised,  suggesting the presence of ordered magnetic fields. \\

During the imaging observations with ASKAP, the Commensal Real-time ASKAP Fast Transients Survey COherent \citep[CRACO;][]{2024arXiv240910316W} signal processor independently detected sub-pulse structures from \lpt\ (see Figure~\ref{fig:J1832_discovery}).
The sub-pulses displayed a frequency-dependent time delay due to propagation through the ionized interstellar medium, from which we measured a charged-particle column density in units of dispersion measure (DM) of $458 \pm 14$\,\pccm. This corresponds to a distance of $4.5 \pm 1.2$\,kpc based on a model of the distribution of ionized gas in the Galaxy  \citep{2017ApJ...835...29Y}. 
The shortest $0.5$\,s duration flux-density variations constrain the emission region to be smaller than $<$150,000\,km, indicating the presence of a compact object: a white dwarf star, neutron star, or black hole.
The implied brightness temperature is $\gtrsim10^{19}$\,K, requiring a non-thermal, coherent emission mechanism (see Extended Data Figure~\ref{fig:transient_phase}). \\

We did not detect any emission from ASKAP J1832-0911 in approximately 40 hours of archival radio data collected by ASKAP, MeerKAT and the Karl G. Jansky Very Large Array (VLA) between 2013 and 2023, suggesting that the source may have activated only after November 2023. In follow-up observations, we observed two peaks in flux density, one in 2023 December and another in 2024 February. These have been followed by a decline.  \\

Follow-up observations with the Australia Telescope Compact Array (ATCA), ASKAP, MeerKAT, and the Giant Metrewave Radio Telescope (GMRT) detected multiple pulses from \lpt\, which exhibit diverse range of phenomenology, including variations in morphology, flux density, radio spectral index, and polarisation (see Extended Data Table~\ref{tab:radio_table} and Figure~\ref{fig:J1832_pol_pa}). The flux density fluctuated from approximately 30\,mJy to 20\,Jy, while the radio spectral indices ($\alpha$) of the pulses varied between $ -2.2$ and $-0.3$ (see Extended Data Table~\ref{tab:radio_table}).
Times of arrival (TOAs) for the pulses were extracted using a simple Gaussian template, and we estimated a period of $P=2656.247\pm0.001$\,s. 
We did not detect any secular or periodic variations in arrival times, and limit the spin-down rate to be $|\dot{P}| \lesssim 9.8\times10^{-10}$\,s\,s$^{-1}$ at 95\% confidence. \\

\hi\ absorption from Galactic foreground gas along the line of sight to \lpt\ was weakly detected in both the GMRT and MeerKAT data (see Extended Data Figure~\ref{fig:hi_absorption}). 
The velocity of the most distant detected absorbing gas component is estimated as $75 \pm 15$ km\,s$^{-1}$, corresponding to a near-side kinematic distance of $4.8 \pm 0.8$ kpc \citep{2018ApJ...856...52W}, consistent with the distance inferred from the DM. Taking both DM implied distance and \hi\ derived lower limit into account, we estimate the distance to \lpt\ to be $4.5^{+1.2}_{-0.5}$\,kpc. \\

\begin{figure}
    \centering
    \includegraphics[width=\linewidth]{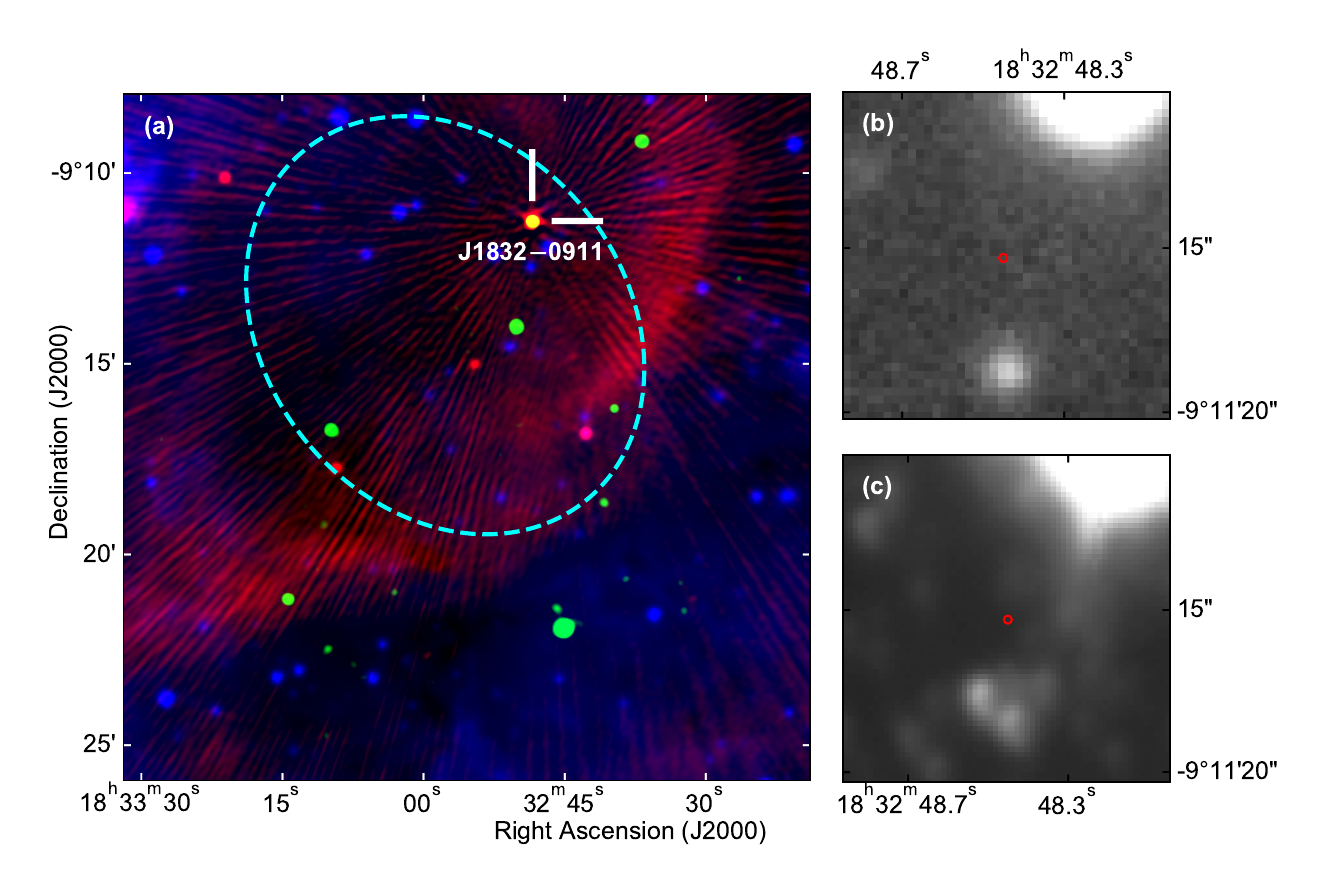}
    \caption{Field of \lpt. Panel (a) shows a composite of radio (MeerKAT 816\,MHz, red), X-ray ({\em Chandra} 1--10\,keV, green), and infrared ({\it WISE} 12\,$\mu$m, blue) emission of the field of \lpt. The {\em Fermi} 95\% positional error ellipse around 4FGL~J1832.9$-$0913 is shown in cyan dashed line. Panel (b) and (c) show the deepest near-infrared images of \lpt\ at $J$- and $K_s$-band, respectively. Red circles show 50 times the systematic uncertainty of the source position ($\sim$5\,mas).}
    \label{fig:J1832_multilambda}
\end{figure}

\lpt\ is located deep in the Galactic Plane ($l=22.64^{\circ}$, $b=-0.08^{\circ}$), in a region dense in stars, gas, and dust. To definitively identify a counterpart, a high-precision localization is needed to avoid misassociation.
Observations with the Very Long Baseline Array (VLBA) refined the position of \lpt\ to right ascension (R.A.) = 18h32m48.4589s and declination (Dec.) = $-09^{\circ}11'15.297''$ at epoch MJD 60400 and equinox J2000 with an estimated uncertainty of 5\,mas (Methods). With this refined position we searched for multi-band counterparts at other wavelengths. No infrared counterpart was detected with $3\sigma$ magnitude limits of $K_s > 19.86$ and $J>19.98$ (see Figure~\ref{fig:J1832_multilambda}), using the FourStar infrared camera on the Magellan telescope and the Wide-field Infrared Camera on the Palomar 200-inch Telescope, respectively. The derived limits allow us to rule out the presence of a main sequence star with a spectral type earlier than M0 (Methods) or a white dwarf star with a temperature greater than $10^5$\,K. \\

\lpt\ is spatially coincident with a known supernova remnant (SNR), SNR~G22.7$-$0.2, located at a distance of 4.74\,kpc \citep{2020A&A...639A..72W}. The similarity in both spatial location and distance suggest a potential association between \lpt\ and G22.7$-$0.2; however, the off-centre position of our target, the high chance coincidence probability of 70\,\%, and inconsistent velocity disfavour this association (Methods).  \lpt\ also resides within the 95\% positional error ellipse of the Fermi source, 4FGL~J1832.0--0913. However, this source has been recently associated with the $\gamma$-ray binary, HESS\,J1832$-$093 \citep{2020A&A...637A..23M}, which is offset from \lpt{} by $\sim11\,$arcmin. \\

\lpt\ was observed in a serendipitous X-ray observation with the {\em Chandra} ACIS instrument for 20\,ks on 2024 February 14, targeting SNR~G22.7$-$0.2. An uncatalogued X-ray source was detected that is positionally coincident with \lpt{} (Methods). A blind Lomb-Scargle periodicity search identified a period of  $2634_{-64}^{+71}$\,s at $3\sigma$ significance (see Extended Data Figure~\ref{fig:Xray_LS}), consistent with the radio periodicity. 
The source spectrum and absorption are not well constrained given the low number of counts, however it can be fit with an absorbed power-law model, adopting a fixed hydrogen column density $N_H = 1.8\times10^{22}$\,cm$^{-2}$, which is the total Galactic column density in the direction of the source (Methods), and a photon index $\Gamma=0.0\pm0.5$ or with a blackbody with temperature $kT = 2.2^{+1.3}_{-0.6}$\,keV with blackbody radius $\lesssim0.1$\,km (see Extended Data Table~\ref{tab:xfit} and Figure~\ref{fig:contour}). 
We estimated a phase-averaged X-ray luminosity between 1 and 10 keV of $L_{\rm x} = 7.4_{-3.5}^{+3.6} \times 10^{32} (\frac{d}{4.5\,{\rm kpc}})^2$\,~erg~s$^{-1}$. 
\lpt\ was in an exceptionally bright radio state at that time, with 10--20\,Jy pulses observed by ASKAP ten days prior, $\sim$8\,Jy pulses recorded by the VLA one day after, and several $\sim$2\,Jy pulses recorded by MeerKAT six days after the X-ray observation. 
In August 2024, the source was re-observed with the Follow-up X-ray Telescope (FXT) on-board of Einstein Probe (EP; \cite{2022hxga.book...86Y}) for 30\,ks and by the {\em Chandra} ACIS instrument for 10\,ks, and was not detected, with a 3$\sigma$ limit of $L_X < 6\times10^{31} (\frac{d}{4.5\,{\rm kpc}})^2$~erg~s$^{-1}$ (assuming the most conservative scenario for the quiescence spectrum: see Methods).  Thus, the X-ray emission had decreased in luminosity by almost one order of magnitude in less than six months.
Simultaneous radio observations were conducted with ASKAP and MeerKAT, revealing that the radio flux density of the source had decreased to $\sim$60\,mJy: three orders of magnitude lower than in February.
The field surrounding \lpt\ had been previously observed by XMM-Newton in 2011 for $\sim17$\,ks and \lpt\ was not detected, with a 3$\sigma$ limit of $L_X < 5.3\times10^{31} (\frac{d}{4.5\,{\rm kpc}})^2$~erg~s$^{-1}$ (but see Methods for the assumptions and corresponding uncertainties on this value). \\

\begin{figure}
    \centering
    \includegraphics[width=0.75\linewidth]{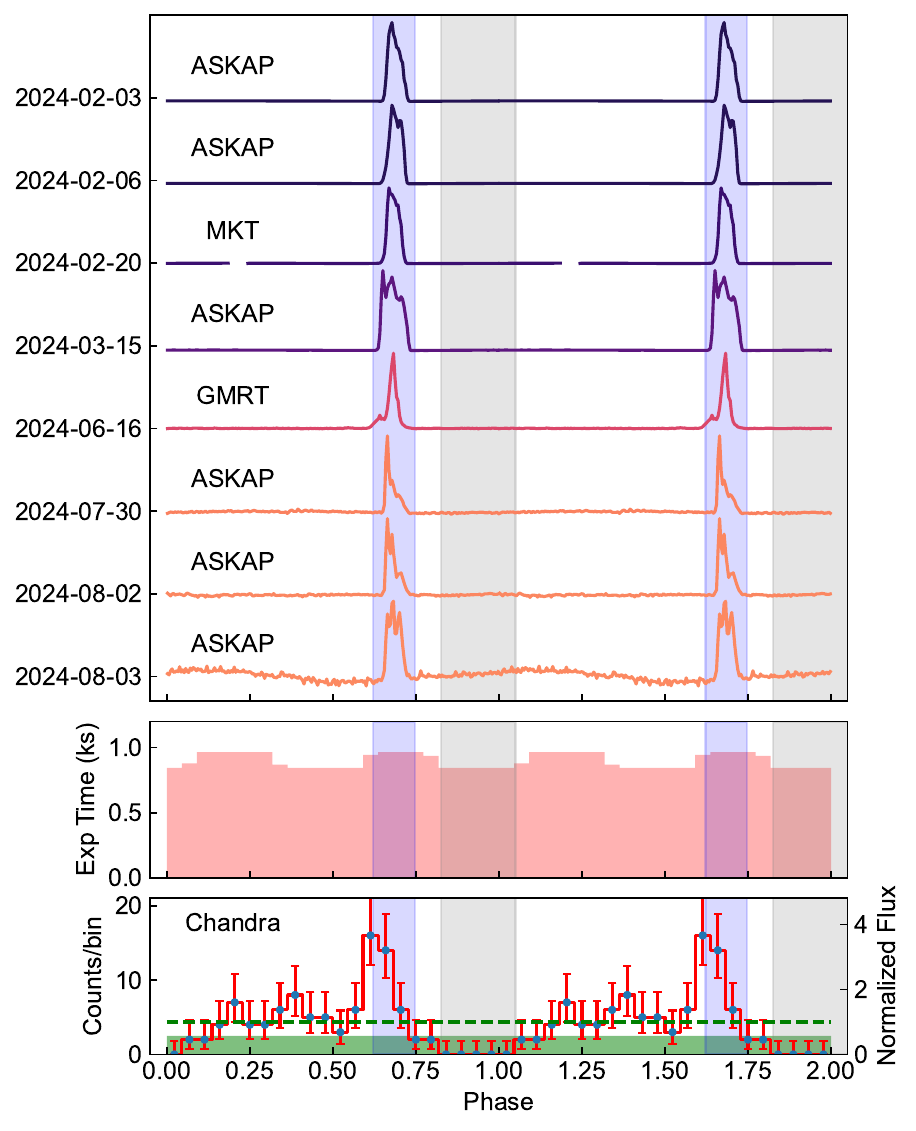}
    \caption{Radio and X-ray detections of \lpt. Radio lightcurves and {\em Chandra} X-ray (on 2024 Feb 14) lightcurves folded with the radio period are shown in the top and bottom panels, respectively, while the binned {\em Chandra} exposure time is shown in the middle panel. Blue shaded areas (at phase 0.3--0.45) indicate the phase when radio pulses were detected, and grey shaded areas (at phase 0.5--0.75) show the phase when no X-ray photons were detected.}
    \label{fig:J1832_xray_radio}
\end{figure}

\lpt\ is currently the only LPT detected with (pulsed) X-ray emission --- perhaps unsurprisingly, given its extreme radio brightness and the potential correlation between the radio and X-ray luminosities. Given the pulse phase alignment, and similar phase widths, the radio and X-ray emission from \lpt\ likely originates from magnetically linked regions within the progenitor system.
The high fractional polarised coherent emission from \lpt\, indicates the presence of a compact object with highly ordered magnetic fields, such as a white dwarf (WD) or neutron star (NS). However, the combination of 2656\,s long radio and X-ray periodicities, and the extreme variability of its luminosity, are unseen in any class of known Galactic compact objects (Methods). \\

\lpt\ is not a rotation-powered pulsar: 
the radio luminosity is 10,000 times greater than the spin-down luminosity,  $|\dot{E}_{\rm spin}| < 4\pi^2I|\dot{P}|/P^3= 5\times10^{27}\,$\,erg\,s$^{-1}$ (assuming a NS with 1.4$M_{\odot}$ and 12\,km radius), and its X-ray luminosity is even higher.
Like other known LPTs, \lpt\ resides in the ``death valley" of the spin vs spin down phase space \citep{1993ApJ...402..264C,2000ApJ...531L.135Z,2011ApJ...726L..10H}, where curvature-radiation-initiated pair production thought to be responsible for the radio emission is expected to cease (see Extended Data Figure~\ref{fig:p_B}). Furthermore, the wild variability of its X-ray emission is distinct from the steady emission from classical rotation-powered pulsars \citep{1997A&A...326..682B}. These properties are at odds with an accretion-powered pulsar or transitional millisecond pulsar (Methods).\\

We can also exclude an isolated white dwarf progenitor on the basis of the observed X-ray emission. 
WD surface thermal X-ray emission is typically fainter, stable over time \citep{2022PhR...988....1S}, and has a softer spectrum \citep{1985SSRv...40...79H}.
An unusual magnetically-powered isolated WD is also readily excluded for \lpt: the likely $\gtrsim 10^{40}$~erg energy output over the active period is almost the total field reservoir energy for such a system \citep{2023MNRAS.520.1872B}.\\

The observed characteristics of \lpt\ qualitatively resemble those of some binary WDs, as similar radio and X-ray pulsations have been observed in AR\,Sco \citep{2016Natur.537..374M} and its sibling J1912--44 \citep{2023NatAs...7..931P}. In these systems, an M dwarf and fast-spinning ($\sim$minutes) WD are in a tight ($\sim$ hours) binary. However, their radio emission is seven orders of magnitude weaker than the emission from \lpt\, and also highly circularly polarised. 
Nevertheless, the recent discovery of two LPTs with M-type stellar counterparts \citep{2024arXiv240811536D, 2024arXiv240815757H}, notably a potential WD identified in ILT~J1101+5521, suggests that WD binaries likely account for at least a subset of the LPT population. 
If the radio emission of \lpt\ comes from the relativistic electron cyclotron maser emission from the magnetic interaction between the two stars \cite{2024arXiv240905978Q}, the required magnetic field strength is $> 5\times10^{9}$\,G (Methods), making it the most magnetised white dwarf known in the Galaxy \citep{2022Msngr.186...14B}.\\

\begin{figure}
    \centering
    \includegraphics[width=0.8\linewidth]{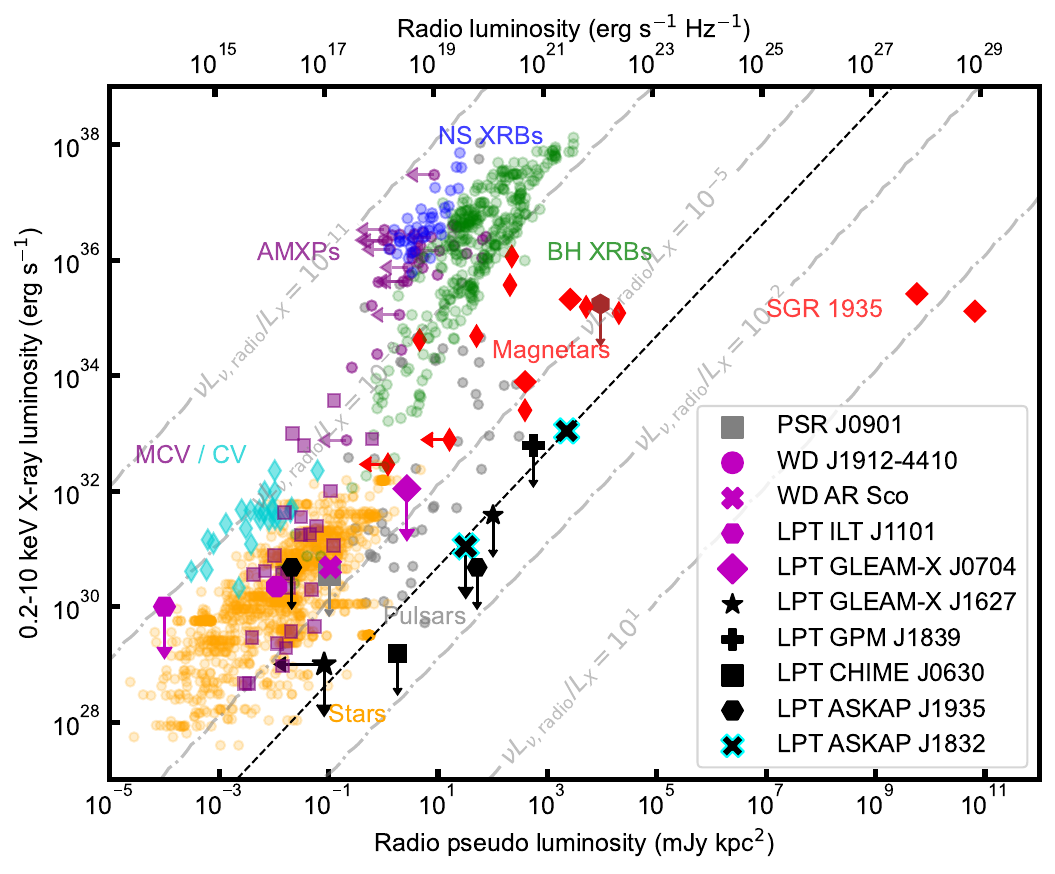}
    \caption{Radio vs. X-ray luminosity for a range of Galactic radio transients. ASKAP J1832 was shown as highlighted crosses. Sources of interest are labelled, including the long period pulsar J0901--4046 in grey; white dwarfs AR Sco, J1912--4410 in magenta; LPTs potentially associated with white dwarfs, ILT J1101+5521 and GLEAM-X J0704--37, also in magenta; the LPTs GLEAM-X J1627--52, GPM J1839, CHIME J0630+25, and ASKAP J1935+2148 in black.}
    \label{fig:J1832_xray_radio_parameter}
\end{figure}

Magnetars, a subclass of isolated neutron stars mostly powered by the decay of their strong magnetic fields ($B \sim10^{13-15}$\,G; see \cite{2017ARA&A..55..261K, 2021ASSL..461...97E} for recent reviews), are considered potential progenitors for some of the LPTs \cite{2020MNRAS.496.3390B,2023MNRAS.520.1872B,2024MNRAS.533.2133C}.
The radio properties of \lpt\ are qualitatively in line with typical radio-loud magnetars --- varied pulse morphology, high linear polarisation fraction, constantly changing radio spectrum (often rather flat compared to the typical radio pulsar population), and high radio flux variabilities \citep{2006Natur.442..892C}. The observed X-ray outburst is also typical of magnetars \citep{2018MNRAS.474..961C}. \\

The combination of low spin-down luminosity, and low quiescent X-ray luminosity would however make \lpt\ unique amongst the known magnetar population (see Methods). Models of magnetar ageing \citep{2013MNRAS.434..123V, 2023MNRAS.518.1222D} suggest that many of the known population would have spin-down and quiescent X-ray luminosities consistent with \lpt\ after $\gtrsim 0.5$\,Myr (see Extended Data Figure~\ref{fig:mag_luminosities_models}). 
However, producing bright and transient radio emission from such old magnetars is challenging ---
models of magnetically powered radio emission require large magnetic fields ($\gtrsim 10^{14}$\,G) \citep{2024MNRAS.533.2133C}, which, if evolving in the crust, imply a brighter quiescent X-ray luminosity than is observed. Alternatively, it could host an unusual core-dominated magnetic field (see Figure~8 in \citep{2022ApJ...940...72R}) similar to what is speculated to power PSR~J0901$-$4046 \citep{2022NatAs...6..828C}, requiring a revision to models of magnetic field evolution in neutron stars \citep{2024arXiv241108020L}.
Therefore, if \lpt\ is an old magnetar, explaining the radio emission challenges existent models.

\backmatter
\section*{Declarations}

\begin{itemize}

\item \textbf{Funding}
N.R. is supported by the European Research Council (ERC) via the Consolidator Grant “MAGNESIA” (No. 817661) and the Proof of Concept ``DeepSpacePulse" (No. 101189496), by the Catalan grant SGR2021-01269, and by the program Unidad de Excelencia Maria de Maeztu CEX2020-001058-M. 
T.B.\ acknowledges financial support from the Framework per l’Attrazione e il Rafforzamento delle Eccellenze (FARE) per la ricerca in Italia (R20L5S39T9).
D.K. is supported by NSF grant AST-1816492.
Z.Wadiasingh, J.H., and G.Y. acknowledge support by NASA under award number 80GSFC21M0002. 
P.B. acknowledges support from a NASA grant 80NSSC24K0770, a grant (no. 2020747) from the United States-Israel Binational Science Foundation (BSF), Jerusalem, Israel and by a grant (no. 1649/23) from the Israel Science Foundation.
A.J.C. acknowledges support from the Oxford Hintze Centre for Astrophysical Surveys which is funded through generous support from the Hintze Family Charitable Foundation.
Basic research in radio astronomy at the U.S. Naval Research Laboratory is supported by 6.1 Base funding. Construction and installation of VLITE was supported by the NRL Sustainment Restoration and Maintenance fund. 
M.G. is supported by the Australian Government through the Australian Research Council’s Discovery Projects funding scheme (DP210102103), and through UK STFC Grant ST/Y001117/1. M.G. acknowledges support from the Inter-University Institute for Data Intensive Astronomy (IDIA). IDIA is a partnership of the University of Cape Town, the University of Pretoria and the University of the Western Cape. For the purpose of open access, the author has applied a Creative Commons Attribution (CC BY) licence to any Author Accepted Manuscript version arising from this submission.
N.H.W.\ is the recipient of an Australian Research Council Future Fellowship (project number FT190100231)
M.C.\ acknowledges the support of an Australian Research Council Discovery Early Career Research Award (project number DE220100819) funded by the Australian Government.
C.W.J.\ acknowledges support by the Australian Government through the Australian Research Council's Discovery Projects funding scheme (project DP210102103).
M.E.L.\ receives support from the ARC Discovery Early Career Research Award DE250100508.
The Chandra X-ray observation presented in this paper and partial funding for K. Mori are supported by SAO grant GO3-24121X.
M.P.T. acknowledges financial support from the Severo Ochoa grant CEX2021-001131-S and from the National grant PID2023-147883NB-C21, funded by MCIU/AEI/ 10.13039/501100011033.
K.R.\ thanks the LSST-DA Data Science Fellowship Program, which is funded by LSST-DA, the Brinson Foundation, and the Moore Foundation; Their participation in the program has benefited this work.
A.T.D., R.M.S., Y.W., J.N.J.S.\ and Y.W.J.L. acknowledge support through Australia Research Council Discovery Project DP220102305. 
R.T. acknowledges support from funding provided by the National Aeronautics and Space Administration (NASA), under award number 80NSSC20M0124, Michigan Space Grant Consortium (MSGC).
F.W.\ was supported by the National Natural Science Foundation of China (grant No. 12273009).
Parts of this research were conducted by the Australian Research Council Centre of Excellence for Gravitational Wave Discovery (OzGrav), through project numbers CE170100004 and CE230100016.

\item \textbf{Author Contributions} Z.W. and N.R. drafted the manuscript with suggestions from co-authors. E.L., Z.W., K.W.B., and Y.W. discovered the source. 
Z.W. is the PI of the MeerKAT data and the GMRT data. A.T.D. is the PI of the VLBA data. T.M. is the PI of the ATCA data (C3363). 
E.L., Z.W., and A.A. further reduced the ASKAP data to produce dynamic spectra for detections and non-detections. Z.W. and N.H.W. reduced and analysed the MeerKAT imaging data. M.C. helped propose the MeerKAT observation and analyse MeerKAT data. M.G. reduced the MeerKAT \hi{} data. Z.W., A.A., K.R., J.P., and Y.W. observed, reduced, and analysed the ATCA data. A.B. and Z.W. reduced and analysed the GMRT data. A.A. reduced the archival VLA data. T.E.C., W.M.P., and E.J.P. developed and maintained the VLITE data archive and performed the VLITE archive search, time slicing, imaging and cataloguing. A.T.D. reduced and analysed the VLBA data.  
D.L.K., S.K.O., V.K. and M.M.K. observed and analysed the infrared data. K.M. is the PI of the {\em Chandra} data. T.B., N.R. reduced and analysed {\em Chandra} data. T.B. and N.R. reduced and analysed archival XMM-Newton and {\em Swift} data. N.R., D.L.K., Z.W., and H.Q. helped in asking for the EP data. N.R., as member of the EP collaboration, reduced and analysed the EP data. 
D.L.K., K.C.D., and R.T. performed multi-wavelength archive searches.
A.Z., S.J.M., D.L.K., and R.M.S. performed the radio timing analysis. A.Z. performed the radio polarimetry data reduction. J.R.D. performed \hi{} absorption line analysis. N.H.W. performed the SNR association analysis.
Z.Wadiasingh, J.H., A.J.C., B.P., G.Y., M.E.L., M.P.T., F.W., and Z.Z. contributed to discussions about the nature and emission mechanism of the source.
K.W.B., A.D., C.W.J., and R.M.S. are the PIs of CRACO. Z.W., M.G., A.J., X.D., J.N.J.S., Y.W.J.L., J.T., and N.T. contributed to the design and commissioning of CRACO. 
T.M., and D.L.K. are the PIs of VAST, and D.D. and L.N.D. are the Project Scientists of VAST. The PIs and builders of VAST and CRACO coordinated the initial investigation of the source.

\item \textbf{Data availability} The data that support the findings of this study will be made available on Zenodo. All the ASKAP data are publicly available via CASDA (\url{https://research.csiro.au/casda/}). The MeerKAT data used in this study are available via the SARAO archive (\url{https://archive.sarao.ac.za}) under project ID DDT-20240213-AW-01. The ATCA data used in this study are available via the Australia Telescope Online Archive (\url{https://atoa.atnf.csiro.au/}) under project ID C3363. Other specific data are available on request from Z.W.

\item \textbf{Code availability} The code that support the findings of this study will be made available on Zenodo. Specific scripts used in the data analysis are available on request from Z.W.

\item \textbf{Acknowledgements} 
We thank Bryan Gaensler, Shi Dai and Francesco Coti Zelati for valuable discussions.
We are grateful to the ASKAP engineering and operations team for their assistance in developing fast radio burst instrumentation for the telescope and supporting the survey. 
This work uses data obtained from Inyarrimanha Ilgari Bundara / the CSIRO Murchison Radio-astronomy Observatory. We acknowledge the Wajarri Yamaji People as the Traditional Owners and native title holders of the Observatory site. CSIRO’s ASKAP radio telescope is part of the Australia Telescope National Facility (https://ror.org/05qajvd42). Operation of ASKAP is funded by the Australian Government with support from the National Collaborative Research Infrastructure Strategy. ASKAP uses the resources of the Pawsey Supercomputing Research Centre. Establishment of ASKAP, Inyarrimanha Ilgari Bundara, the CSIRO Murchison Radio-astronomy Observatory and the Pawsey Supercomputing Research Centre are initiatives of the Australian Government, with support from the Government of Western Australia and the Science and Industry Endowment Fund.
CRACO was funded through Australian Research Council Linkage Infrastructure Equipment, and Facilities grant LE210100107.
We thank the staff of the GMRT that made these observations possible. GMRT is run by the National Centre for Radio Astrophysics of the Tata Institute of Fundamental Research.
We thank SARAO for the approval of the MeerKAT DDT request DDT-20240213-AW-01.The MeerKAT telescope is operated by the South African Radio Astronomy Observatory, which is a facility of the National Research Foundation, an agency of the Department of Science and Innovation.
The National Radio Astronomy Observatory is a facility of the National Science Foundation operated under cooperative agreement by Associated Universities, Inc.

This research has made use of data obtained from the Chandra Data Archive provided by the Chandra X-ray Center (CXC). We acknowledge the use of public data from the Swift data archive. This research is based on observations obtained with XMM-Newton, an ESA science mission with instruments and contributions directly funded by ESA Member States and NASA. We thank the Einstein Probe PI (Prof. Weimin Yuan) for accepting our ToO observation, Dr. Yong Chen as the FXT PI, and the Einstein Probe Science Center for performing the observations. Einstein Probe is a space mission supported
by the Strategic Priority Program of the Space Science of the Chinese Academy of Sciences (Grant No. XDB0550200), in collaboration with ESA, MPE and CNES (Grant No. XDA15310000), and the National
Key R\&D Program of China (2022YFF0711500). 

This paper includes data gathered with the 6.5 meter Magellan Telescope located at Las Campanas Observatory, Chile.

Part of this work was performed on the OzSTAR national facility at Swinburne University of Technology. The OzSTAR program receives funding in part from the Astronomy National Collaborative Research Infrastructure Strategy (NCRIS) allocation provided by the Australian Government, and from the Victorian Higher Education State Investment Fund (VHESIF) provided by the Victorian Government.
We acknowledge the use of the ilifu cloud computing facility - www.ilifu.ac.za, a partnership between the University of Cape Town, the University of the Western Cape, Stellenbosch University, Sol Plaatje University, the Cape Peninsula University of Technology and the South African Radio Astronomy Observatory. The ilifu facility is supported by contributions from the Inter-University Institute for Data Intensive Astronomy (IDIA - a partnership between the University of Cape Town, the University of Pretoria and the University of the Western Cape), the Computational Biology division at UCT and the Data Intensive Research Initiative of South Africa (DIRISA).
This work was carried out using the data processing pipelines developed at the Inter-University Institute for Data Intensive Astronomy (IDIA) and available at https://idia-pipelines.github.io. IDIA is a partnership of the University of Cape Town, the University of Pretoria and the University of the Western Cape.
This work made use of the CARTA (Cube Analysis and Rendering Tool for Astronomy) software (DOI 10.5281/zenodo.3377984 – https://cartavis.github.io). This research has made use of the NASA Astrophysics Data System.

\item \textbf{Conflict of interest/Competing interests} The authors declare no competing interests.

\item \textbf{Ethics approval and consent to participate} Not applicable

\item \textbf{Consent for publication} Not applicable

\end{itemize}


\setcounter{figure}{0}
\captionsetup[figure]{name={\bf Extended Data Fig.}}
\setcounter{table}{0}
\captionsetup[table]{name={\bf Extended Data Table}}

\newpage
\section*{METHODS}


\subsection*{ASKAP Observations}

ASKAP is a $36\times12$-meter antenna radio interferometer located at Inyarrimanha Ilgari Bundara, the CSIRO Murchison Radio-astronomy Observatory. The ASKAP VAST survey is one of the ASKAP Surveys Science Projects aiming at detecting transient polarised radio sources. The VAST survey is conducted at a central frequency of 887.5\,MHz with a bandwidth of 288\,MHz. All four polarisation products (XX, YY, XY, YX) were recorded to allow images to be made in full Stokes parameters (I, Q, U, V).  The survey has multiple components.  One is targeting the Galactic plane.  Each Galactic field is observed every two weeks with a total integration time of 12\,minutes. The data are sampled at 9.95\,s time resolution and 1\,MHz frequency resolution.
We use both variable search \citep{2021PASA...38...54M} and circular polarisation search \citep{2021MNRAS.502.5438P} to identify interesting candidates for further investigation.

In addition to the  ASKAP VAST data, there have been 10 ASKAP Director's Discretionary Time requests to follow up \lpt. There have also been several archival ASKAP observations (totalling $\sim35$\, hours) of the field.
For all ASKAP data, we downloaded calibrated ASKAP measurement sets from \textsc{CASDA}, where \textsc{ASKAPSoft} \citep{2019ascl.soft12003G} was used for the data processing. We flagged the data where pulses were predicted to occur for each observation, and made deep field images using \textsc{WSClean} to produce deep models. We then subtracted the models, phase-rotated the visibilities, and averaged the data across baselines to produce full-resolution lightcurve and dynamic spectra. We used a simple circular Gaussian model as an approximation to correct for the primary beam attenuation. 
In the archival observations prior to the discovery, no pulsed emission was detected, despite the coverage of approximately 40 expected pulses. A $3\sigma$ upper limit of $\sim15$\,\mJy{} was placed on these non-detections.

\begin{figure}
    \centering
    \includegraphics[width=0.8\linewidth]{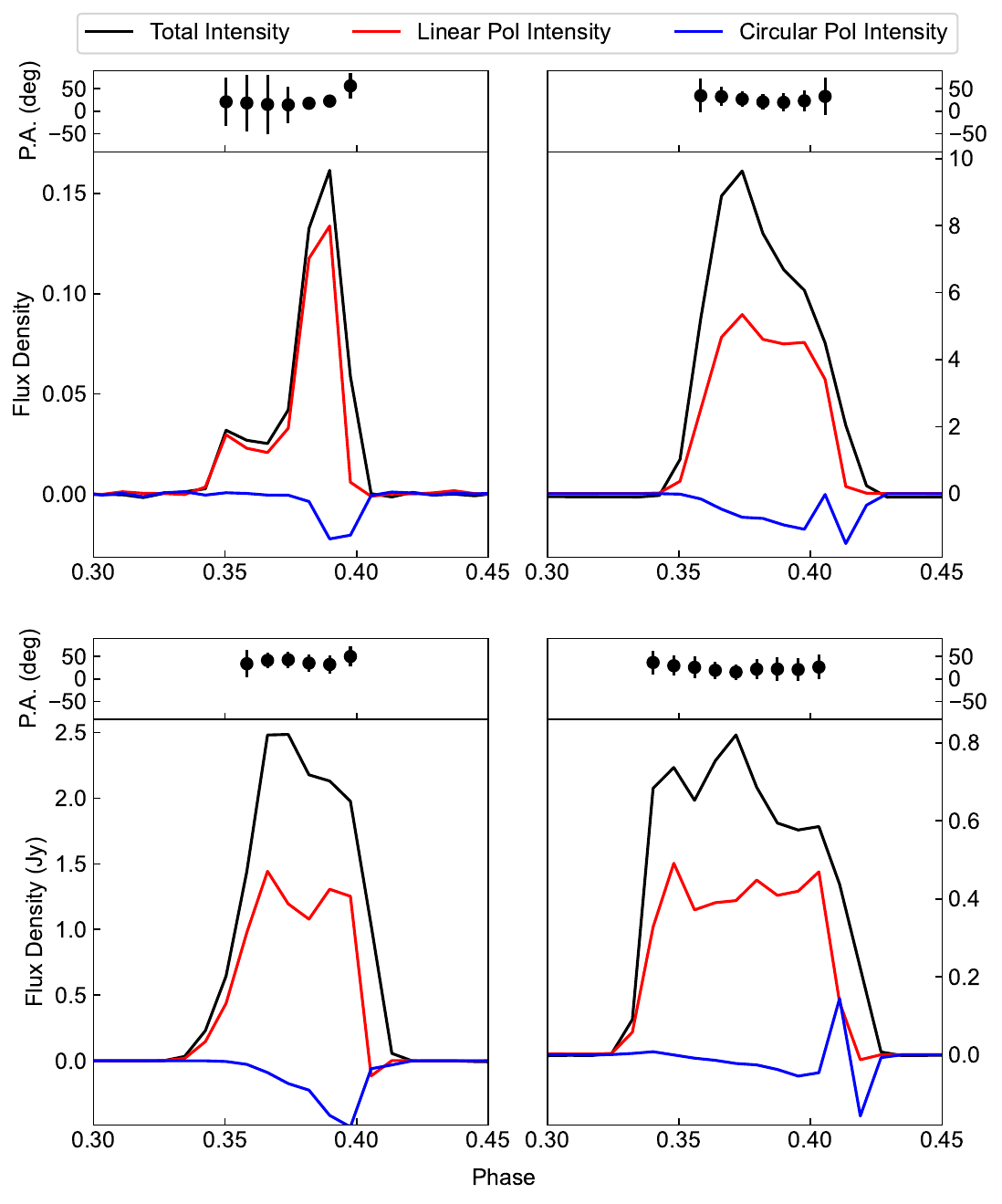}
    \caption{Folded pulse profiles from selected long observations ($>3$\,hours) with ASKAP. In each subplot, the upper panel shows the polarisation position angle (P.A.) swing of the pulse in degrees, and the lower panel shows the total intensity (black), the linear polarisation intensity (red), and the circular polarisation intensity (blue). In order to show the errors on P.A. measurement, we show $3\sigma$ error in the plot.}
    \label{fig:J1832_pol_pa}
\end{figure}

We picked up the brightest observed pulse of \lpt{} (ASKAP SB58609) to determine its rotation measure (RM). We used \textsc{RM-Tools}\footnote{\url{https://github.com/CIRADA-Tools/RM-Tools}} \citep{2020ascl.soft05003P} to perform RM-synthesize, and derived ${\rm RM} = +90.5\pm0.1\,$rad\,m$^{-2}$. We show folded full polarisation pulse profiles from selected long observations in Extended Figure~\ref{fig:J1832_pol_pa}. Detailed discussions on its polarisation properties will be presented in a later paper.  

CRACO, the newest ASKAP fast transient search backend \citep{2024arXiv240910316W}, has been operating commensally with most of the ASKAP surveys, including VAST survey since the beginning of 2024. CRACO can record visibility data at millisecond time resolution (from 1.728\,ms to 110.592\,ms). At the time of the observation when \lpt\ was discovered, CRACO was observed with a time resolution of 13.824\,ms and a frequency resolution of 1\,MHz. We used the self-calibration technique to derive the calibration solutions based on the field model generated from Rapid ASKAP Continuum Survey \citep[RACS;][]{2020PASA...37...48M} Low DR1 catalogue \citep{2021PASA...38...58H}. We phase-rotated the calibrated CRACO visibilities and then averaged them across baselines to produce dynamic spectra (see Figure~\ref{fig:J1832_discovery}). From the dynamic spectrum, we measured a dispersion measure of $458\pm14$\,\pccm{} with \textsc{pdmp} script in the \textsc{PsrCHIVE} suite.

\subsection*{ATCA Observations}

The Australia Telescope Compact Array (ATCA) is an array of six 22-m antennas at the Paul Wild Observatory. 
Following the discovery, we observed \lpt\ with ATCA at 2.1\,GHz for 6 hours on 2023~Dec~26 and 2023~Dec~29 each under the project code C3363 (see Extended Data Table \ref{tab:radio_table}). The observations were calibrated using PKS~B1934–-638 for the flux density scale and the instrumental bandpass, and PMN~J1822--0939 was used for phase calibration. We used \textsc{DsTools}\footnote{\url{https://github.com/askap-vast/dstools.git}} to reduce the visibility data and produce the dynamic spectra, while \textsc{Miriad} \citep{1995ASPC...77..433S} was used for calibration and continuum imaging, and \textsc{CasaCore}\footnote{\url{https://github.com/casacore/casacore}} was used for manipulating the visibility.
We identified 15~pulses in the observations, with an average pulse flux density of $\sim12\,$\mJy.

\subsection*{GMRT Observations}

The Giant Metrewave Radio Telescope (GMRT) is an array of thirty 45-m radio telescopes, which can observe over a wide frequency range from $\approx$ $100$\,MHz to $1.6$\,GHz. We conducted an observation under project code ddTC341 for \lpt\ on 2024~Jun~17 in sub-array mode. We divided the whole array equally into three sub-arrays, with each sub-array observed at different bands, simultaneously observing the source from 300\,MHz to 1.5\,GHz (from band3 to band5). The data are sampled at 2.6\,s time resolution and 0.195\,MHz frequency resolution.
We used 3C48 for the flux density scale and the instrumental bandpass, and PMN~J1822--0939 for the phase calibration. 
We used \textsc{CASA} for calibration and continuum imaging.
The band 5 observation (1060--1460\,MHz) also covered the frequency range of Galactic neutral hydrogen (H{\sc i}) emission and absorption through the 21-cm transition. 
For H{\sc i} line analysis, instead of averaging the visibilities, we made spectral line cubes for each integration to get the dynamic spectra (with a \textsc{tclean} parameter \texttt{robust}=0.5, and all baselines).

\subsection*{MeerKAT Observations}

The MeerKAT radio telescope is an array of 64 13.5-m radio telescopes. We carried out two observations for \lpt{} with MeerKAT under project code DDT-20240213-AW-01.

On 2024 Feb 13, we conducted an 8-hour observation in sub-array mode. We divided the whole array equally into two sub-arrays, split equally between the UHF- (from 565 to 1060\,MHz) and S0-band (from 1750 to 2625\,MHz), respectively. The data are sampled at 2\,s time resolution.
On 2024 Aug 11, we conducted another 4-hour observation in `32K' spectral-line mode (32768 channels with width 26.123~kHz) at L-band (856--1712\,MHz). The data are sampled at 4\,s time resolution for the second observation. PKS1934-638 and J1833-2103 were used as the bandpass/flux and phase calibrator, respectively.

For continuum data reduction, we reduced the data using {\sc Oxkat}\footnote{\url{https://github.com/IanHeywood/oxkat}} \citep[v0.3;][]{2020ascl.soft09003H}, where the Common Astronomy Software Applications \citep[CASA;][]{2007ASPC..376..127M} package and {\sc Tricolour}\footnote{\url{https://github.com/ska-sa/tricolour}} \citep{2022ASPC..532..541H} were used for measurement sets splitting, cross-calibration, flagging, {\sc CubiCal}\footnote{\url{https://github.com/ratt-ru/CubiCal}} \citep{2018MNRAS.478.2399K} was used for self-calibration, and {\sc Wsclean} \citep{2014MNRAS.444..606O} was used for continuum imaging. All processes were executed with {\sc Oxkat} default settings. 
We subtracted the sky model derived from {\sc Wsclean} and generated the dynamic spectra of the target by averaging all baselines longer than 800\,m. 

For H{\sc i} absorption, we processed only the calibrator data and on-pulse time bins. We chose the time range when the pulse flux density $>20\,$\mJy{}, totalling $\sim14\,$mins of data.
{\sc processMeerKAT} \citep{2021ursi.confE...4C} was used for calibration and flagging with default settings, and CASA for spectral-line imaging. Spectral cube imaging was performed without baselines $<$ 500~m, to filter out extended \hi\ emission and absorption, with a robust weighting of 0.5 and cleaned to 3$\times$ the measured RMS of the continuum cube (4.5\,mJy). 

\begin{figure}
    \centering
    \includegraphics[width=0.9\linewidth]{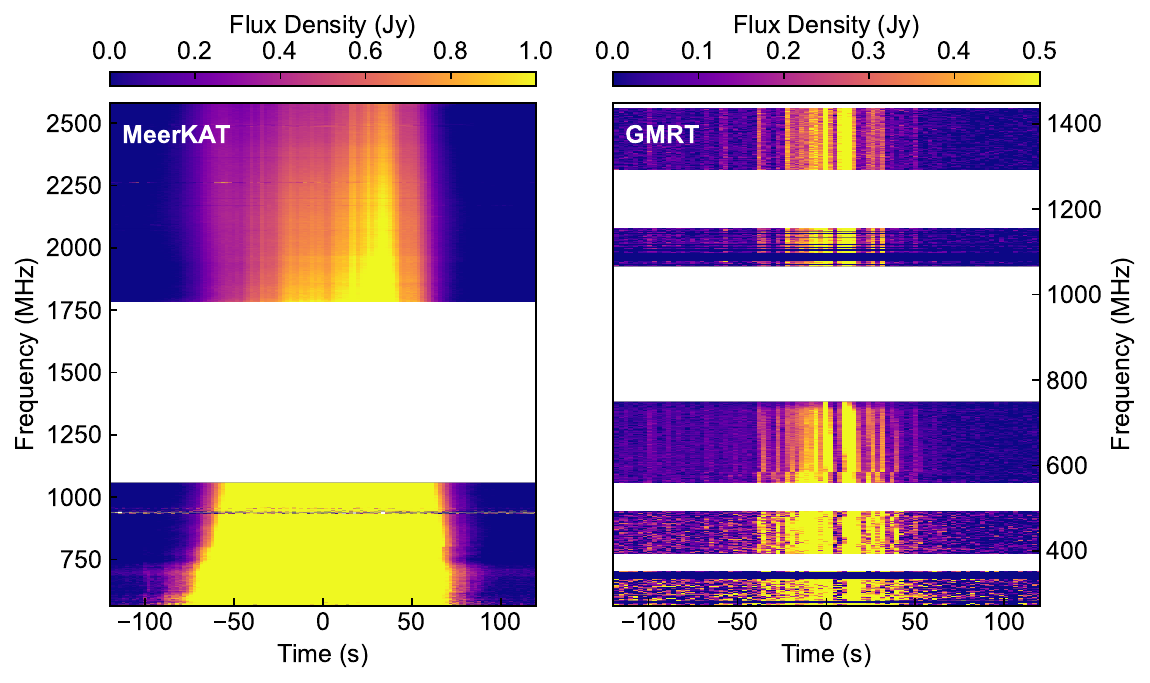}
    \caption{Simultaneous Wide-Frequency Coverage Observation of \lpt. Left: MeerKAT observation from 565--2625\,MHz; Right: GMRT observation from 272--1450\,MHz. Both dynamic spectra are de-dispersed with a DM of 458\,\pccm{}.}
    \label{fig:wide_radio}
\end{figure}

\subsection*{VLBA Observations}

In February 2024, we were awarded $10\times$1 hours of Director's Discretionary time on the Very Long Baseline Array (VLBA) under project code BD275.  
Making use of the dynamic scheduling capabilities of the VLBA, we ensure that each observation commenced 5 -- 10 minutes before one pulse would arrive, ensuring that the telescope would be on-source during the pulse, and that a second pulse would be observed close to the end of the observation.
We used ICRF J182537.6$-$073730 (hereafter J1825$-$0737) as a primary phase reference calibrator, with a cycle time of 7 minutes, including 3.8 minutes on-target, centred on the predicted pulse arrival time. We also observed ICRF J183519.5$-$111559 (hereafter J1835$-$1115) as a secondary calibrator. In the $\sim$30 minutes of time during the centre of the one-hour observation during which no emission from \lpt\ is expected to be visible, we observed the bandpass calibrator ICRF J180024.7+384830 (hereafter J1800+3848), along with six other calibrator sources within $\sim$5 degrees of \lpt. Data was correlated using the DiFX software correlator \citep{2011PASP..123..275D} with 1 second time resolution and 62 kHz frequency resolution.

As both the spectral index of \lpt\ and the degree of angular broadening along this (low Galactic latitude) sightline were poorly constrained at the time shortly after the source discovery, we made two initial observations, one at 1.5 GHz and one at 4.5 GHz. 
Despite showing a steep spectral index, \lpt\ was strongly detected in both the 1.5 GHz and 4.5 GHz observations. However, the heavy scatter broadening at 1.5 GHz (deconvolved source size $\sim$25 mas) led us to choose 4.5 GHz as the preferred frequency for subsequent observations. Observations were made approximately monthly between 2024 March and 2024 September (with several observations scheduled to occur until 2025 January).

Data reduction was performed using the \verb+psrvlbireduce+ pipeline\footnote{\url{https://github.com/dingswin/psrvlbireduce}} that makes use of AIPS via the ParselTongue interface \citep{2006ASPC..351..497K}. Briefly, we applied {\em a priori} amplitude calibration, and performed time-invariant delay and bandpass calibration using J1800+3848 followed by delay, phase, and amplitude calibration refinement using simple phase referencing to J1825$-$0737. As J1825$-$0737 and J1835$-$1115 are quasi-colinear with \lpt, future work could employ 1D phase interpolation \citep[e.g.,][]{2024ApJ...971L..13D}, but this has not been employed in the reduction presented here. Similarly, no attempt was made to make use of calibrator scans between the two pulses of \lpt\ at this stage. We applied a simple top-hat weighting to the radio pulse of \lpt, first imaging the data in 60s increments, and then flagging all increments where the source was not detected. Future work could further improve the detection significance by compensating the visibility amplitudes and weights for the variation of the pulse strength during the unflagged time regions.

We detect \lpt\ in every observing epoch, with significances ranging from 35\,$\sigma$ at the earliest epochs to 8\,$\sigma$ at the most recent. We note that \lpt\ is still moderately scatter-broadened at 4.5 GHz (angular size 3 -- 4 mas, consistent with expectations of $\lambda^2$ scaling from the 1.5 GHz measurement due to multipath propagation, and slightly larger than the synthesised beam size of the VLBA at this frequency).

In order to estimate the transverse velocity of the source, we fix the parallax to 0.22 mas, corresponding to the approximate DM-based distance of 4.5\,kpc. We note however that the fitted proper motion is only weakly sensitive to the assumed parallax for any reasonable distance ($>$1 kpc, corresponding to a parallax in the range 0 to 1 mas). Based on the six detections between February and September, we measure a best-fit proper motion of $-$1.3 mas yr$^{-1}$ in right ascension and $-$1.3 mas yr$^{-1}$ in declination, corresponding to a relatively low transverse velocity of 40 km\,s$^{-1}$ at an assumed distance $D=4.5$\,kpc. However, the scatter in the position residuals is of order 1 mas, considerably larger than the statistical astrometric precision (which ranges from 0.05 to 0.3 mas at each epoch). While improved calibration (as suggested above) may be able to reduce this scatter in a future analysis, for the purposes of using our current dataset to estimate an upper limit to the transverse velocity we added an assumed systematic uncertainty term of 1 mas in both right ascension and declination. This resulted in a $\chi^2$ of 8.4 for 8 degrees of freedom, indicating a reasonable fit. The 95\% confidence interval for the total proper motion ranges from 0 to 8 mas\,yr$^{-1}$, allowing a maximum transverse velocity of $180\left({D}/{4.5\,{\rm kpc}}\right)$\,km\,s$^{-1}$. If \lpt\ remains bright, a highly significant measurement of the proper motion (and possibly also a constraint on the parallax) should become possible within the next 12 months.

Finally, the best-fit reference position for \lpt\ on MJD 60400 is 18h32m48.45889s $-09^{\circ}11'15.2973''$. The formal uncertainty on this position is approximately 0.5 mas, and the position of the calibrator source J1825$-$0737 is known to 0.25 mas in the Radio Fundamental Catalog\footnote{\url{https://astrogeo.org/sol/rfc/rfc_2024c/}}; however, residual phase referencing errors and the possibility for frequency-dependent intrinsic structure and/or time-dependent changes in the (frequency dependent) angular broadening of J1825$-$0737 mean that we take a more conservative estimate on the reference position uncertainty of 5 mas, at least until cross-checks against additional nearby calibrators are made.

\subsection*{VLA}

VLA observations that contain the position of \lpt{} within the half-with of the maximum power have been selected. This has resulted in the observations taken under the project codes 13A-120, 19B-255, and 21A-285. There were five different pointings, each two minutes in duration that were taken at L-band (1-2\,GHz) in 13A-120, two observations each spanning half an hour at L-band under 19B-255 and one observation lasting 42\,minutes at S, X bands in 21A-285 (although this observation was broken into two parts, with observations carried at each band for 21\,minutes). Data for these is calibrated using 3C286 as the bandpass calibrator and J1822$-$0938 as the phase calibrator. Calibrated observations are then cleaned using \texttt{tclean} procedure in \textsc{CASA}. During imaging, we excluded baselines shorter than 250\,m to mitigate extended emission in the final image and sky model.

We did not detect a point source in any of the time-integrated images. We then phase-rotated the observations to the target location and generated the dynamic spectra of the target by averaging all the baselines. The time resolution for the VLA observations is 5\,s and hence we formed the dynamic spectra at this resolution and examined the data for any fainter burst-like emission. However, \lpt{} remained undetected, with a $3\sigma$ upper limit of $\sim6\,$\mJy{}.

\subsection*{VLITE}
 
The VLA Low-band Ionosphere and Transient Experiment (VLITE) \cite{2016ApJ...832...60P,2016SPIE.9906E..5BC} operates
commensally on the National Radio Astronomy Observatory's Karl G.\
Jansky Very Large Array (VLA). VLITE records and correlates data with a sample time of 2 seconds across an effective bandwidth of 40 MHz centred on 340 MHz. It has been operational on up to 18 VLA antennas during nearly all VLA observations since 2017-07-20. VLITE accumulates roughly 6,000 hours of commensal data each year across all VLA configurations. VLITE data are processed on a daily basis using a customized and automated Obit-based \cite{2008PASP..120..439C} calibration and imaging pipeline \cite{2016ApJ...832...60P} which produces flagged and calibrated visibility datasets and self-calibrated images. These VLITE images and associated META data are then passed through the VLITE Database Pipeline (VDP) \cite{2019ASPC..523..441P} to populate a Structured Query Language (SQL) database containing the archive of catalogued sources and image information.

We queried the VDP database for archival images in which \lpt\ lies within 2.0$^\circ$ of the center of the field of view for VLITE observations taken in the VLA A, B, or C configurations. The half-power radius of the VLITE primary beam response is roughly 1.23$^\circ$ at the VLITE center frequency but the system is sensitive and well characterized for emission from sources that are at much larger distances from the field center. We identified 342 data sets that matched our search criteria, of which 29 were from observations more recent than the \lpt\ discovery date of 2023-12-08. No detections of the target were identified in VLITE data for observations prior to the initial VAST detection. For this paper, we concentrate on the data from 2024-02-15 where we have detected two consecutive bursts close in time to the {\em Chandra} observations. These observations were taken in the VLA's C configuration and have a spatial resolution of roughly 60 arcsec $\times$ 30 arcsec.

The consecutive bursts of \lpt\ detected by VLITE occurred within the field of view of two different VLA pointings where \lpt\ was located at offsets of 0.49$^\circ$ and 1.40$^\circ$ from the centre of the two fields of view. For the two VLITE datasets, a model of the sky around the target position which had been generated by the standard processing pipeline was subtracted from the visibilities to produce a residual dataset. The residual visibilities were then phase-shifted to the position of \lpt\ and a small field of view centered on the source was imaged on time steps of 2 seconds, 10 seconds, and 30 seconds. Each image was catalogued at the source position using the PyBDSF \cite{2015ascl.soft02007M} and the fitted measurements were corrected for the instrumental response appropriate to the specific image. The instrumental response-corrected noise level in the 2-second images was on average 251 mJy/beam while for the 10- and 30-second images the average noise levels were 211 mJy/beam and 76 mJy/beam, respectively.

VLITE observations on 2024-02-15 missed the initial portion of the first of the two consecutive pulses. The maximum peak flux densities on 10-second images are 7427 $\pm$ 1159 mJy beam$^{-1}$ and 8766 $\pm$ 1370 mJy beam$^{-1}$ where a 15\% flux calibration uncertainty is included. 

\begin{landscape}
\begin{table}[]
  \centering
    \begin{tabular}{ccccccccc}
    \hline\hline
    Start Time & Duration & Telescope & Frequency & $S_{\rm peak, max}$ & $\alpha$ & Npulse & ObsID & Note \\
    (UTC) & (min) & & (MHz) & (\mJy{}) & & & & \\
    \hline
    2023-12-08 6:50:37 & 12 & ASKAP & 744--1032 & $1870\pm6$ & $-1.5\pm0.1$ & 1 & 55237 & \\
    2023-12-26 23:33:34 & 360 & ATCA & 1076--3124 & $17\pm1$ & $-2.6\pm0.6$ & 8 & C3363 & \\
    2023-12-29 21:36:15 & 360 & ATCA & 1076--3124 & $14\pm1$ & $<-1.3$ & 7 & C3363 & \\
    2024-02-01 2:46:53 & 120 & ASKAP & 788--1076 & $254\pm7$ & $-0.6\pm0.1$ & 3 & 58387 & \\
    2024-02-03 21:03:04 & 480 & ASKAP & 788--1076 & $17956\pm5$ & $-1.4\pm0.1$ & 11 & 58609 & \\
    2024-02-06 20:51:42 & 480 & ASKAP & 788--1076 & $4271\pm5$ & $-1.6\pm0.1$ & 10 & 58753 & \\
    2024-02-15 14:03:08 & 2 & VLA & 320--360 & $7877\pm1231$ & ... & ... & & VLITE\\
    2024-02-15 14:46:31 & 3 & VLA & 320--360 & $9234\pm1443$ & ... & 1 & & VLITE\\
    2024-02-20 2:55:55 & 440 & MeerKAT & 565--1060 & $4371.6\pm0.3$ & $-0.4\pm0.1$ & 10 & 1708396251 & \\
     & & & 1750--2625 & $1184\pm1$ & $-1.4\pm0.1$ & & 1708396132 & \\
    2024-02-29 2:47:43 & 60 & ASKAP & 788--1076 & $1688\pm3$ & $-1.2\pm0.1$ & 1 & 59605 & \\
    2024-03-01 21:34:51 & 60 & ASKAP & 788--1076 & $1518\pm2$ & $-1.2\pm0.2$ & 1 & 59659 & \\
    2024-03-15 18:18:01 & 510 & ASKAP & 788--1076 & $1297\pm11$ & $-1.6\pm0.1$ & 11 & 60091 & \\
    2024-03-18 21:21:52 & 12 & ASKAP & 744--1032 & $1233\pm2$ & $-1.5\pm0.1$ & 1 & 60182 & \\
    2024-03-24 22:29:50 & 12 & ASKAP & 744--1032 & $1169\pm2$ & $-2.2\pm0.1$ & ... & 60333 & \\
    2024-04-04 22:03:15 & 12 & ASKAP & 744--1032 & $1522\pm4$ & $-1.4\pm0.1$ & 1 & 60804 & \\
    2024-05-19 19:16:38 & 12 & ASKAP & 744--1032 & $422\pm2$ & $-1.4\pm0.1$ & ... & 62473 & \\
    2024-06-17 20:52:22 & 190 & GMRT & 272--500 & $987\pm18$ & $-0.3\pm0.1$ & 4 & 16878 & \\
     &  & GMRT & 550--760 & $990\pm3$ & $-1.1\pm0.1$ & & & \\
     &  & GMRT & 1050--1450 & $772\pm4$ & $-0.6\pm0.1$ & & & \\
    2024-06-26 11:45:58 & 180 & ASKAP & 788--1076 & $105\pm3$ & $-0.5\pm0.2$ & 4 & 63296 & \\
    2024-07-06 16:45:42 & 12 & ASKAP & 744--1032 & $52\pm4$ & ... & 1 & 63498 & \\
    2024-07-30 9:30:00 & 540 & ASKAP & 788--1076 & $121\pm2$ & $-0.7\pm0.1$ & 12 & 64280 & \\
    2024-08-02 9:30:03 & 540 & ASKAP & 788--1076 & $104\pm4$ & $-1.2\pm0.1$ & 12 & 64328 & \\
    2024-08-03 9:00:01 & 480 & ASKAP & 788--1076 & $58\pm5$ & $-0.6\pm0.2$ & 10 & 64345 & \\
    2024-08-11 17:19:17 & 215 & MeerKAT & 856--1712 & $63\pm2$ & $-1.4\pm0.1$ & 5 & 1723396099 & \\
    \hline\hline
    \end{tabular}
    \caption{All of the radio observations with \lpt{} detections. $S_{\rm peak, max}$ is the maximum peak flux density of all pulses detected in the given observation. $\alpha$ is the radio spectral index of the pulse for each detection. The missing values mean that we could not establish a good fit of the spectral index due to the not enough signal-to-noise ratio. Npulse is the number of pulses detected in the observation. The rows with missing values correspond to the observation when no complete pulse was detected. }
    \label{tab:radio_table}
\end{table}
\end{landscape}


\begin{figure}
    \centering
    \includegraphics[width=0.8\linewidth]{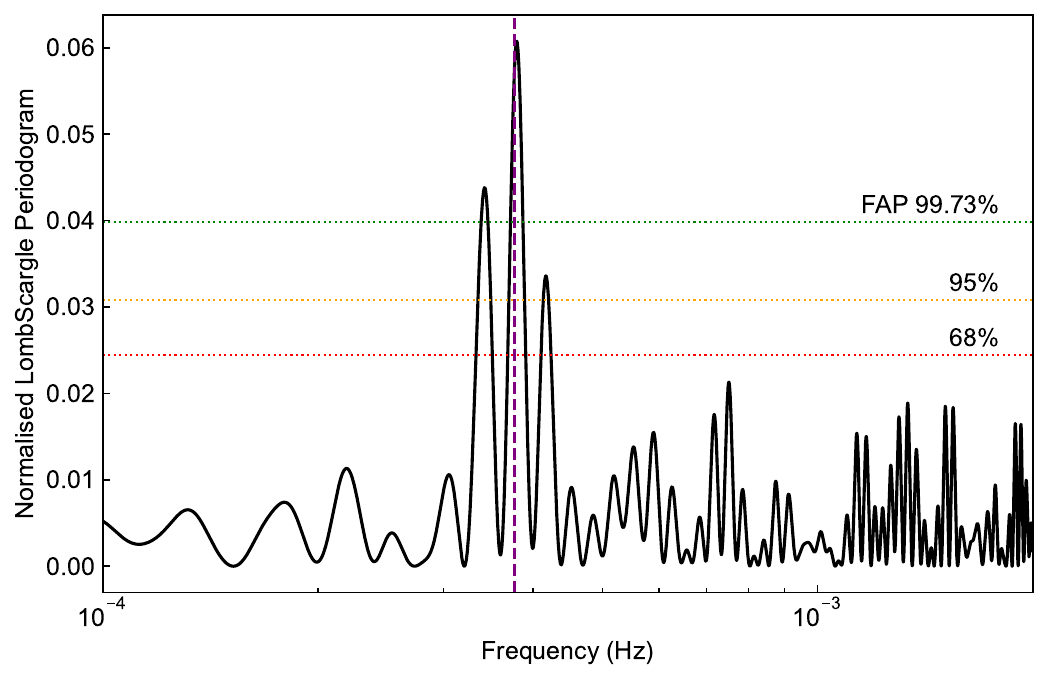}
    \caption{Normalised LombScargle Periodogram for {\em Chandra} Observations on 2024 February 14. Horizontal lines show the false alarm probabilities at 3$\sigma$ (green), 2$\sigma$ (orange), and 1$\sigma$ (red). The purple vertical line shows the best frequency we fit from radio observations.}
    \label{fig:Xray_LS}
\end{figure}

\subsection*{Chandra Observations}

\lpt\ was observed 3 times by the {\it Chandra} X-ray Observatory, for approximately 30 ksec between February and August 2024. All observations were performed with the ACIS imaging array (ACIS-I; obsIDs 29265, 26682, and 29266). For all Chandra data, we downloaded and uniformly reprocessed the archival data with CIAO v4.15 and CALDB v4.10.2, and the level 2 event ﬁle was created for each ObsID, with the arrival time of each event corrected to the Solar system barycentre (i.e. Temps Dynamique Barycentrique time) by using the CIAO tool axbary. Then we generated for each observation an exposure map as well as point spread function (PSF) maps with an enclosed count fraction (ECF) of 90 per cent. 
For each ObsID, source counts were extracted from the 90 per cent enclosed counts radius (ECR), while background counts were extracted from a concentric annulus with inner-to-outer radii of 2–4 times the 90 per cent ECR, excluding any pixel falling within two times the 90 per cent ECR of neighbouring sources.

\lpt\ was detected in the first two observations and then became too faint to be detected in the last observation. We extract the light curves for the two observations and perform a blind-period search using the Lomb-Scargle periodogram \cite{1976Ap&SS..39..447L}. A periodic signal at $2634_{-64}^{+71}$ seconds ($\sim$ 43.9 mins) with significance higher than $3\sigma$ (see Extended Data Figure~\ref{fig:Xray_LS}), the period of which is close to the radio period ($\sim$ 44.3 mins).

The source and background spectra are extracted from the same aperture by using the CIAO tool specextract. Then we binned the spectra to one count per bin for analysis and C statistics to fit the resultant spectra using XSPEC v12.12.0. We adopt a spectral model consisting of a power-law continuum and an unknown line-of-sight absorption (tbabs in XSPEC) to fit the spectra. 
For all the fits we used photoelectric cross-sections derived from \citet{1992ApJ...400..699B}, and assuming solar abundances from \citet{2003ApJ...591.1220L}.

We fit the X-ray spectra with two different models. And the fitting results do not change significantly when assuming different abundances and cross-sections, or binning differently.  
Neither of these models gives a reasonable constraint for the Galactic absorption, mainly due to the lack of photons $\lesssim$ 2 keV, as shown in Figure~\ref{fig:contour}. The fitting results of $N_{\rm H}$ have large uncertainties, way beyond the total $N_{\rm H}$ estimated in the direction of the source throughout the Galaxy, which is about $1.8 \times 10^{22} \rm cm^{-2}$ from HI4PI \citep{2016A&A...594A.116H}. This value is also consistent with that implied from the  $N_{\rm H}$-DM relation \citep{2013ApJ...768...64H}, $1.4^{+0.6}_{-0.1}\times10^{22}\,{\rm cm}^{-2}$.

Thus, we present two groups of best-fit results in Extended Data Table \ref{tab:xfit}, one with $N_{\rm H}$ fixed at $\rm 1.8 \times 10^{22}~cm^{-2}$ and the other with $N_{\rm H}$ treated as a free parameter. 
Based on the best-fit model with fixed $\rm N_{H}$, we estimate a phase-average luminosity of $L_{\rm X} = 7.4_{-3.5}^{+3.6} \times 10^{32} (\frac{d}{4.5\,{\rm kpc}})^2$\,~erg~s$^{-1}$ between 1 to 10 keV, and an X-ray luminosity of $L_{\rm X, peak} = 2.8 \times 10^{33} \left(\frac{d}{4.5\,{\rm kpc}}\right)^2$ergs$^{-1}$ at the peak of the X-ray pulse profile.

In the third observation, the source was not detected anymore, falling below the detection limit. We derived a 3$\sigma$ upper limit on the source counts of $<8.53\times10^{-4}$ c/s (see the following section for upper limit estimation).

\begin{figure}[htbp]
    \centering
    \includegraphics[width=0.92\textwidth]{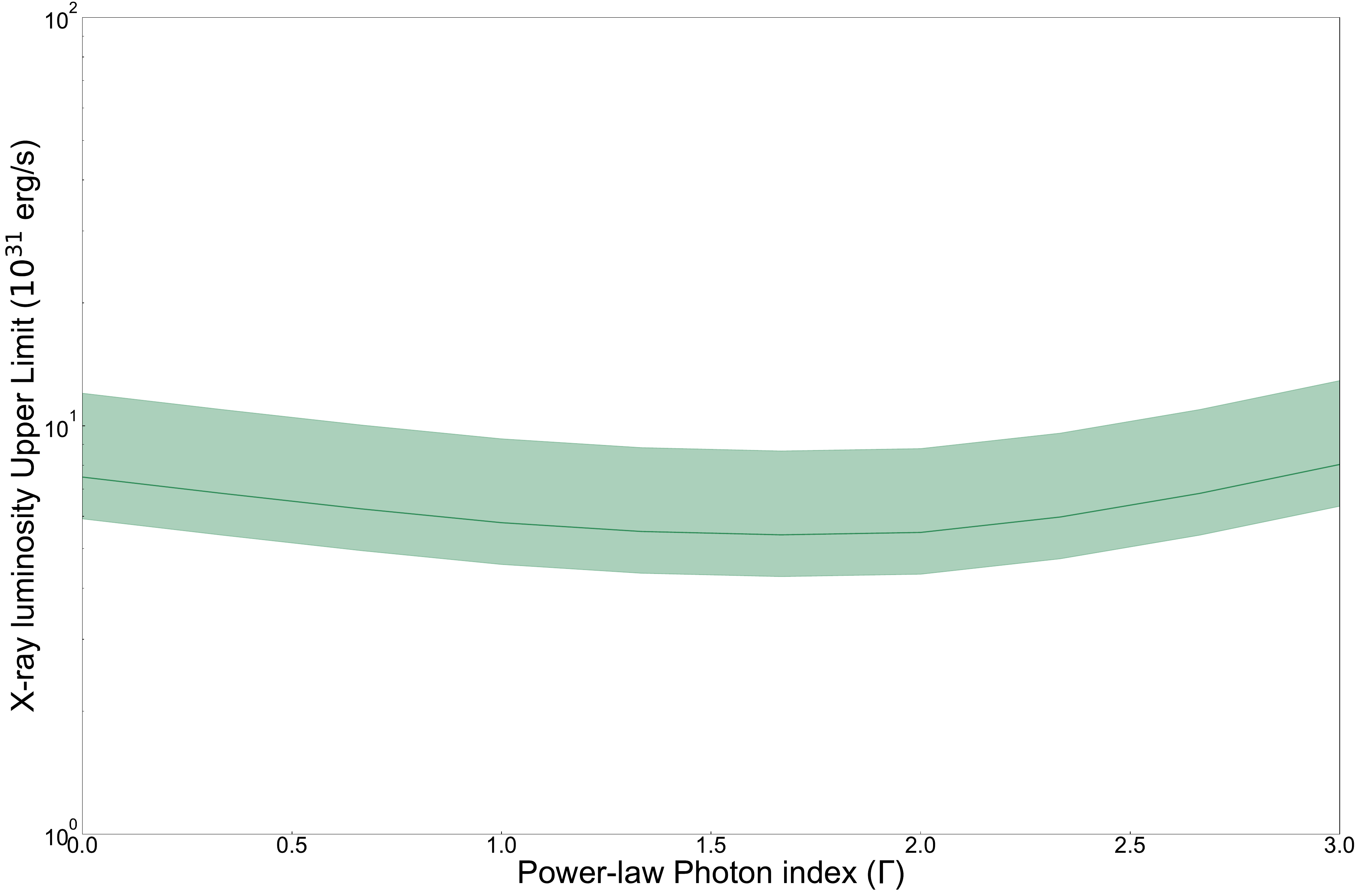} 
    \hspace{0.01\textwidth}
    \includegraphics[width=0.92\textwidth]{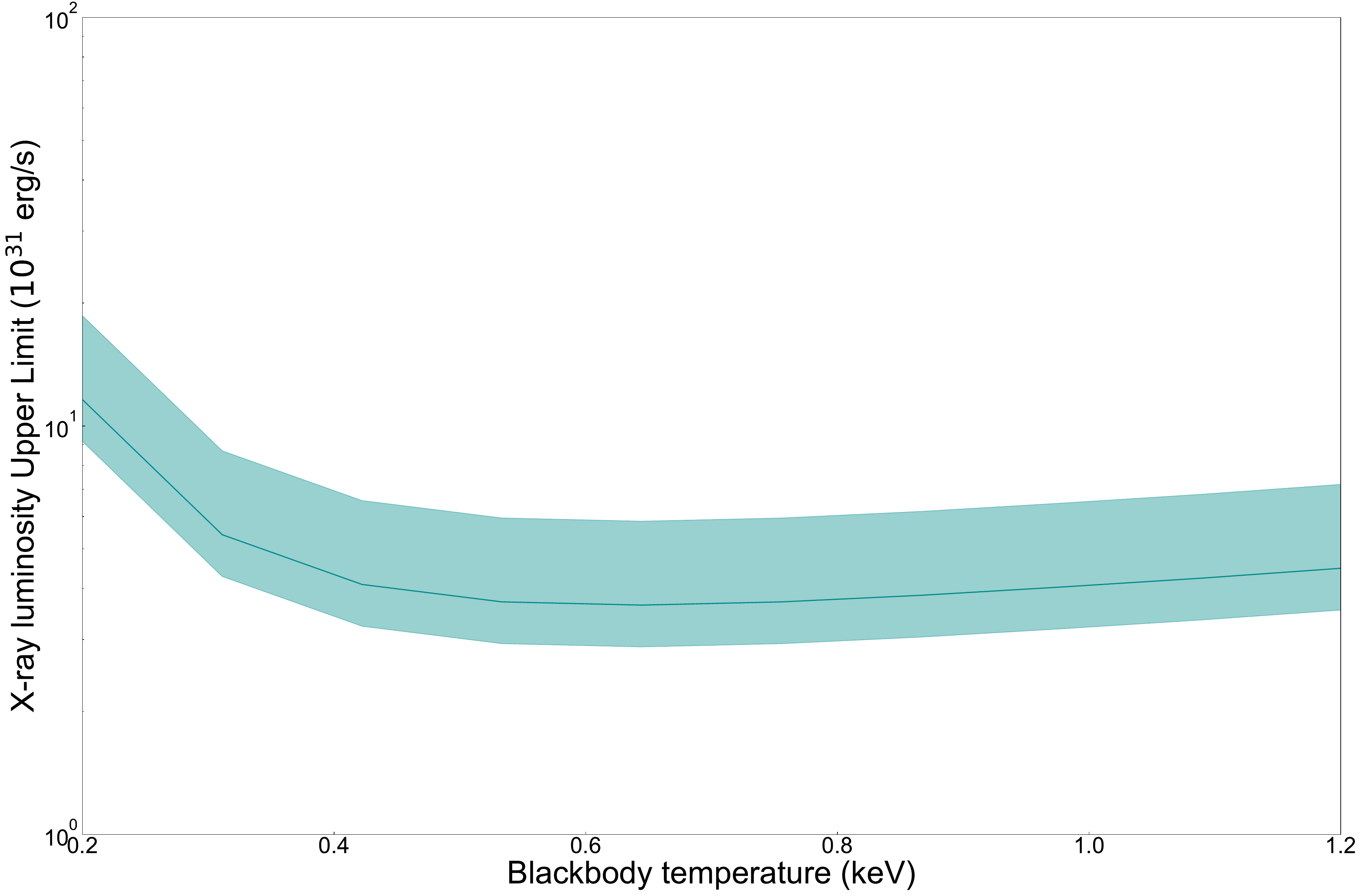}  
    \caption{Upper limits on the quiescent X-ray luminosity as derived by {\em XMM-Newton}. The 3$\sigma$ X-ray luminosity limits are calculated as a function of photon index $\Gamma$ (top panel) and black body temperature (kT). We assume $N_{\rm H} = 1.8\times10^{22}$\,cm$^{-2}$, which is the Galactic column density in the direction of the source, consistent with the X-ray spectral fits during the outburst. The shaded region assumes the error in the distance ($4.5^{+1.2}_{-0.5}$\,kpc).}
    \label{fig:luminosity_limits}
\end{figure}

\begin{figure}[htbp]
    \centering
    \includegraphics[width=0.9\textwidth]{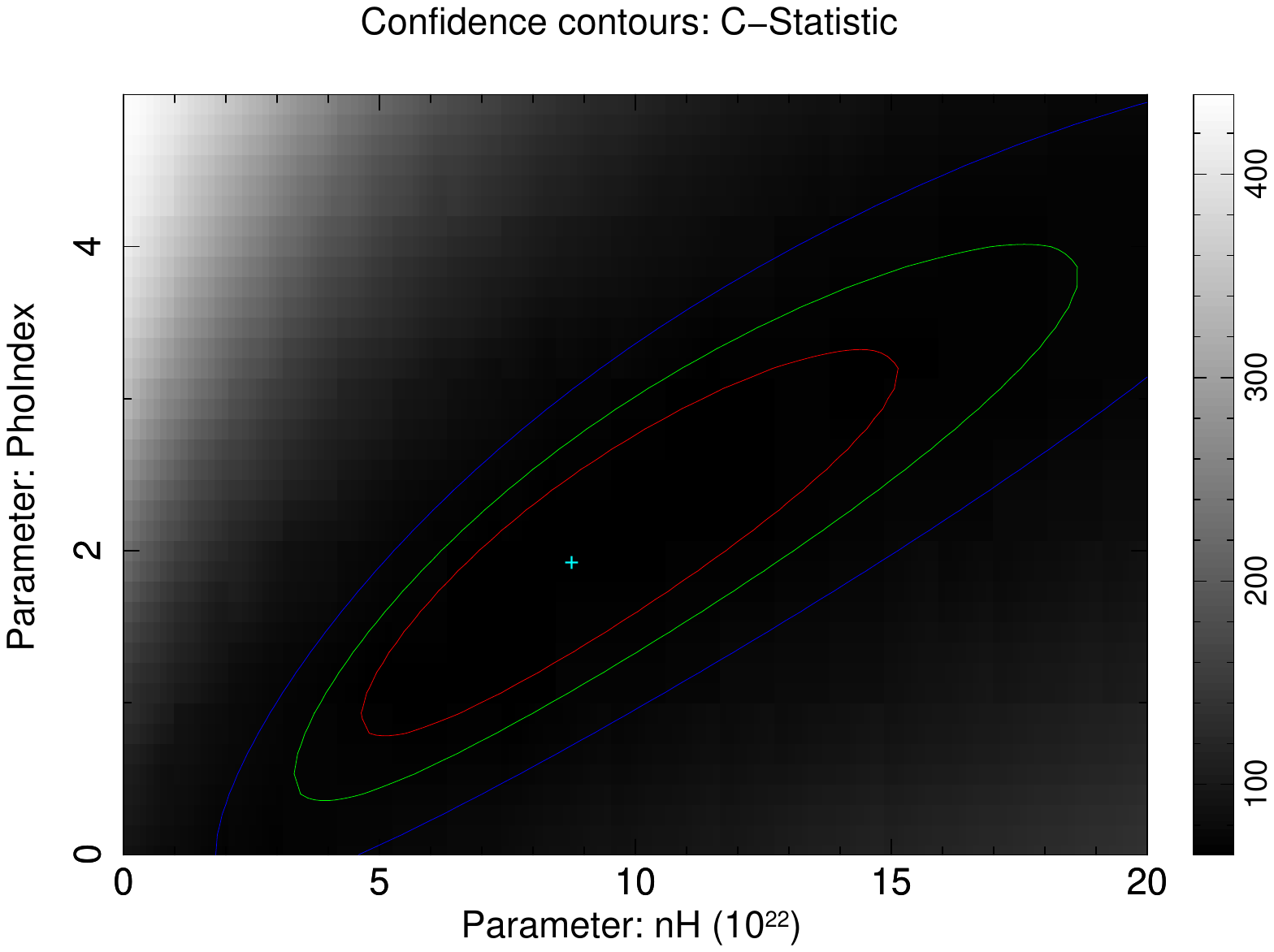} 
    \hspace{0.01\textwidth}
    \includegraphics[width=0.9\textwidth]{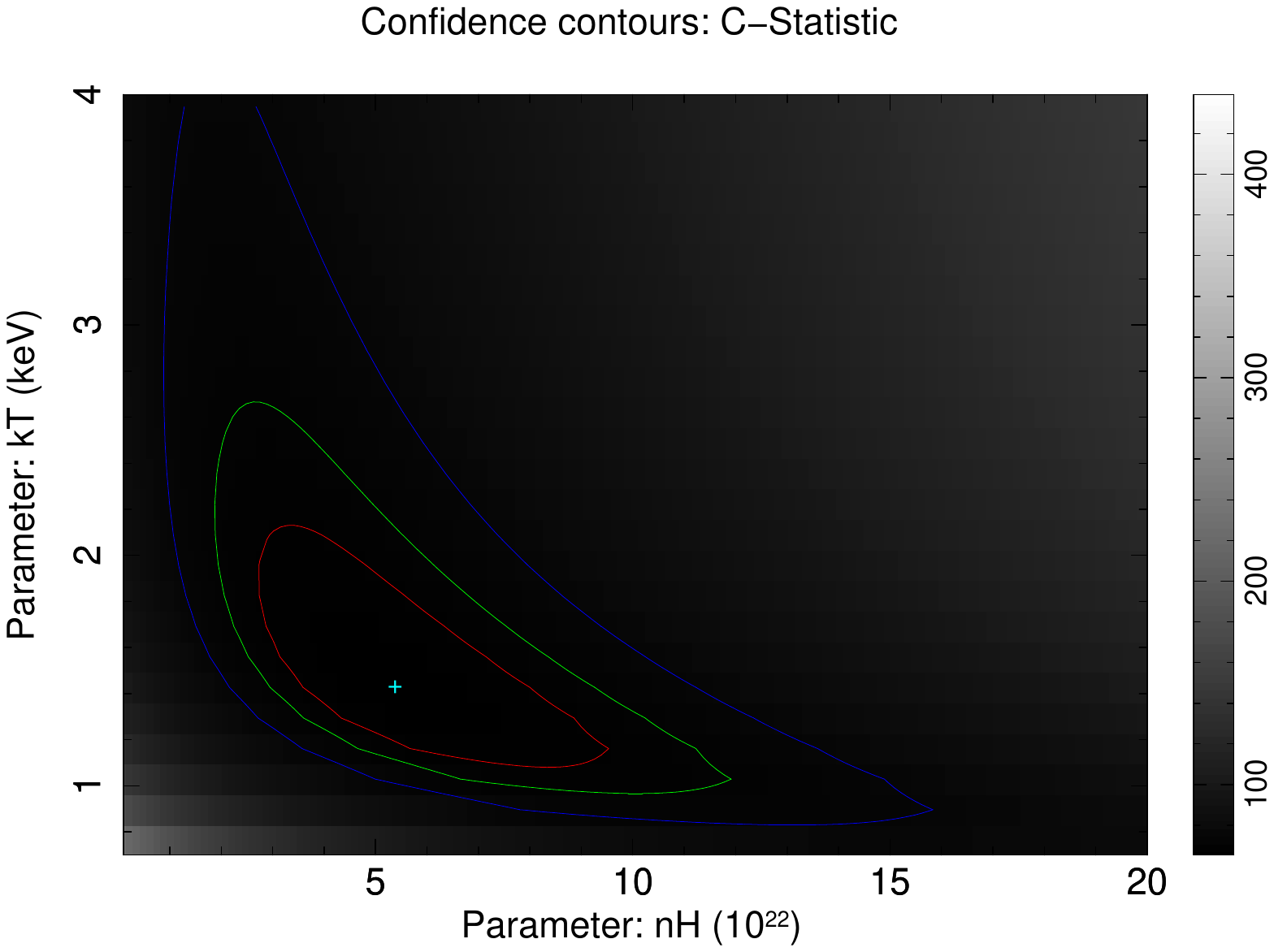}  
    \caption{Upper panel: $\Gamma$ as a function of $N_{\rm H}$;
    Lower panel: kT as a function of $N_{\rm H}$ The red, green and blue lines in both panels represent the 1, 2 and 3$\sigma$ confidence levels, respectively.}
    \label{fig:contour}
\end{figure}


\begin{figure}
    \centering
    \includegraphics[width=\linewidth]{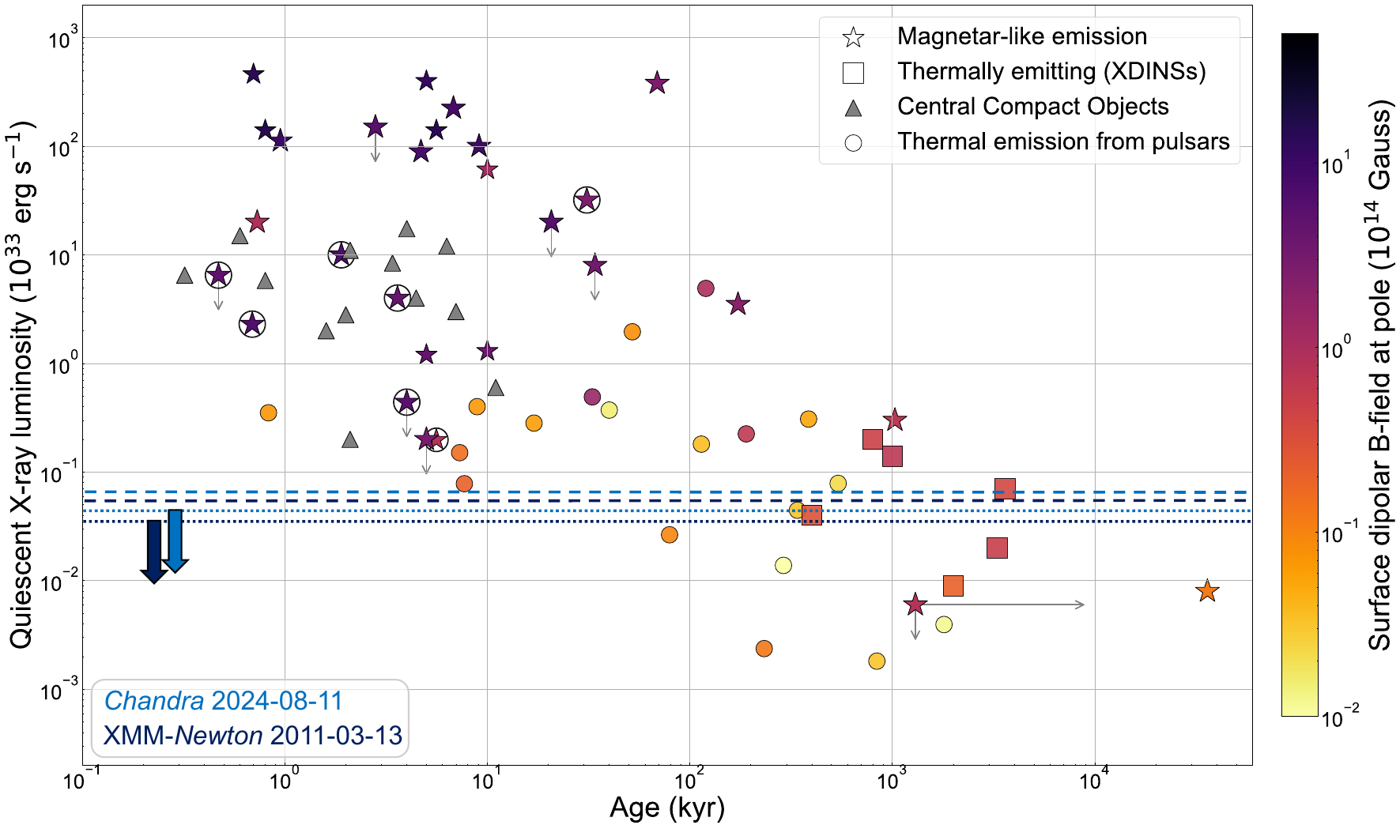}
    \caption{Quiescent X-ray luminosity as emitted thermally from the surface for different neutron star classes plotted as a function of age. Circled stars are radio-emitting magnetars. See \cite{2022ApJ...940...72R} for more details on the data extraction and sources. We assume as a luminosity limit for the quiescent emission of \lpt\, that is derived from the deepest XMM-Newton limits on 2011-03-13 (dark blue lines; see also  Fig.\,\ref{fig:luminosity_limits} and Methods), and the most recent limits derived after the outburst by {\em Chandra} on 2024-08-11 (light blue lines). Dashed lines assume an X-ray spectrum modelled by a power law with $\Gamma=2$, while dotted lines assume a blackbody with $kT=0.5$\,keV. The age refers to age of the SNR when available, and to the characteristic age otherwise.}
    \label{fig:mag_luminosities}
\end{figure}


\begin{figure}
    \centering  \includegraphics[width=\linewidth]{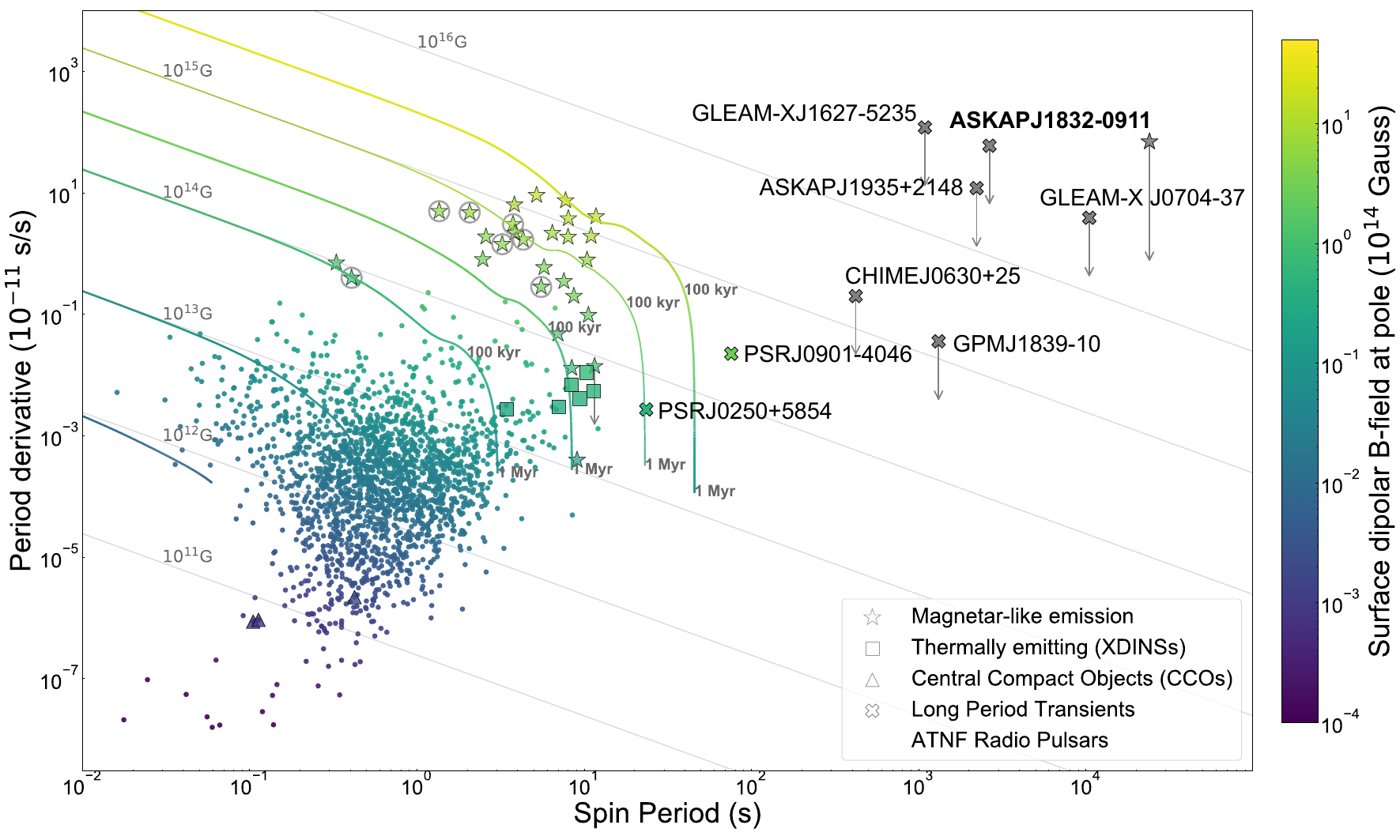}
    \includegraphics[width=\linewidth]{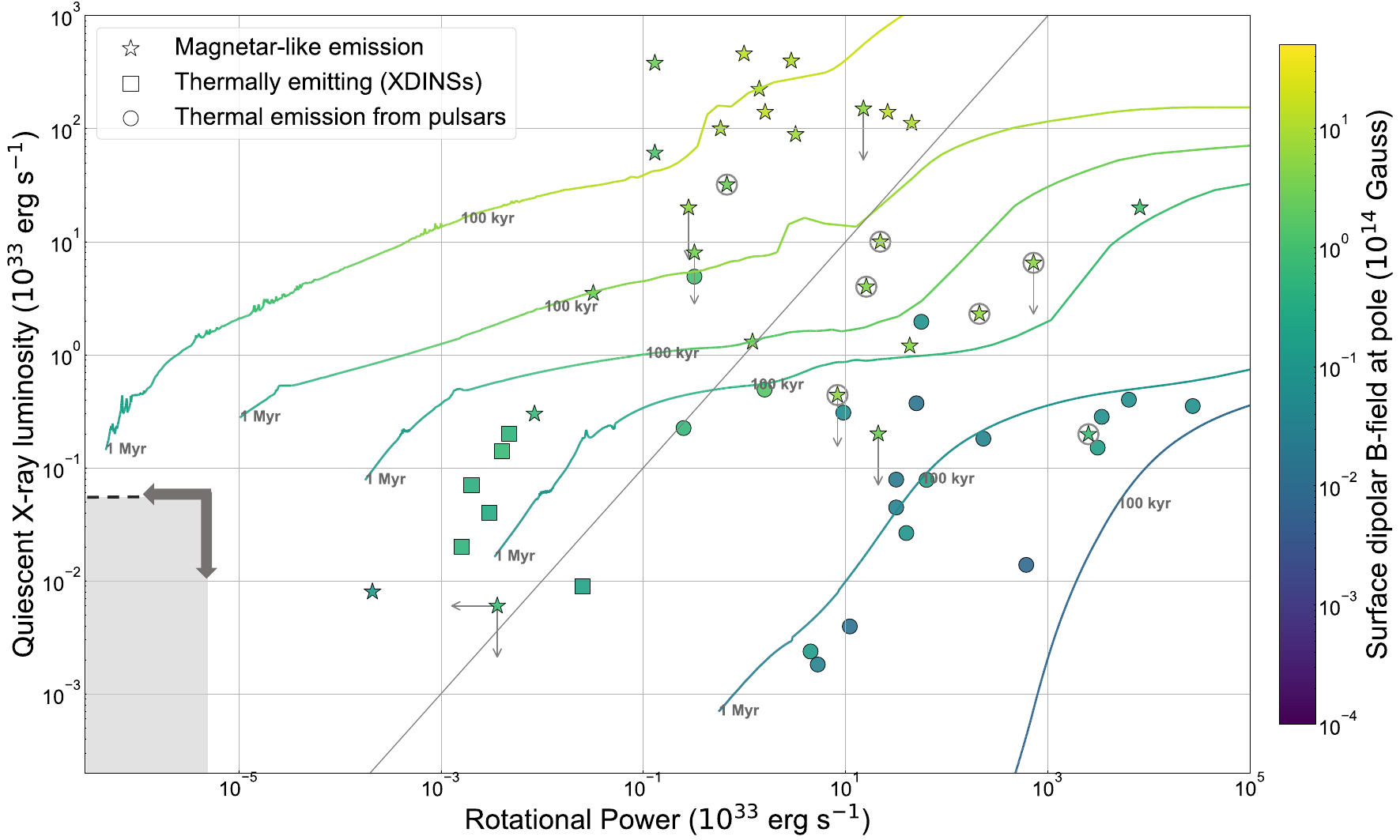}
    \caption{Magneto-thermal evolutionary models for neutron stars assuming different initial magnetic-field strengths and configurations. Top panel: evolution in the P-Pdot plane. Bottom panel: Evolution of the quiescent X-ray luminosity as a function of the rotational power. The grey line is $L_{X} = \dot{E}$ while the grey shaded area is limited by the upper limit derived for \lpt\, during quiescence. We report here the XMM-Newton upper limit on 2011-03-13 as derived assuming a power-law spectrum of $\Gamma=2$ (see also Extended Data Figure~\ref{fig:luminosity_limits} and Methods) See \cite{2022ApJ...940...72R, 2024NatAs...8.1020M, 2021CoPhC.26508001V} and reference therein for details on the theoretical cooling models and the plotted sources. }
    \label{fig:mag_luminosities_models}
\end{figure}

\begin{table}
\fontsize{8}{10}\selectfont
\renewcommand{\arraystretch}{1.2}
\centering
\caption{The best-fit spectral parameters and their errors for the combined Chandra X-ray spectrum of \lpt.}
\label{tab:xfit}
\begin{tabular*}{\textwidth}{@{\extracolsep{\fill}} llcc @{}}
\hline
\textbf{Model} & \textbf{Parameter} & \textbf{Best-fit value} & \textbf{Best-fit value} \\ 
\hline
\multirow{4}{*}{\texttt{tbabs $\times$ powerlaw}} 
& $N_{\rm H}$ ($\rm \times 10^{22}\ cm^{-2}$) & $1.8$ (fixed) & $8.82^{+6.99}_{-4.46}$ \\
& $\Gamma$ & $0.02^{+0.46}_{-0.48}$ &  $1.93^{+1.53}_{-1.24}$ \\
& $F_{\rm 1-10 keV }$ ($\rm \times 10^{-13}\ erg\ s^{-1} cm^{-2}$) & $3.29^{+0.97}_{-0.71}$ & $6.51^{+2.72}_{-1.53}$ \\
& $\chi^2_{\rm red}$ (dof) & $1.23$ (78) & $0.97 (77)$ \\
\hline
\multirow{5}{*}{\texttt{tbabs $\times$ bbodyrad}} 
& $N_{\rm H}$ ($\rm \times 10^{22}\ cm^{-2}$) & $1.8$ (fixed) & $5.45^{+4.58}_{-2.91} $\\
& $\rm kT$ (keV) & $2.21^{+1.28}_{-0.56}$ & $1.43^{+0.78}_{-0.40}$ \\
& $\rm R$ (km) & $0.017^{+0.007}_{-0.006}$ & $< 0.08$ \\
& $F_{\rm 1-10 keV }$ ($\rm \times 10^{-13}\ erg\ s^{-1} cm^{-2}$) & $2.51^{+0.89}_{-0.60}$ & $2.70^{+1.06}_{-0.60}$ \\
& $\chi^2_{\rm red}$ (dof) & $1.12$ (78) & $0.99 (77)$ \\
\hline
\end{tabular*}
\flushleft
\footnotesize{Notes: Errors represent 90\% confidence intervals. Reported fluxes are unabsorbed.}
\end{table}

\subsection*{Einstein-Probe Observations}

Einstein Probe \citep{2022hxga.book...86Y} is an X-ray satellite launched on January 9, 2024 into a low-Earth circular orbit, carrying two main instruments: The Wide X-ray Telescope (WXT) and  Follow-up X-ray Telescope (FXT). \lpt\, was observed twice by the FXT instrument (by both model A and B) in full frame mode: on 2024 July 30th between 09:44:58 -- 18:36:21 UTC (total live time of 17.1\,ks), and on 2024 August 02 between 09:56:04 -- 18:47:19 UTC (total live time of 16.7\,ks). Data were processed using the \texttt{fxtchain} tool available within the FXT Data Analysis Software (\texttt{FXTDAS}). 
We merged the event files of the two modelus (A and B), and of both observations. No source was detected at $>3\sigma$ at the refined radio position. We extracted background photons using a circle of radius 128$^{\prime\prime}$ centred on the source position to calculate the upper limit on the source X-ray emission. During the total live-time of 33.8\,ks we find a 3$\sigma$ upper limit on the source 0.5--10 kev count rate of $1.173
\times10^{-3}$\,cts/s.

\subsection*{XMM-Newton \& Swift Observations}

The location of \lpt has also been monitored by XMM-Newton Observatory in 2011, with ObsID=0654480101 (total observing time of 17\,ks).
We used the task {\it evselect} in the XMM-Newton Science Analysis System (SAS v19.1.0) to construct a high energy background light curve (energy between 10–12 keV for EPIC-pn and above 10 keV for EPIC-MOS1/MOS2) by selecting only PATTERN==0. 
The background light curve was then used to filter out periods of high-background flaring activity, enabling the creation of good time intervals (GTIs). After applying the GTI filtering, the effective live times were approximately 8370 seconds for EPIC-pn and 12360 seconds for EPIC-MOS2. We chose a circular region of 25'' radius for the source product extraction. The background products were extracted from an annular region centred on the source position with inner and outer radii of 50'' and 100'', respectively. 
We derived a $3\sigma$ upper limits on the source counts of $<1.072\times10^{-3}$ c/s for the MOS2 camera, and $<2.174\times10^{-3}$ c/s for the pn camera. These limits consider the 9$^{\prime}$ off-axis position of the source with respect to the aim-point.

This source has also been observed multiple times by the Swift Observatory. We select nine observations with exposure times longer than 3000 seconds, however Swift did not detect any significant X-ray emission at the position of \lpt\ in either of these observations. 

\subsection*{Upper limits on the X-ray luminosity in quiescence}

We used the deepest observations before and after the observed X-ray outburst 
to derive limits on the quiescent X-ray emission of \lpt. We used the XMM--Newton EPIC-pn limit in 2011, years before the observed outburst in Feb 2024, and also the {\em Chandra} observation performed in August 2024.

In Extended Data Figure~\ref{fig:luminosity_limits} we report the 3$\sigma$ luminosity limits in 2011 as a function of the assumed spectral shape, assuming an $N_{H} = 1.8\times10^{22}$\,cm$^{-2}$. Note that this value is the whole Galactic value despite the source distance is constrained at 4.5\,kpc. If we assume a lower $N_{H}$ the derived limits on the luminosities would be even lower. The shaded region reports the calculations considering the error in the distance.

To compare these limits with the quiescent emission of magnetars we take into account two possible spectral shapes, a power-law with $\Gamma=2$ (PL) and a black-body (BB) with temperature $kT=0.5$\,keV \citep{2018MNRAS.474..961C}. For the 2011 XMM-Newton observation we derive 3$\sigma$ luminosity limits of $L_X < 3.6\times10^{31}$\,erg\,s$^{-1}$ and $L_X < 5.3\times10^{31}$\,erg\,s$^{-1}$, for the BB and PL, respectively. Following a similar approach for the post-outburst {\em Chandra} observation on August 2024 deriving 3$\sigma$ luminosity limits of $L_X < 4.6\times10^{31}$\,erg\,s$^{-1}$ and $L_X < 6.5\times10^{31}$\,erg\,s$^{-1}$, for the BB and PL, respectively.

\begin{figure}
    \centering
    \includegraphics[width=0.8\linewidth]{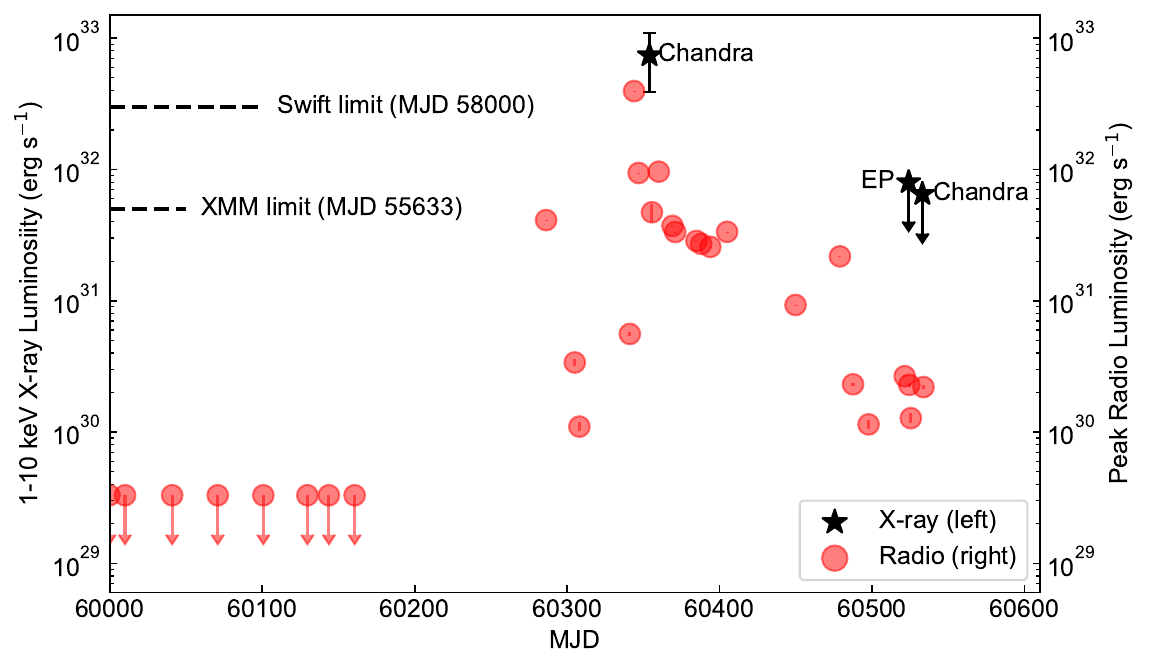}
    \caption{X-ray and radio luminosity evolution for \lpt{}. We showed two early X-ray luminosity limits from {\em Swift} and {XMM} as black dashed lines. For all detections, $1\sigma$ errorbars are shown.}
    \label{fig:xray_radio_longterm_lc}
\end{figure}

In Extended Data Figure~\ref{fig:mag_luminosities} and \ref{fig:mag_luminosities_models} (bottom panel) we report the luminosity limits compared with what observed in other isolated neutron stars, in particular with magnetars. Similarly in Extended Data Figure~\ref{fig:xray_radio_longterm_lc} we show the upper limits (derived considering a PL with $\Gamma=2$ spectrum) for all the available observations, compared to the radio detections.

\begin{table}
\centering
\caption{X-ray Observation log} 
\label{tab:obsinfo}
\centering
\begin{tabular}{lcccr}
\hline
\hline
ObsID & Date & R.A. & Decl. & Exposure Time (ks)  \\
(1) & (2) & (3) & (4) & (5)   \\ 
\hline
\multicolumn{5}{c}{\it Chandra} \\
\hline
26682    & 2024-02-14  & 18 32 39.70 & -09 16 08.7 & 9.9  \\
29265   & 2024-02-14  & 18 32 39.70 & -09 16 08.7 & 9.9 \\
29266   & 2024-08-11  & 18 32 48.41 & -09 11 15.8 & 9.9\\
\hline
\multicolumn{5}{c}{\it Einstein Probe} \\
\hline
 06800000022 & 2024-07-30 & 18 32 48.46 & -09 11 15.3 & 17.1  \\
06800000023 & 2024-08-02 & 18 32 48.46 & -09 11 15.3 & 16.7  \\
\hline
\multicolumn{5}{c}{\it Swift} \\
\hline
00034056022 & 2016-07-28  & 18 32 31.03  & -09 20 14.4 & 3.6\\
00034056038 & 2016-11-02  & 18 32 42.49  & -09 18 34.6 & 3.7\\
00087519001 & 2017-11-01  & 18 32 21.43  & -09 04 15.1 & 4.9\\
00010378001 & 2018-03-26  & 18 32 49.04  & -09 20 52.4 & 4.7\\
00010378007 & 2018-04-23  & 18 32 51.50  & -09 21 00.1 & 4.9\\
00010378008 & 2018-04-30  & 18 32 48.44  & -09 21 10.9 & 4.6\\
00010378011 & 2018-05-21  & 18 32 39.60  & -09 20 55.3 & 4.1\\
00010378012 & 2018-05-28  & 18 32 48.15  & -09 20 43.2 & 4.6\\
00010378015 & 2018-06-18  & 18 32 48.73  & -09 20 59.8 & 4.7\\
\hline
\multicolumn{5}{c}{\it XMM-Newton} \\
\hline
0654480101 & 2011-03-13  & 18 32 42.47  & -09 19 48.0 & 16.9\\
\hline

\hline
\hline
\end{tabular}
\begin{tablenotes}
      \small
      \item 
      Notes:
      (1)-(2) Observation ID and date.
      (3)-(4) J2000 sky coordinates of the telescope aimpoint. (5): Effective exposure time, in units of ks.
\end{tablenotes}
\end{table}

\subsection*{Period Measurements}

We used the dynamic spectra extracted from radio observations dating to October 2024 to measure pulse times of arrival (ToAs). As seen in Fig. \ref{fig:J1832_multilambda}, the radio pulse shape varies over time, preventing reliable measurement of ToAs using a fixed template. We instead fit a triangular pulse shape with variable height, width, and central pulse phase to determine the time of arrival of each pulse. While this is a simplistic choice, any ToA measurement is limited by the changing pulse shape.

These limitations with ToA measurements can be mitigated by adjusting the ToA uncertainties with EQUAD, a parameter added in quadrature to the ToA uncertainties as $\sqrt{\sigma_{\mathrm{ToA}}^2 + \mathrm{EQUAD}^2}$. This parameter adjusts the ToA uncertainties to account for the scatter in the ToAs induced by profile variability. We included an independent EQUAD parameter for every observing epoch, which is fitted simultaneously with the overall spin frequency and spin frequency derivative. Based on an initial timing solution derived using \textsc{tempo2} \cite{2006MNRAS.369..655H}, we observe a scatter in timing residuals of $\sim 50$\,s. We used PINT \cite{2021ApJ...911...45L} and \texttt{emcee} \cite{2013PASP..125..306F} to fit the spin frequency and frequency derivative along with EQUAD per epoch. For epochs with less than four ToAs, we assume an EQUAD of 30\,s.

\begin{figure}
    \centering
    \includegraphics[width=0.9\linewidth]{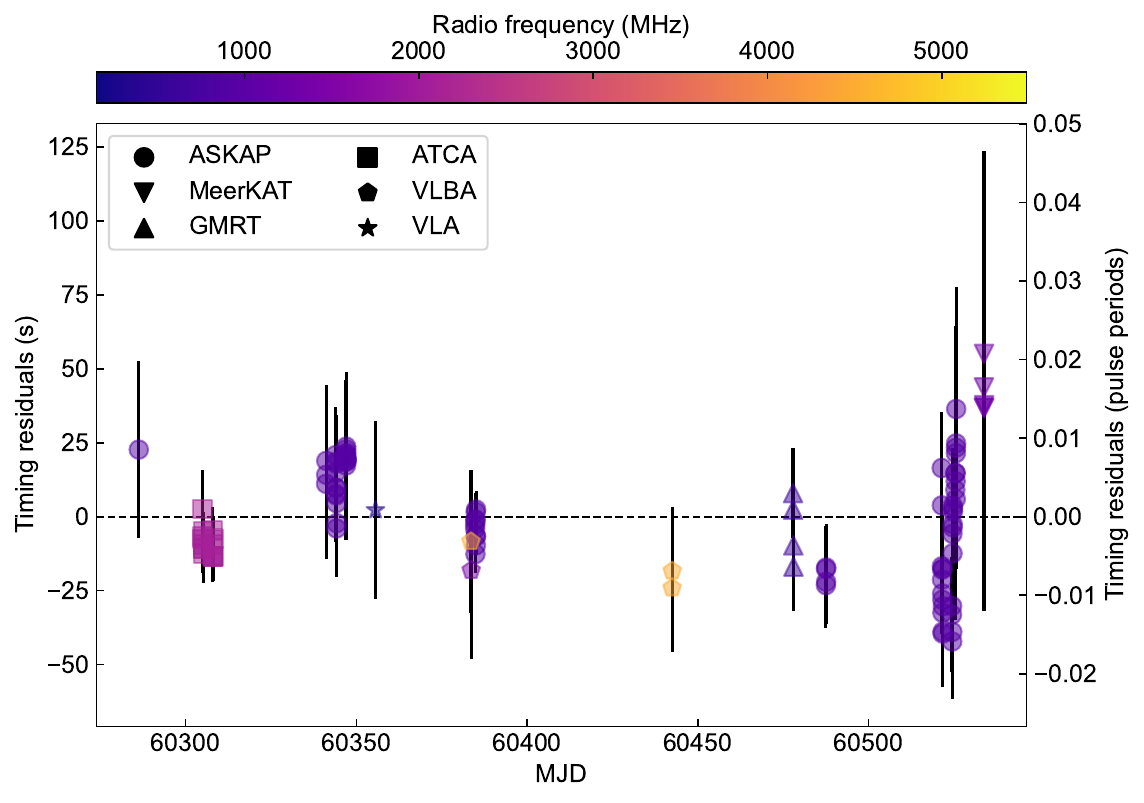}
    \caption{Timing residuals for J1832-0911 assuming $f = 3.764709(2)\times 10^{-4}\,{\text{s}^{-1}}$ and $\dot{f} = 0$. Uncertainties on the times of arrival have been adjusted in quadrature by EQUAD, fitted per-epoch, with values ranging from 6 -- 69\,s. The points are coloured by radio frequency, and the telescopes used for each measurement are indicated by the markers.}
    \label{fig:timing_resids}
\end{figure}

\begin{figure}
    \centering
    \includegraphics[width=\textwidth]{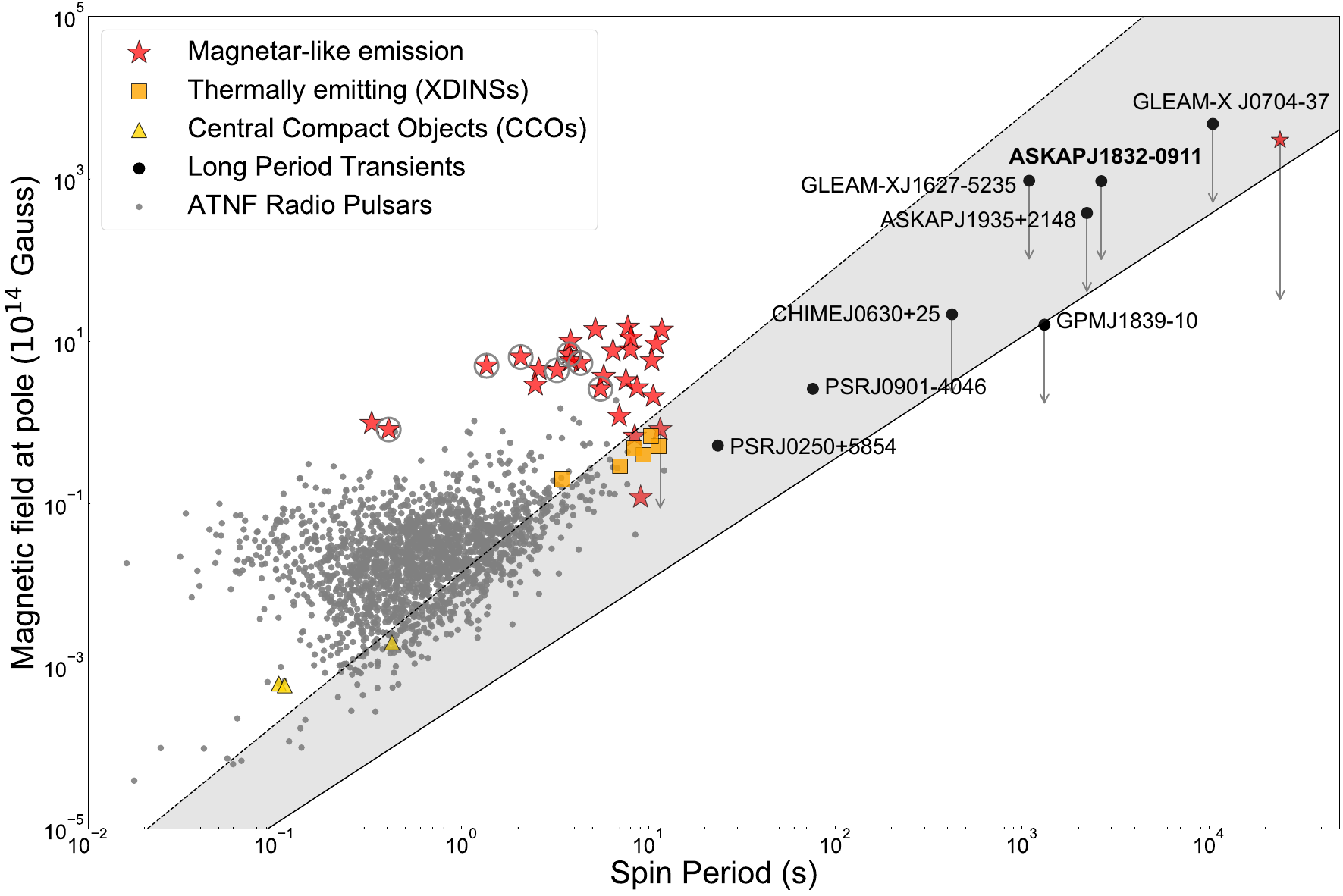}
    \caption{Surface dipolar magnetic field at the pole, $B$, against spin period, $P$, for observed isolated NSs. Arrows represent upper $B$-field limits. We show isolated ATNF radio pulsars \citep{2005AJ....129.1993M} (gray dots), pulsars with magnetar-like X-ray emission (red stars; gray circles highlight the radio magnetars), including the long-period magnetar 1E~161348-5055 \citep{2006Sci...313..814D,2016ApJ...828L..13R,2016MNRAS.463.2394D}, X-ray Dim Isolated NSs (XDINSs; orange squares) and Central Compact Objects (CCOs; gold triangles) \citep{2014ApJS..212....6O, 2018MNRAS.474..961C}. Other long-period radio pulsars are reported as black circles \citep{Tan-etal2018,2022NatAs...6..828C,2022Natur.601..526H, 2023Natur.619..487H, 2024NatAs.tmp..107C, 2024arXiv240707480D, 2024arXiv240811536D}. Dashed (solid) lines correspond to theoretical death lines for a pure dipole (highly multipolar) configuration \citep{1993ApJ...402..264C, 2000ApJ...531L.135Z}. The grey-shaded region indicates the radio pulsar ``death valley" between the two extreme configurations.}
\label{fig:p_B}
\end{figure}

We found posterior median EQUAD values range from 6 -- 69\,s, consistent with the scatter seen in the ToAs. With these EQUAD values, we find a spin frequency of $f = 3.764709(2)\times 10^{-4}\,{\text{s}^{-1}}$ and place a 95\% spin frequency derivative upper limit of $|\dot{f}| < 1.4\times 10^{-16}\,{\text{s}^{-2}}$, with a period epoch MJD 60344. From these values we derive a spin period $P = 2656.247(1)$\,s and spin period derivative limit $\dot{P} = < 9.8\times 10^{-10}$. We show the resultant timing residuals, assuming $\dot{f} = 0$, in Extended Data Figure \ref{fig:timing_resids}. The reduced chi-squared for this fit is close to unity, indicating that the timing model and adjusted uncertainties adequately explain the data.

\subsection*{Distance Measurements}

Galactic electron density models can be used to infer the distance to the source based on the DM. We used the most recent model \citep{2017ApJ...835...29Y}, to obtain the estimated distance (and its associated uncertainty) of $4.47\pm0.04$\,kpc. According to \citep{2017ApJ...835...29Y}, the standard deviation of the distribution of relative distance errors in their model is 0.398, based on the 189 pulsars with independent distance measurements. We also used the same method used by \cite{2022Natur.601..526H} to estimate the relative distance error based only on pulsars that are near \lpt\ on the sky. For 18~pulsars within 10$^\circ$ of \lpt, we extracted their inferred distances based on \citep{2017ApJ...835...29Y}, as well as independent distance measurements from version 2.5.1 of the Australia Telescope National Facility Pulsar Catalogue \citep{2005AJ....129.1993M}, and estimated a standard deviation on the relative distance error of 0.282. The standard deviation on the relative distance error for nearby pulsars and all pulsars is therefore broadly consistent. Conservatively, we conclude there is about 30\% uncertainty on the estimated distance of 4.47\,kpc.

\begin{figure}
    \centering
    \includegraphics[width=0.95\linewidth]{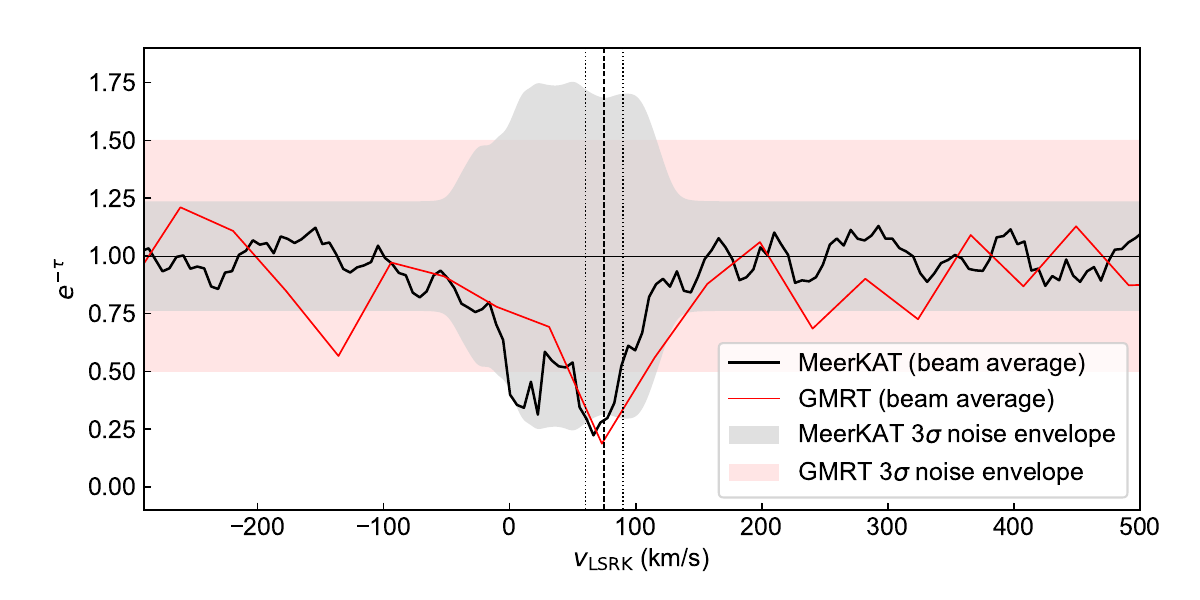}
    \caption{21-cm \hi\ absorption against ASKAP J1832-911 in units of $e^{-\tau}$, where $\tau$ is the line optical depth. The MeerKAT data have a native channel width of 5.5 km\,s$^{-1}$ and are smoothed with a 5-channel boxcar kernel. The GMRT data have a channel width of 41 km\,s$^{-1}$ and have not been smoothed. The 3$\sigma$ noise envelopes are computed as described in the text. The vertical dashed line shows the estimated centroid velocity of the absorbing component with the farthest near-side distance (corresponding to $d\sim4.8$ kpc, \citep{2018ApJ...856...52W}), and the vertical dotted lines show the associated uncertainty estimate of $\pm15$ km\,s$^{-1}$.}
    \label{fig:hi_absorption}
\end{figure}

Absorption from intervening H\,{\sc i} gas along the line of sight may also be used to constrain the distance. Figure \ref{fig:hi_absorption} shows absorption spectra from the MeerKAT and GMRT datasets, extracted over the area of their respective synthesized beams, centred on the source position. In both cases, the absorption depth is expressed in units of $e^{-\tau}$, where $\tau$ is the optical depth, obtained as $e^{-\tau} = S_v/S_{\rm cont}$, where $S_v$ is the absorption profile in units of flux density (including the continuum), and $S_{\rm cont}$ is the continuum flux density of the source, as measured from the off-line portions of the spectrum.

The spectral noise is in general frequency dependent, since \hi\ emission contributes to the system temperature in channels where it is present. The noise envelope is computed as $\sigma_\nu=\sigma_{\rm cont}~(T_{\rm HI}(v)+T_{\rm sys}+T_{\rm sou})~/~(T_{\rm sys}+T_{\rm sou})$. Here, $\sigma_{\rm cont}$ is the measured noise outside of the \hi\ emission velocity range, $T_{\rm sys}$ is the system temperature including the instrumental contribution and sky brightness at 1.4 GHz in the primary beam, $T_{\rm sou}$ is the correlated self-noise of the source \citep{2021PASA...38...13M}, and $T_{\rm HI}(v)$ is the HI emission brightness temperature in the primary beam. In the case of the GMRT data, the spectral noise is dominated by correlated self-noise such that $\sigma_\nu\approx\sigma_{\rm cont}$. For MeerKAT we estimate the continuum and \hi\ emission within the $\sim1$ degree primary beam using the CHIPASS \citep{2014PASA...31....7C}, and HI4PI \citep{2016A&A...594A.116H} surveys, respectively, and assume the correlated self-noise is negligible (reasonable given the relatively small collecting area of MeerKAT relative to the GMRT, and the low source flux density in the MeerKAT observations). The effective instrumental system temperature is taken as 20~K, and we intentionally make no primary beam correction for the \hi\ emission, to produce the most extreme plausible \hi\ contribution to the noise. 

As can be seen in Figure \ref{fig:hi_absorption}, both the MeerKAT and GMRT spectra show detections just above the $3\sigma$ threshold at $\sim75$ km\,s$^{-1}$. While either detection might be considered marginal alone, the agreement between the two is convincing. We estimate an uncertainty of $\pm15$ km\,s$^{-1}$ on the velocity of the absorbing component, based on the effective resolution of the smoothed MeerKAT data ($\sim$30 km\,s$^{-1}$). To convert velocity to distance we use the Monte Carlo kinematic distance method of Wenger et al. 2018 \citep{2018ApJ...856...52W}, which assumes the Galactic Rotation model of Reid et al. \citep{2014ApJ...783..130R}. The resulting lower limit on the distance to \lpt\, based on the near-side distance to the $\sim75$ km\,s$^{-1}$ absorbing component, is $4.8 \pm 0.8$ kpc.
Combining the constraints from DM and H\,{\sc i}, we conservatively estimated the distance to \lpt\ of $4.5^{+1.2}_{-0.5}\,$kpc.

\subsection*{Radio Luminosity Calculation}

The radio luminosity of \lpt\ highly depends on the emission mechanism and the system geometry. For the simplest model, if we assume the emission is isotropic and a flat spectral model, the estimated peak radio luminosity $L_{\rm rad} = 4\pi d^2\nu S_{\nu} = 2.4\times10^{31} \left(\cfrac{d}{4.5\,{\rm kpc}}\right)^2 \left(\cfrac{\nu}{1\,{\rm GHz}}\right) \left(\cfrac{S_\nu}{1\,{\rm Jy}}\right)\,$erg\,s$^{-1}$. For the brightest pulse we detected, it has a radio luminosity of $L_{\rm rad, max} = 4\times10^{32}\,$erg\,s$^{-1}$.
If the radio emission from \lpt\ is similar to pulsars or magnetars, according to \citep{2004hpa..book.....L}, we can write the radio luminosity as
\begin{equation}
    L_{\rm rad} = \cfrac{2\pi d^2}{\delta}\left(1-\cos\rho\right)S\left(\nu_0\right)\cfrac{\nu_0^{-\alpha}}{\alpha+1}\left(\nu_2^{\alpha+1}-\nu_1^{\alpha+1}\right)
\end{equation}
in which $\delta$ is the pulse duty cycle, $\rho$ is the opening angle of the emission cone, $\alpha$ is the radio spectral index, $S\left(\nu_0\right)$ is flux density at the reference frequency, and $\nu_1$ and $\nu_2$ are frequency boundaries where the pulses were detected. The pulse widths are roughly 2-4 mins, which means a duty cycle of 5-10\%. The opening angle depends on the geometry of the system. However, with a relatively flat polarisation position angle swing, it is hard to model the geometry precisely. We adopt a typical opening angle of $\rho=6^\circ$. We did not detect any emission in any MWA observation, we therefore use $\nu_1=200\,$MHz as the lower-end cut-off frequency. With a spectral index of $\alpha=-1.4$, for the brightest pulse we detected, the radio luminosity will be $L_{\rm rad} = 5\times10^{31}\,$erg\,s$^{-1}$. This is lower but similar to the radio luminosity we derived from the simplest model.

\subsection*{Potential SNR association}

\lpt\ is spatially coincident with a known supernova remnant (SNR), SNR~G22.7$-$0.2, located at a distance of 4.74\,kpc \cite{2018AJ....155..204R,2020A&A...639A..72W}, with a shell diameter of 30$'$ \cite{1992AJ....103..943K,2021A&A...651A..86D}. Since the number density of long-period radio transients and their relationship (if any) with supernova remnants is currently unknown, it is difficult to estimate the false positive rate. An analogous calculation is the false positive rate of a pulsar/SNR association in this area. We used version~2.5.1 of the Australia Telescope National Facility Pulsar Catalogue \cite{2005AJ....129.1993M} to estimate the spatial density of pulsars in this region, finding an average of 3~pulsars per square degree; given the 0.22\,sq.\,deg. area subtended by SNR~G22.7$-$0.2, there is a $\sim$70\,\% chance that a pulsar would lie within its shell purely by chance.

With a physical radius of $\sim20$\,pc, and assuming a local ISM density of $\sim$0.5\,cm$^{-3}$ appropriate for the low Galactic latitude, standard SNR age calculations \cite{2017AJ....153..239L} yield an age of 25\,kyr (consistent with the estimation from \citep{2014ApJ...796..122S}), with the SNR just exiting the Sedov-Taylor expansion stage, consistent with its morphology. If \lpt{} was formed at the same time, it would therefore have been travelling at $\sim$400\,km\,s$^{-1}$ (in the plane of the sky) to reach its current location, which is consistent with the kick velocity distribution of pulsars \cite{2002ApJ...568..289A,2017A&A...608A..57V}. However, it is inconsistent with the upper limit on the transverse velocity measured by the VLBA (190\,km\,s$^{-1}$ at the same distance as the SNR). 

Additionally, if \lpt\ is a neutron star associated with the SNR~G22.7$-$0.2, both the low X-ray quiescent emission and spin-down luminosity would imply an old system ($\gtrsim 0.5\,$Myr, see Extended Data Figure~\ref{fig:mag_luminosities_models}). The age of such an old system is not consistent with the age of the SNR.
We therefore rule out an association between \lpt{} and SNR~G22.7$-$0.2.

\subsection*{Optical/Infrared Constraints}

\begin{figure}
    \centering
    \includegraphics[width=0.8\linewidth]{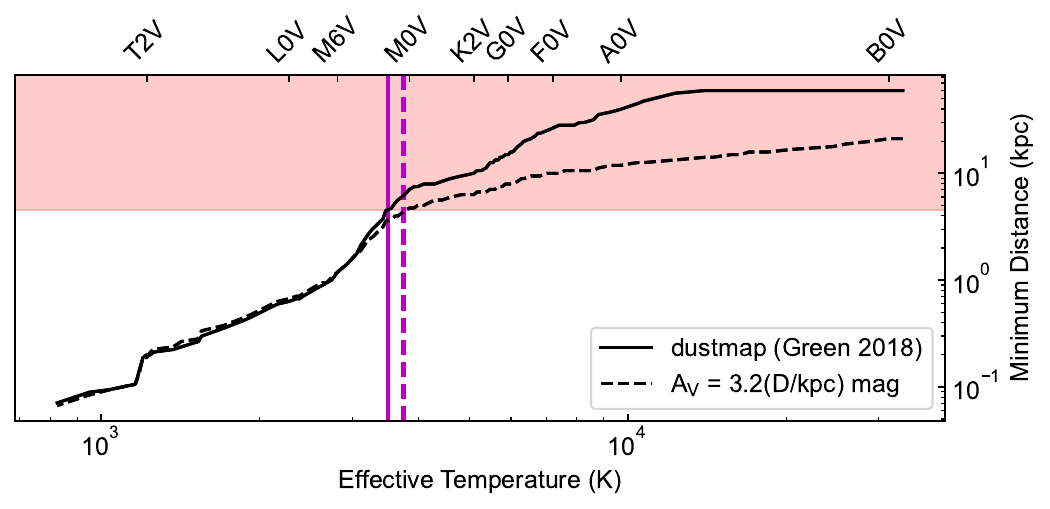}
    \caption{Constraints on the minimum distance for different stellar spectral types (black line). The purple line shows the spectral type limit we can constrain based on the current infrared observation.}
    \label{fig:star_limit}
\end{figure}

Previous infrared observations of ASKAP~J1832 limit the maximum brightness of its infrared counterpart to $K_s=19.86$ (Vega) and $J=19.98$ (Vega) at $3\sigma$. 
We retrieved the color and absolute magnitude for main-sequence and ultra-cool stars from \citep{2013ApJS..208....9P}, and used \textsc{dustmaps} \citep{2018JOSS....3..695G} to estimate the extinction. For each stellar type, we estimated the minimum distance required for a non-detection with the limiting magnitude (solid black line in Extended Data Figure~\ref{fig:star_limit}). We can currently rule out any stellar object with spectral type earlier than M1 at 4.5\,kpc for \lpt. 
We note that the optical extinction may be underestimated. The derived optical extinction from \textsc{dustmaps} is $A_V=5.8\,$mag, and the one inferred from DM is $A_V=6.2\,$mag. 
\citep{2024MNRAS.529L.102V} found an $A_V \sim 21.6$ at a distance of $6.7$\,kpc towards HESS~J1832$-$093, which is $\sim10$\,arcmin away from \lpt. We used a simple linear model to conservatively estimate the extinction based on this measurement, i.e., $A_V = 3.2(D/{\rm kpc})\,$mag, which corresponds to $A_V=16.1\,$mag at 4.5\,kpc. With this more conservative estimation, we can rule out any stellar object with spectral earlier than M0 at 4.5\,kpc for \lpt.

\subsection*{Interpretation of the X-ray pulse profile}
\label{sec:Xrayinterp}

A curious feature of the X-ray emission, which is not related to biased exposure, is the dip to zero counts between phases 0.55 and 0.75 (beginning at the trailing edge of the pulse at phase 0.45 and ending at phase 0.85). This might be reminiscent of an eclipse and would establish that \lpt\, is in a binary viewed edge-on. However, the interpretation of this dip as an eclipse is problematic. This putative eclipse is generally much longer than encountered in compact binaries — indeed, a 44.2-minute period circular binary with a compact object (WD or NS) would have an orbital size of $3-6\times 10^{11}$ cm. The upper bound on the largest a low mass companion could be in such a system is the Roche lobe at $\sim 1-2 \times 10^{11}$ cm. However, this maximum size is still not large enough to produce an eclipse duration as long as the observed dip. Moreover, an emitting region of comparable size to a WD would not be point-like and perhaps not fully eclipsed. In principle, a large eclipse phase duration can also be used to place constraints on the size of the companion, which can then be compared to the established photometric non-detection \citep[e.g.,][]{2021MNRAS.503.5600B}.

This high of a pulsed fraction (above 2 keV) is generally not observed in IPs, as they typically they show pulsed fractions $\lesssim$ 60\% \citep[see, e.g.,][]{2012A&A...542A..22B,2021ApJ...914...85H}. However, polars with such high X-ray pulsed fractions have been found  \citep[e.g.,][]{2014MNRAS.445.1403B}. The X-ray emission in IPs and Polars is generally attributed to thermal bremsstrahlung radiated at accretion shocks at magnetic poles, involving high plasma densities and optically thick conditions. It is unclear how bright coherent emission, demanding non-thermal particle acceleration and unscreened electric fields, would be produced in such plasma rich environment, or how the radio would escape even if not colocated but tied to regions of an accretion stream (as demanded by the phase alignment of radio and X-rays). The radio pulses of \lpt\, exhibit bright broadband sub-pulse structure (Figure~\ref{fig:J1832_discovery}) throughout, at odds with an accretion stream. In fact, it has been suggested that the observed radio emission from WD systems (including Ar Sco \citep[][]{2018A&A...611A..66S} and possibly J191213.72--441045.1 \citep{2023NatAs.tmp..120P}) comes from electron cyclotron maser emission from the corona of the donor star, and is generally orders of magnitude lower in radio luminosity than \lpt\, (see e.g., \cite{2020AdSpR..66.1226B}). These magnetic CVs' radio emission are also always circularly polarised \citep{2020AdSpR..66.1226B}. This is at odds with the broadband and high linear polarisation of \lpt{}, which requires that the local cyclotron frequency (i.e. magnetic field) where the radio emission escapes, is much larger than the wave frequency or plasma frequency.

The observed ratio of on-pulse X-ray to radio luminosities is also orders of magnitude off the G\"{u}del-Benz law of active stars \citep{1993ApJ...405L..63G,1994A&A...285..621B,2024arXiv240915507C}, extrapolation of which predicts a radio luminosity of $\sim 10^{27}$ erg s$^{-1}$ for an X-ray luminosity of $10^{33}$ erg s$^{-1}$. 

Another scenario is a magnetar, either isolated or in a binary with a spin period of 44.2 min. The low Galactic $b$ suggests a low-kick birth or not exceptionally old age. In a magnetar scenario, the X-ray emission must originate to the same field lines tied to the radio pulsations, presumably produced by pair cascades from a mild field twist either plastically or thermoelectrically driven \citep{2024MNRAS.533.2133C} or by low-level crustally-driven twists invoked in magnetar outbursts \citep{2018MNRAS.474..961C}. The statistical quality of the X-ray data do not allow us yet to firmly distinguish the thermal or non-thermal nature of the emission; a small hot spot of radius $\sim 100$ m with temperature $kT\sim1$ keV and anisotropic radiation transport in magnetized atmosphere of a magnetar can also produce a high pulsed fraction \citep{2022ApJ...928...82H}. Such a small hot spot is not unprecedented and has been observed magnetar Swift J1818.0–1607 \citep{2022MNRAS.512.1687R,2023ApJ...943...20I}. However, we cannot distinguish such a hot spot from non-thermal emission with the current data. It is apparent there is a broader component to the pulse profile, so additional ingredients are necessary to fully describe the system. \lpt\, as a magnetar would be different from classical magnetars, perhaps related to differing possible field evolution paths in neutron star cores \citep{2024MNRAS.tmp.2386L,2024arXiv241108020L} or an old analogue of the 6.7 hour spin period magnetar in SNR RCW 103 \citep{DeLuca2006,2016MNRAS.463.2394D,2016ApJ...828L..13R,2017MNRAS.464L..65H,2018MNRAS.478..741B}. 

\subsection*{Further Model Discussion}

\subsubsection*{NS Binary System}

The possibility of a normal neutron star in a binary with a low mass companion star (constrained by the absence of a bright optical counterpart) is also inconsistent with the observed data. The pulse morphology (see Extended Data Fig.~\ref{fig:wide_radio}) and the pulse phase alignment exclude rotation-powered recycled binary systems (i.e. redbacks or black windows \cite{2023MNRAS.525.3963K}). The quasi-simultaneous occurrence of radio and X-ray flares in \lpt\ contradicts the typical behaviour of low mass X-ray binaries (LMXBs). During LMXB accretion outbursts, the magnetospheric radius contracts, engulfing the light cylinder and suppressing radio pulsar activity. Radio emission from jets might be present, but it is not highly polarised, does not vary on these short timescales, nor does it have coherent brightness temperatures (see Extended Figure~\ref{fig:transient_phase}). 
Additionally, LMXBs usually exhibit rapid spin periods due to angular momentum transfer from the companion to the neutron star during past accretion phases. However, the 44-minute periodicity observed in \lpt\ is far too long compared to the millisecond spin periods typical of neutron stars in LMXBs. Similarly, the slow spin period and large radio luminosity is inconsistent with the transitional millisecond pulsars class \citep{Papitto2022}. If the periodicity was to represent the system's orbital period, it would imply a highly unusual geometry and viewing angle. The sharp pulse edges further argue against an orbital origin.

\subsubsection*{WD Binary System}

The recent discovery of two LPTs with M-type star counterparts (ILT~J1101+5521 and GLEAM-X~J0704--37) suggests that WD binaries are progenitors for some of the LPTs. However, their radio observational properties make them different from the previously discovered WD pulsars -- AR\,Sco and J1912--44. They may represent different evolutional stages of WD binaries (see \citep{2021NatAs...5..648S}).
Similarly, \lpt{} does not fit the properties of WD pulsars --- seven orders of magnitude higher radio luminosity and high fractional linear polarisation, but is more similar to ILT~J1101+5521 and GLEAM-X~J0704--37 (though still four orders of magnitude higher in radio luminosity).

The polarisation angle of \lpt\, across a pulse (Extended Data Fig.~\ref{fig:J1832_pol_pa}) varies $\lesssim 50^\circ$ demanding the observer is sampling a stable ordered magnetic field on the timescale of $\sim 3$ minutes. This disfavours the existence of a shorter unresolved underlying spin periodicity from a compact object, for instance from a compact asynchronous binary where the observed 44.2 min period is the orbital one. This severely limits the energy budget available in an asynchronous binary scenario \citep[e.g.,][]{1983ApJ...274L..71L,2024arXiv240905978Q}.
Circular polarisation towards the trailing edge of the pulse is suggestive of a propagation effect but could be the result of a variety of causes \citep[e.g.,][]{1997A&A...327..155G, 2024NatAs...8..606L}. The high linear polarisation demands small electron pitch angles in the emitting plasma, either intrinsic from injection or by strong cyclotron/synchrotron cooling to radiate them away; this requires a combination of either strong magnetic fields or extreme particle acceleration  \citep{2019ApJ...887...44D}. 

The radio radiation emission mechanism from the WD binary system is still poorly understood. There have been only a limited number of models explaining radio emission from such systems \citep[e.g.,][]{2024arXiv240905978Q}. We semi-quantitively estimated the magnetic field of the WD for \lpt{} under the assumption that the radio emission arises from the relativistic electron cyclotron maser emission (ECME) as discussed in \citep{2024arXiv240905978Q}. With the highest observed radio luminosity of $L_{\rm rad, max} = 5\times10^{31}\,$erg\,s$^{-1}$, a period of $P = 44.3\,$mins, and a radiation efficiency of $10^{-2}$, using the Eq.(11) in \citep{2024arXiv240905978Q}, we estimated a WD magnetic field of $B\approx5\times10^9\,$G.
However, it is well possible that the model does not apply to \lpt{}. \lpt\, might be a similar system but in a different evolutionary stage when accretion episodes are still viable, leading to the observed X-ray luminosity increase. We note that a 44\,min periodicity is generally too short to be the synchronized spin and orbit of a polar-like system, while it is more compatible with an IP stage \citep{2015SSRv..191..111F}. In this scenario, future observations might reveal an orbital modulation of the radio pulses in a $\sim$1--50\,hr timescale.

\subsection*{Comparison to other LPTs}

\begin{figure}
    \centering
    \includegraphics[width=\linewidth]{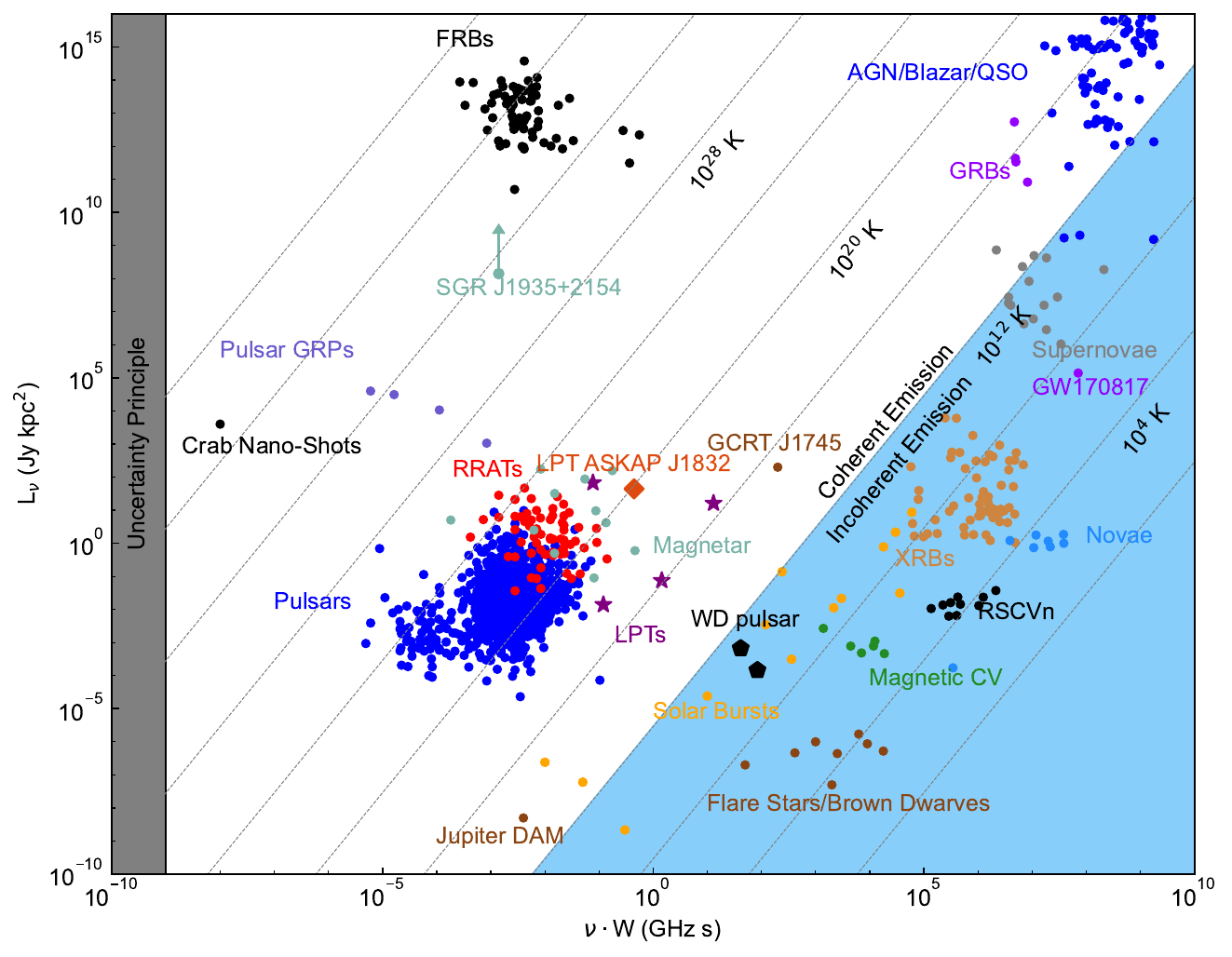}
    \caption{Radio transient parameter phase space plot of different types of radio transients, adapted from \citep{2022ApJ...940...72R, 2018NatAs...2..865K}. The brightest sub-pulse from \lpt\ is indicated with the orange diamond.}
    \label{fig:transient_phase}
\end{figure}

There have been seven LPTs discovered so far: 
GCRT~J1745--3009 with a period of $\sim$77~minutes \citep{2005Natur.434...50H}; 
GLEAMX~J162759.5--523504.3 with a period of $\sim$18~minutes \citep{2022Natur.601..526H}; GPM~J1839$-$10 with a period of $\sim$21~minutes \citep{2023Natur.619..487H}; ASKAP~J1935$+$2148 with a period of $\sim$54~minutes \citep{2024NatAs.tmp..107C}; CHIME~J0630+25 with a period of $\sim$7~minutes \citep{2024arXiv240707480D}; ILT~J1101+5521 with a period of $\sim$126 minutes \citep{2024arXiv240811536D}; and GLEAM-X~J0704$-$37 with a period of $\sim$2.9 hours \citep{2024arXiv240815757H}.
\lpt{} shares many similar properties as other LPTs including the period, high fractional linear polarisation, and radio spectral index (though \lpt{} has a relatively flatter radio spectrum compared to others). \lpt{} also locates at a similar region in the transient phase plot (see Extended Data Figure~\ref{fig:transient_phase}).

Many attempts have been made to observe LPTs at X-ray energies, while none of them have been successfully detected so far. Few observations have occurred when the LPTs were radio bright. GLEAM-X\,J1627$-52$ has the deepest limits, set by {\em Chandra}, when the source was radio-quiet, with a limit on its X-ray luminosity of $L_X \lesssim 10^{29} - 10^{30}\,$erg\,s$^{-1}$ \citep{2022ApJ...940...72R}. GPM\,J1839$-10$ had a deep XMM-Newton observation, with no detection, during which a bright radio burst was detected by MeerKAT \citep{2023Natur.619..487H}. Similarly, all other sources had short {\em Swift}-XRT observations that did not detect any counterparts with limits ranging between $L_X\lesssim 10^{30}-10^{33}$\,erg\,s$^{-1}$. 
However, this is not surprising because all other LPTs were detected with a low radio luminosity (see Figure~\ref{fig:J1832_xray_radio_parameter}).

\bibliography{sn-bibliography}


\begin{thebibliography}{116}
\ifx \bisbn   \undefined \def \bisbn  #1{ISBN #1}\fi
\ifx \binits  \undefined \def \binits#1{#1}\fi
\ifx \bauthor  \undefined \def \bauthor#1{#1}\fi
\ifx \batitle  \undefined \def \batitle#1{#1}\fi
\ifx \bjtitle  \undefined \def \bjtitle#1{#1}\fi
\ifx \bvolume  \undefined \def \bvolume#1{\textbf{#1}}\fi
\ifx \byear  \undefined \def \byear#1{#1}\fi
\ifx \bissue  \undefined \def \bissue#1{#1}\fi
\ifx \bfpage  \undefined \def \bfpage#1{#1}\fi
\ifx \blpage  \undefined \def \blpage #1{#1}\fi
\ifx \burl  \undefined \def \burl#1{\textsf{#1}}\fi
\ifx \doiurl  \undefined \def \doiurl#1{\url{https://doi.org/#1}}\fi
\ifx \betal  \undefined \def \betal{\textit{et al.}}\fi
\ifx \binstitute  \undefined \def \binstitute#1{#1}\fi
\ifx \binstitutionaled  \undefined \def \binstitutionaled#1{#1}\fi
\ifx \bctitle  \undefined \def \bctitle#1{#1}\fi
\ifx \beditor  \undefined \def \beditor#1{#1}\fi
\ifx \bpublisher  \undefined \def \bpublisher#1{#1}\fi
\ifx \bbtitle  \undefined \def \bbtitle#1{#1}\fi
\ifx \bedition  \undefined \def \bedition#1{#1}\fi
\ifx \bseriesno  \undefined \def \bseriesno#1{#1}\fi
\ifx \blocation  \undefined \def \blocation#1{#1}\fi
\ifx \bsertitle  \undefined \def \bsertitle#1{#1}\fi
\ifx \bsnm \undefined \def \bsnm#1{#1}\fi
\ifx \bsuffix \undefined \def \bsuffix#1{#1}\fi
\ifx \bparticle \undefined \def \bparticle#1{#1}\fi
\ifx \barticle \undefined \def \barticle#1{#1}\fi
\bibcommenthead
\ifx \bconfdate \undefined \def \bconfdate #1{#1}\fi
\ifx \botherref \undefined \def \botherref #1{#1}\fi
\ifx \url \undefined \def \url#1{\textsf{#1}}\fi
\ifx \bchapter \undefined \def \bchapter#1{#1}\fi
\ifx \bbook \undefined \def \bbook#1{#1}\fi
\ifx \bcomment \undefined \def \bcomment#1{#1}\fi
\ifx \oauthor \undefined \def \oauthor#1{#1}\fi
\ifx \citeauthoryear \undefined \def \citeauthoryear#1{#1}\fi
\ifx \endbibitem  \undefined \def \endbibitem {}\fi
\ifx \bconflocation  \undefined \def \bconflocation#1{#1}\fi
\ifx \arxivurl  \undefined \def \arxivurl#1{\textsf{#1}}\fi
\csname PreBibitemsHook\endcsname

\bibitem[\protect\citeauthoryear{{Hurley-Walker} et~al.}{2022}]{2022Natur.601..526H}
\begin{barticle}
\bauthor{\bsnm{{Hurley-Walker}}, \binits{N.}},
\bauthor{\bsnm{{Zhang}}, \binits{X.}},
\bauthor{\bsnm{{Bahramian}}, \binits{A.}},
\bauthor{\bsnm{{McSweeney}}, \binits{S.J.}},
\bauthor{\bsnm{{O'Doherty}}, \binits{T.N.}},
\bauthor{\bsnm{{Hancock}}, \binits{P.J.}},
\bauthor{\bsnm{{Morgan}}, \binits{J.S.}},
\bauthor{\bsnm{{Anderson}}, \binits{G.E.}},
\bauthor{\bsnm{{Heald}}, \binits{G.H.}},
\bauthor{\bsnm{{Galvin}}, \binits{T.J.}}:
\batitle{{A radio transient with unusually slow periodic emission}}.
\bjtitle{\nat}
\bvolume{601}(\bissue{7894}),
\bfpage{526}--\blpage{530}
(\byear{2022})
\doiurl{10.1038/s41586-021-04272-x}
\end{barticle}
\endbibitem

\bibitem[\protect\citeauthoryear{{Hurley-Walker} et~al.}{2023}]{2023Natur.619..487H}
\begin{barticle}
\bauthor{\bsnm{{Hurley-Walker}}, \binits{N.}},
\bauthor{\bsnm{{Rea}}, \binits{N.}},
\bauthor{\bsnm{{McSweeney}}, \binits{S.J.}},
\bauthor{\bsnm{{Meyers}}, \binits{B.W.}},
\bauthor{\bsnm{{Lenc}}, \binits{E.}},
\bauthor{\bsnm{{Heywood}}, \binits{I.}},
\bauthor{\bsnm{{Hyman}}, \binits{S.D.}},
\bauthor{\bsnm{{Men}}, \binits{Y.P.}},
\bauthor{\bsnm{{Clarke}}, \binits{T.E.}},
\bauthor{\bsnm{{Coti Zelati}}, \binits{F.}},
\bauthor{\bsnm{{Price}}, \binits{D.C.}},
\bauthor{\bsnm{{Horv{\'a}th}}, \binits{C.}},
\bauthor{\bsnm{{Galvin}}, \binits{T.J.}},
\bauthor{\bsnm{{Anderson}}, \binits{G.E.}},
\bauthor{\bsnm{{Bahramian}}, \binits{A.}},
\bauthor{\bsnm{{Barr}}, \binits{E.D.}},
\bauthor{\bsnm{{Bhat}}, \binits{N.D.R.}},
\bauthor{\bsnm{{Caleb}}, \binits{M.}},
\bauthor{\bsnm{{Dall'Ora}}, \binits{M.}},
\bauthor{\bsnm{{de Martino}}, \binits{D.}},
\bauthor{\bsnm{{Giacintucci}}, \binits{S.}},
\bauthor{\bsnm{{Morgan}}, \binits{J.S.}},
\bauthor{\bsnm{{Rajwade}}, \binits{K.M.}},
\bauthor{\bsnm{{Stappers}}, \binits{B.}},
\bauthor{\bsnm{{Williams}}, \binits{A.}}:
\batitle{{A long-period radio transient active for three decades}}.
\bjtitle{\nat}
\bvolume{619}(\bissue{7970}),
\bfpage{487}--\blpage{490}
(\byear{2023})
\doiurl{10.1038/s41586-023-06202-5}
\end{barticle}
\endbibitem

\bibitem[\protect\citeauthoryear{{Caleb} et~al.}{2024}]{2024NatAs.tmp..107C}
\begin{barticle}
\bauthor{\bsnm{{Caleb}}, \binits{M.}},
\bauthor{\bsnm{{Lenc}}, \binits{E.}},
\bauthor{\bsnm{{Kaplan}}, \binits{D.L.}},
\bauthor{\bsnm{{Murphy}}, \binits{T.}},
\bauthor{\bsnm{{Men}}, \binits{Y.P.}},
\bauthor{\bsnm{{Shannon}}, \binits{R.M.}},
\bauthor{\bsnm{{Ferrario}}, \binits{L.}},
\bauthor{\bsnm{{Rajwade}}, \binits{K.M.}},
\bauthor{\bsnm{{Clarke}}, \binits{T.E.}},
\bauthor{\bsnm{{Giacintucci}}, \binits{S.}},
\bauthor{\bsnm{{Hurley-Walker}}, \binits{N.}},
\bauthor{\bsnm{{Hyman}}, \binits{S.D.}},
\bauthor{\bsnm{{Lower}}, \binits{M.E.}},
\bauthor{\bsnm{{McSweeney}}, \binits{S.}},
\bauthor{\bsnm{{Ravi}}, \binits{V.}},
\bauthor{\bsnm{{Barr}}, \binits{E.D.}},
\bauthor{\bsnm{{Buchner}}, \binits{S.}},
\bauthor{\bsnm{{Flynn}}, \binits{C.M.L.}},
\bauthor{\bsnm{{Hessels}}, \binits{J.W.T.}},
\bauthor{\bsnm{{Kramer}}, \binits{M.}},
\bauthor{\bsnm{{Pritchard}}, \binits{J.}},
\bauthor{\bsnm{{Stappers}}, \binits{B.W.}}:
\batitle{{An emission-state-switching radio transient with a 54-minute period}}.
\bjtitle{Nature Astronomy}
(\byear{2024})
\doiurl{10.1038/s41550-024-02277-w}
{\href{https://arxiv.org/abs/2407.12266}{{arXiv:2407.12266}}}
{[astro-ph.HE]}
\end{barticle}
\endbibitem

\bibitem[\protect\citeauthoryear{{Dong} et~al.}{2024}]{2024arXiv240707480D}
\begin{botherref}
\oauthor{\bsnm{{Dong}}, \binits{F.A.}},
\oauthor{\bsnm{{Clarke}}, \binits{T.}},
\oauthor{\bsnm{{Curtin}}, \binits{A.P.}},
\oauthor{\bsnm{{Kumar}}, \binits{A.}},
\oauthor{\bsnm{{Stairs}}, \binits{I.}},
\oauthor{\bsnm{{Chatterjee}}, \binits{S.}},
\oauthor{\bsnm{{Cook}}, \binits{A.M.}},
\oauthor{\bsnm{{Fonseca}}, \binits{E.}},
\oauthor{\bsnm{{Gaensler}}, \binits{B.M.}},
\oauthor{\bsnm{{Hessels}}, \binits{J.W.T.}},
\oauthor{\bsnm{{Kaspi}}, \binits{V.M.}},
\oauthor{\bsnm{{Lazda}}, \binits{M.}},
\oauthor{\bsnm{{Masui}}, \binits{K.W.}},
\oauthor{\bsnm{{McKee}}, \binits{J.W.}},
\oauthor{\bsnm{{Meyers}}, \binits{B.W.}},
\oauthor{\bsnm{{Pearlman}}, \binits{A.B.}},
\oauthor{\bsnm{{Ransom}}, \binits{S.M.}},
\oauthor{\bsnm{{Scholz}}, \binits{P.}},
\oauthor{\bsnm{{Shin}}, \binits{K.}},
\oauthor{\bsnm{{Smith}}, \binits{K.M.}},
\oauthor{\bsnm{{Tan}}, \binits{C.M.}}:
{The discovery of a nearby 421\raisebox{-0.5ex}\textasciitilde transient with CHIME/FRB/Pulsar}.
arXiv e-prints,
2407--07480
(2024)
\doiurl{10.48550/arXiv.2407.07480}
{\href{https://arxiv.org/abs/2407.07480}{{arXiv:2407.07480}}}
{[astro-ph.HE]}
\end{botherref}
\endbibitem

\bibitem[\protect\citeauthoryear{{de Ruiter} et~al.}{2024}]{2024arXiv240811536D}
\begin{botherref}
\oauthor{\bsnm{{de Ruiter}}, \binits{I.}},
\oauthor{\bsnm{{Rajwade}}, \binits{K.M.}},
\oauthor{\bsnm{{Bassa}}, \binits{C.G.}},
\oauthor{\bsnm{{Rowlinson}}, \binits{A.}},
\oauthor{\bsnm{{Wijers}}, \binits{R.A.M.J.}},
\oauthor{\bsnm{{Kilpatrick}}, \binits{C.D.}},
\oauthor{\bsnm{{Stefansson}}, \binits{G.}},
\oauthor{\bsnm{{Callingham}}, \binits{J.R.}},
\oauthor{\bsnm{{Hessels}}, \binits{J.W.T.}},
\oauthor{\bsnm{{Clarke}}, \binits{T.E.}},
\oauthor{\bsnm{{Peters}}, \binits{W.}},
\oauthor{\bsnm{{Wijnands}}, \binits{R.A.D.}},
\oauthor{\bsnm{{Shimwell}}, \binits{T.W.}},
\oauthor{\bsnm{{ter Veen}}, \binits{S.}},
\oauthor{\bsnm{{Morello}}, \binits{V.}},
\oauthor{\bsnm{{Zeimann}}, \binits{G.R.}},
\oauthor{\bsnm{{Mahadevan}}, \binits{S.}}:
{A white dwarf binary showing sporadic radio pulses at the orbital period}.
arXiv e-prints,
2408--11536
(2024)
\doiurl{10.48550/arXiv.2408.11536}
{\href{https://arxiv.org/abs/2408.11536}{{arXiv:2408.11536}}}
{[astro-ph.HE]}
\end{botherref}
\endbibitem

\bibitem[\protect\citeauthoryear{{Cooper} and {Wadiasingh}}{2024}]{2024MNRAS.533.2133C}
\begin{barticle}
\bauthor{\bsnm{{Cooper}}, \binits{A.J.}},
\bauthor{\bsnm{{Wadiasingh}}, \binits{Z.}}:
\batitle{{Beyond the Rotational Deathline: Radio Emission from Ultra-long Period Magnetars}}.
\bjtitle{\mnras}
\bvolume{533}(\bissue{2}),
\bfpage{2133}--\blpage{2155}
(\byear{2024})
\doiurl{10.1093/mnras/stae1813}
{\href{https://arxiv.org/abs/2406.04135}{{arXiv:2406.04135}}}
{[astro-ph.HE]}
\end{barticle}
\endbibitem

\bibitem[\protect\citeauthoryear{{Katz}}{2022}]{2022Ap&SS.367..108K}
\begin{barticle}
\bauthor{\bsnm{{Katz}}, \binits{J.I.}}:
\batitle{{GLEAM-X J162759.5‑523504.3 as a white dwarf pulsar}}.
\bjtitle{\apss}
\bvolume{367}(\bissue{11}),
\bfpage{108}
(\byear{2022})
\doiurl{10.1007/s10509-022-04146-2}
{\href{https://arxiv.org/abs/2203.08112}{{arXiv:2203.08112}}}
{[astro-ph.SR]}
\end{barticle}
\endbibitem

\bibitem[\protect\citeauthoryear{{Qu} and {Zhang}}{2024}]{2024arXiv240905978Q}
\begin{botherref}
\oauthor{\bsnm{{Qu}}, \binits{Y.}},
\oauthor{\bsnm{{Zhang}}, \binits{B.}}:
{Magnetic Interaction in White Dwarf Binaries as Mechanism for Long-Period Radio Transients}.
arXiv e-prints,
2409--05978
(2024)
\doiurl{10.48550/arXiv.2409.05978}
{\href{https://arxiv.org/abs/2409.05978}{{arXiv:2409.05978}}}
{[astro-ph.HE]}
\end{botherref}
\endbibitem

\bibitem[\protect\citeauthoryear{{Schwope} et~al.}{2023}]{2023A&A...674L...9S}
\begin{barticle}
\bauthor{\bsnm{{Schwope}}, \binits{A.}},
\bauthor{\bsnm{{Marsh}}, \binits{T.R.}},
\bauthor{\bsnm{{Standke}}, \binits{A.}},
\bauthor{\bsnm{{Pelisoli}}, \binits{I.}},
\bauthor{\bsnm{{Potter}}, \binits{S.}},
\bauthor{\bsnm{{Buckley}}, \binits{D.}},
\bauthor{\bsnm{{Munday}}, \binits{J.}},
\bauthor{\bsnm{{Dhillon}}, \binits{V.}}:
\batitle{{X-ray properties of the white dwarf pulsar eRASSU J191213.9{\ensuremath{-}}441044}}.
\bjtitle{\aap}
\bvolume{674},
\bfpage{9}
(\byear{2023})
\doiurl{10.1051/0004-6361/202346589}
{\href{https://arxiv.org/abs/2306.09732}{{arXiv:2306.09732}}}
{[astro-ph.HE]}
\end{barticle}
\endbibitem

\bibitem[\protect\citeauthoryear{{Rea} et~al.}{2022}]{2022ApJ...940...72R}
\begin{barticle}
\bauthor{\bsnm{{Rea}}, \binits{N.}},
\bauthor{\bsnm{{Coti Zelati}}, \binits{F.}},
\bauthor{\bsnm{{Dehman}}, \binits{C.}},
\bauthor{\bsnm{{Hurley-Walker}}, \binits{N.}},
\bauthor{\bsnm{{de Martino}}, \binits{D.}},
\bauthor{\bsnm{{Bahramian}}, \binits{A.}},
\bauthor{\bsnm{{Buckley}}, \binits{D.A.H.}},
\bauthor{\bsnm{{Brink}}, \binits{J.}},
\bauthor{\bsnm{{Kawka}}, \binits{A.}},
\bauthor{\bsnm{{Pons}}, \binits{J.A.}},
\bauthor{\bsnm{{Vigan{\`o}}}, \binits{D.}},
\bauthor{\bsnm{{Graber}}, \binits{V.}},
\bauthor{\bsnm{{Ronchi}}, \binits{M.}},
\bauthor{\bsnm{{Pardo Araujo}}, \binits{C.}},
\bauthor{\bsnm{{Borghese}}, \binits{A.}},
\bauthor{\bsnm{{Parent}}, \binits{E.}},
\bauthor{\bsnm{{Galvin}}, \binits{T.J.}}:
\batitle{{Constraining the Nature of the 18 min Periodic Radio Transient GLEAM-X J162759.5-523504.3 via Multiwavelength Observations and Magneto-thermal Simulations}}.
\bjtitle{\apj}
\bvolume{940}(\bissue{1}),
\bfpage{72}
(\byear{2022})
\doiurl{10.3847/1538-4357/ac97ea}
{\href{https://arxiv.org/abs/2210.01903}{{arXiv:2210.01903}}}
{[astro-ph.HE]}
\end{barticle}
\endbibitem

\bibitem[\protect\citeauthoryear{{Hotan} et~al.}{2021}]{2021PASA...38....9H}
\begin{barticle}
\bauthor{\bsnm{{Hotan}}, \binits{A.W.}},
\bauthor{\bsnm{{Bunton}}, \binits{J.D.}},
\bauthor{\bsnm{{Chippendale}}, \binits{A.P.}},
\bauthor{\bsnm{{Whiting}}, \binits{M.}},
\bauthor{\bsnm{{Tuthill}}, \binits{J.}},
\bauthor{\bsnm{{Moss}}, \binits{V.A.}},
\bauthor{\bsnm{{McConnell}}, \binits{D.}},
\bauthor{\bsnm{{Amy}}, \binits{S.W.}},
\bauthor{\bsnm{{Huynh}}, \binits{M.T.}},
\bauthor{\bsnm{{Allison}}, \binits{J.R.}},
\bauthor{\bsnm{{Anderson}}, \binits{C.S.}},
\bauthor{\bsnm{{Bannister}}, \binits{K.W.}},
\bauthor{\bsnm{{Bastholm}}, \binits{E.}},
\bauthor{\bsnm{{Beresford}}, \binits{R.}},
\bauthor{\bsnm{{Bock}}, \binits{D.C.-J.}},
\bauthor{\bsnm{{Bolton}}, \binits{R.}},
\bauthor{\bsnm{{Chapman}}, \binits{J.M.}},
\bauthor{\bsnm{{Chow}}, \binits{K.}},
\bauthor{\bsnm{{Collier}}, \binits{J.D.}},
\bauthor{\bsnm{{Cooray}}, \binits{F.R.}},
\bauthor{\bsnm{{Cornwell}}, \binits{T.J.}},
\bauthor{\bsnm{{Diamond}}, \binits{P.J.}},
\bauthor{\bsnm{{Edwards}}, \binits{P.G.}},
\bauthor{\bsnm{{Feain}}, \binits{I.J.}},
\bauthor{\bsnm{{Franzen}}, \binits{T.M.O.}},
\bauthor{\bsnm{{George}}, \binits{D.}},
\bauthor{\bsnm{{Gupta}}, \binits{N.}},
\bauthor{\bsnm{{Hampson}}, \binits{G.A.}},
\bauthor{\bsnm{{Harvey-Smith}}, \binits{L.}},
\bauthor{\bsnm{{Hayman}}, \binits{D.B.}},
\bauthor{\bsnm{{Heywood}}, \binits{I.}},
\bauthor{\bsnm{{Jacka}}, \binits{C.}},
\bauthor{\bsnm{{Jackson}}, \binits{C.A.}},
\bauthor{\bsnm{{Jackson}}, \binits{S.}},
\bauthor{\bsnm{{Jeganathan}}, \binits{K.}},
\bauthor{\bsnm{{Johnston}}, \binits{S.}},
\bauthor{\bsnm{{Kesteven}}, \binits{M.}},
\bauthor{\bsnm{{Kleiner}}, \binits{D.}},
\bauthor{\bsnm{{Koribalski}}, \binits{B.S.}},
\bauthor{\bsnm{{Lee-Waddell}}, \binits{K.}},
\bauthor{\bsnm{{Lenc}}, \binits{E.}},
\bauthor{\bsnm{{Lensson}}, \binits{E.S.}},
\bauthor{\bsnm{{Mackay}}, \binits{S.}},
\bauthor{\bsnm{{Mahony}}, \binits{E.K.}},
\bauthor{\bsnm{{McClure-Griffiths}}, \binits{N.M.}},
\bauthor{\bsnm{{McConigley}}, \binits{R.}},
\bauthor{\bsnm{{Mirtschin}}, \binits{P.}},
\bauthor{\bsnm{{Ng}}, \binits{A.K.}},
\bauthor{\bsnm{{Norris}}, \binits{R.P.}},
\bauthor{\bsnm{{Pearce}}, \binits{S.E.}},
\bauthor{\bsnm{{Phillips}}, \binits{C.}},
\bauthor{\bsnm{{Pilawa}}, \binits{M.A.}},
\bauthor{\bsnm{{Raja}}, \binits{W.}},
\bauthor{\bsnm{{Reynolds}}, \binits{J.E.}},
\bauthor{\bsnm{{Roberts}}, \binits{P.}},
\bauthor{\bsnm{{Roxby}}, \binits{D.N.}},
\bauthor{\bsnm{{Sadler}}, \binits{E.M.}},
\bauthor{\bsnm{{Shields}}, \binits{M.}},
\bauthor{\bsnm{{Schinckel}}, \binits{A.E.T.}},
\bauthor{\bsnm{{Serra}}, \binits{P.}},
\bauthor{\bsnm{{Shaw}}, \binits{R.D.}},
\bauthor{\bsnm{{Sweetnam}}, \binits{T.}},
\bauthor{\bsnm{{Troup}}, \binits{E.R.}},
\bauthor{\bsnm{{Tzioumis}}, \binits{A.}},
\bauthor{\bsnm{{Voronkov}}, \binits{M.A.}},
\bauthor{\bsnm{{Westmeier}}, \binits{T.}}:
\batitle{{Australian square kilometre array pathfinder: I. system description}}.
\bjtitle{\pasa}
\bvolume{38},
\bfpage{009}
(\byear{2021})
\doiurl{10.1017/pasa.2021.1}
{\href{https://arxiv.org/abs/2102.01870}{{arXiv:2102.01870}}}
{[astro-ph.IM]}
\end{barticle}
\endbibitem

\bibitem[\protect\citeauthoryear{{Murphy} et~al.}{2013}]{2013PASA...30....6M}
\begin{barticle}
\bauthor{\bsnm{{Murphy}}, \binits{T.}},
\bauthor{\bsnm{{Chatterjee}}, \binits{S.}},
\bauthor{\bsnm{{Kaplan}}, \binits{D.L.}},
\bauthor{\bsnm{{Banyer}}, \binits{J.}},
\bauthor{\bsnm{{Bell}}, \binits{M.E.}},
\bauthor{\bsnm{{Bignall}}, \binits{H.E.}},
\bauthor{\bsnm{{Bower}}, \binits{G.C.}},
\bauthor{\bsnm{{Cameron}}, \binits{R.A.}},
\bauthor{\bsnm{{Coward}}, \binits{D.M.}},
\bauthor{\bsnm{{Cordes}}, \binits{J.M.}},
\bauthor{\bsnm{{Croft}}, \binits{S.}},
\bauthor{\bsnm{{Curran}}, \binits{J.R.}},
\bauthor{\bsnm{{Djorgovski}}, \binits{S.G.}},
\bauthor{\bsnm{{Farrell}}, \binits{S.A.}},
\bauthor{\bsnm{{Frail}}, \binits{D.A.}},
\bauthor{\bsnm{{Gaensler}}, \binits{B.M.}},
\bauthor{\bsnm{{Galloway}}, \binits{D.K.}},
\bauthor{\bsnm{{Gendre}}, \binits{B.}},
\bauthor{\bsnm{{Green}}, \binits{A.J.}},
\bauthor{\bsnm{{Hancock}}, \binits{P.J.}},
\bauthor{\bsnm{{Johnston}}, \binits{S.}},
\bauthor{\bsnm{{Kamble}}, \binits{A.}},
\bauthor{\bsnm{{Law}}, \binits{C.J.}},
\bauthor{\bsnm{{Lazio}}, \binits{T.J.W.}},
\bauthor{\bsnm{{Lo}}, \binits{K.K.}},
\bauthor{\bsnm{{Macquart}}, \binits{J.-P.}},
\bauthor{\bsnm{{Rea}}, \binits{N.}},
\bauthor{\bsnm{{Rebbapragada}}, \binits{U.}},
\bauthor{\bsnm{{Reynolds}}, \binits{C.}},
\bauthor{\bsnm{{Ryder}}, \binits{S.D.}},
\bauthor{\bsnm{{Schmidt}}, \binits{B.}},
\bauthor{\bsnm{{Soria}}, \binits{R.}},
\bauthor{\bsnm{{Stairs}}, \binits{I.H.}},
\bauthor{\bsnm{{Tingay}}, \binits{S.J.}},
\bauthor{\bsnm{{Torkelsson}}, \binits{U.}},
\bauthor{\bsnm{{Wagstaff}}, \binits{K.}},
\bauthor{\bsnm{{Walker}}, \binits{M.}},
\bauthor{\bsnm{{Wayth}}, \binits{R.B.}},
\bauthor{\bsnm{{Williams}}, \binits{P.K.G.}}:
\batitle{{VAST: An ASKAP Survey for Variables and Slow Transients}}.
\bjtitle{\pasa}
\bvolume{30},
\bfpage{006}
(\byear{2013})
\doiurl{10.1017/pasa.2012.006}
{\href{https://arxiv.org/abs/1207.1528}{{arXiv:1207.1528}}}
{[astro-ph.IM]}
\end{barticle}
\endbibitem

\bibitem[\protect\citeauthoryear{{Murphy} et~al.}{2021}]{2021PASA...38...54M}
\begin{barticle}
\bauthor{\bsnm{{Murphy}}, \binits{T.}},
\bauthor{\bsnm{{Kaplan}}, \binits{D.L.}},
\bauthor{\bsnm{{Stewart}}, \binits{A.J.}},
\bauthor{\bsnm{{O'Brien}}, \binits{A.}},
\bauthor{\bsnm{{Lenc}}, \binits{E.}},
\bauthor{\bsnm{{Pintaldi}}, \binits{S.}},
\bauthor{\bsnm{{Pritchard}}, \binits{J.}},
\bauthor{\bsnm{{Dobie}}, \binits{D.}},
\bauthor{\bsnm{{Fox}}, \binits{A.}},
\bauthor{\bsnm{{Leung}}, \binits{J.K.}},
\bauthor{\bsnm{{An}}, \binits{T.}},
\bauthor{\bsnm{{Bell}}, \binits{M.E.}},
\bauthor{\bsnm{{Broderick}}, \binits{J.W.}},
\bauthor{\bsnm{{Chatterjee}}, \binits{S.}},
\bauthor{\bsnm{{Dai}}, \binits{S.}},
\bauthor{\bsnm{{d'Antonio}}, \binits{D.}},
\bauthor{\bsnm{{Doyle}}, \binits{G.}},
\bauthor{\bsnm{{Gaensler}}, \binits{B.M.}},
\bauthor{\bsnm{{Heald}}, \binits{G.}},
\bauthor{\bsnm{{Horesh}}, \binits{A.}},
\bauthor{\bsnm{{Jones}}, \binits{M.L.}},
\bauthor{\bsnm{{McConnell}}, \binits{D.}},
\bauthor{\bsnm{{Moss}}, \binits{V.A.}},
\bauthor{\bsnm{{Raja}}, \binits{W.}},
\bauthor{\bsnm{{Ramsay}}, \binits{G.}},
\bauthor{\bsnm{{Ryder}}, \binits{S.}},
\bauthor{\bsnm{{Sadler}}, \binits{E.M.}},
\bauthor{\bsnm{{Sivakoff}}, \binits{G.R.}},
\bauthor{\bsnm{{Wang}}, \binits{Y.}},
\bauthor{\bsnm{{Wang}}, \binits{Z.}},
\bauthor{\bsnm{{Wheatland}}, \binits{M.S.}},
\bauthor{\bsnm{{Whiting}}, \binits{M.}},
\bauthor{\bsnm{{Allison}}, \binits{J.R.}},
\bauthor{\bsnm{{Anderson}}, \binits{C.S.}},
\bauthor{\bsnm{{Ball}}, \binits{L.}},
\bauthor{\bsnm{{Bannister}}, \binits{K.}},
\bauthor{\bsnm{{Bock}}, \binits{D.C.-J.}},
\bauthor{\bsnm{{Bolton}}, \binits{R.}},
\bauthor{\bsnm{{Bunton}}, \binits{J.D.}},
\bauthor{\bsnm{{Chekkala}}, \binits{R.}},
\bauthor{\bsnm{{Chippendale}}, \binits{A.P.}},
\bauthor{\bsnm{{Cooray}}, \binits{F.R.}},
\bauthor{\bsnm{{Gupta}}, \binits{N.}},
\bauthor{\bsnm{{Hayman}}, \binits{D.B.}},
\bauthor{\bsnm{{Jeganathan}}, \binits{K.}},
\bauthor{\bsnm{{Koribalski}}, \binits{B.}},
\bauthor{\bsnm{{Lee-Waddell}}, \binits{K.}},
\bauthor{\bsnm{{Mahony}}, \binits{E.K.}},
\bauthor{\bsnm{{Marvil}}, \binits{J.}},
\bauthor{\bsnm{{McClure-Griffiths}}, \binits{N.M.}},
\bauthor{\bsnm{{Mirtschin}}, \binits{P.}},
\bauthor{\bsnm{{Ng}}, \binits{A.}},
\bauthor{\bsnm{{Pearce}}, \binits{S.}},
\bauthor{\bsnm{{Phillips}}, \binits{C.}},
\bauthor{\bsnm{{Voronkov}}, \binits{M.A.}}:
\batitle{{The ASKAP Variables and Slow Transients (VAST) Pilot Survey}}.
\bjtitle{\pasa}
\bvolume{38},
\bfpage{054}
(\byear{2021})
\doiurl{10.1017/pasa.2021.44}
{\href{https://arxiv.org/abs/2108.06039}{{arXiv:2108.06039}}}
{[astro-ph.HE]}
\end{barticle}
\endbibitem

\bibitem[\protect\citeauthoryear{{Wang} et~al.}{2024}]{2024arXiv240910316W}
\begin{botherref}
\oauthor{\bsnm{{Wang}}, \binits{Z.}},
\oauthor{\bsnm{{Bannister}}, \binits{K.W.}},
\oauthor{\bsnm{{Gupta}}, \binits{V.}},
\oauthor{\bsnm{{Deng}}, \binits{X.}},
\oauthor{\bsnm{{Pilawa}}, \binits{M.}},
\oauthor{\bsnm{{Tuthill}}, \binits{J.}},
\oauthor{\bsnm{{Bunton}}, \binits{J.D.}},
\oauthor{\bsnm{{Flynn}}, \binits{C.}},
\oauthor{\bsnm{{Glowacki}}, \binits{M.}},
\oauthor{\bsnm{{Jaini}}, \binits{A.}},
\oauthor{\bsnm{{Lee}}, \binits{Y.W.J.}},
\oauthor{\bsnm{{Lenc}}, \binits{E.}},
\oauthor{\bsnm{{Lucero}}, \binits{J.}},
\oauthor{\bsnm{{Paek}}, \binits{A.}},
\oauthor{\bsnm{{Radhakrishnan}}, \binits{R.}},
\oauthor{\bsnm{{Thyagarajan}}, \binits{N.}},
\oauthor{\bsnm{{Uttarkar}}, \binits{P.}},
\oauthor{\bsnm{{Wang}}, \binits{Y.}},
\oauthor{\bsnm{{Bhat}}, \binits{N.D.R.}},
\oauthor{\bsnm{{James}}, \binits{C.W.}},
\oauthor{\bsnm{{Moss}}, \binits{V.A.}},
\oauthor{\bsnm{{Murphy}}, \binits{T.}},
\oauthor{\bsnm{{Reynolds}}, \binits{J.E.}},
\oauthor{\bsnm{{Shannon}}, \binits{R.M.}},
\oauthor{\bsnm{{Spitler}}, \binits{L.G.}},
\oauthor{\bsnm{{Tzioumis}}, \binits{A.}},
\oauthor{\bsnm{{Caleb}}, \binits{M.}},
\oauthor{\bsnm{{Deller}}, \binits{A.T.}},
\oauthor{\bsnm{{Gordon}}, \binits{A.C.}},
\oauthor{\bsnm{{Marnoch}}, \binits{L.}},
\oauthor{\bsnm{{Ryder}}, \binits{S.D.}},
\oauthor{\bsnm{{Simha}}, \binits{S.}},
\oauthor{\bsnm{{Anderson}}, \binits{C.S.}},
\oauthor{\bsnm{{Ball}}, \binits{L.}},
\oauthor{\bsnm{{Brodrick}}, \binits{D.}},
\oauthor{\bsnm{{Cooray}}, \binits{F.R.}},
\oauthor{\bsnm{{Gupta}}, \binits{N.}},
\oauthor{\bsnm{{Hayman}}, \binits{D.B.}},
\oauthor{\bsnm{{Ng}}, \binits{A.}},
\oauthor{\bsnm{{Pearce}}, \binits{S.E.}},
\oauthor{\bsnm{{Phillips}}, \binits{C.}},
\oauthor{\bsnm{{Voronkov}}, \binits{M.A.}},
\oauthor{\bsnm{{Westmeier}}, \binits{T.}}:
{The CRAFT Coherent (CRACO) upgrade I: System Description and Results of the 110-ms Radio Transient Pilot Survey}.
arXiv e-prints,
2409--10316
(2024)
\doiurl{10.48550/arXiv.2409.10316}
{\href{https://arxiv.org/abs/2409.10316}{{arXiv:2409.10316}}}
{[astro-ph.HE]}
\end{botherref}
\endbibitem

\bibitem[\protect\citeauthoryear{{Yao} et~al.}{2017}]{2017ApJ...835...29Y}
\begin{barticle}
\bauthor{\bsnm{{Yao}}, \binits{J.M.}},
\bauthor{\bsnm{{Manchester}}, \binits{R.N.}},
\bauthor{\bsnm{{Wang}}, \binits{N.}}:
\batitle{{A New Electron-density Model for Estimation of Pulsar and FRB Distances}}.
\bjtitle{\apj}
\bvolume{835}(\bissue{1}),
\bfpage{29}
(\byear{2017})
\doiurl{10.3847/1538-4357/835/1/29}
{\href{https://arxiv.org/abs/1610.09448}{{arXiv:1610.09448}}}
{[astro-ph.GA]}
\end{barticle}
\endbibitem

\bibitem[\protect\citeauthoryear{{Wenger} et~al.}{2018}]{2018ApJ...856...52W}
\begin{barticle}
\bauthor{\bsnm{{Wenger}}, \binits{T.V.}},
\bauthor{\bsnm{{Balser}}, \binits{D.S.}},
\bauthor{\bsnm{{Anderson}}, \binits{L.D.}},
\bauthor{\bsnm{{Bania}}, \binits{T.M.}}:
\batitle{{Kinematic Distances: A Monte Carlo Method}}.
\bjtitle{\apj}
\bvolume{856}(\bissue{1}),
\bfpage{52}
(\byear{2018})
\doiurl{10.3847/1538-4357/aaaec8}
{\href{https://arxiv.org/abs/1802.04203}{{arXiv:1802.04203}}}
{[astro-ph.GA]}
\end{barticle}
\endbibitem

\bibitem[\protect\citeauthoryear{{Wang} et~al.}{2020}]{2020A&A...639A..72W}
\begin{barticle}
\bauthor{\bsnm{{Wang}}, \binits{S.}},
\bauthor{\bsnm{{Zhang}}, \binits{C.}},
\bauthor{\bsnm{{Jiang}}, \binits{B.}},
\bauthor{\bsnm{{Zhao}}, \binits{H.}},
\bauthor{\bsnm{{Chen}}, \binits{B.}},
\bauthor{\bsnm{{Chen}}, \binits{X.}},
\bauthor{\bsnm{{Gao}}, \binits{J.}},
\bauthor{\bsnm{{Liu}}, \binits{J.}}:
\batitle{{Distances to the supernova remnants in the inner disk}}.
\bjtitle{\aap}
\bvolume{639},
\bfpage{72}
(\byear{2020})
\doiurl{10.1051/0004-6361/201936868}
{\href{https://arxiv.org/abs/2005.08270}{{arXiv:2005.08270}}}
{[astro-ph.GA]}
\end{barticle}
\endbibitem

\bibitem[\protect\citeauthoryear{{Mart{\'\i}-Devesa} and {Reimer}}{2020}]{2020A&A...637A..23M}
\begin{barticle}
\bauthor{\bsnm{{Mart{\'\i}-Devesa}}, \binits{G.}},
\bauthor{\bsnm{{Reimer}}, \binits{O.}}:
\batitle{{X-ray and {\ensuremath{\gamma}}-ray orbital variability from the {\ensuremath{\gamma}}-ray binary HESS J1832-093}}.
\bjtitle{\aap}
\bvolume{637},
\bfpage{23}
(\byear{2020})
\doiurl{10.1051/0004-6361/202037442}
{\href{https://arxiv.org/abs/2001.02701}{{arXiv:2001.02701}}}
{[astro-ph.HE]}
\end{barticle}
\endbibitem

\bibitem[\protect\citeauthoryear{{Yuan} et~al.}{2022}]{2022hxga.book...86Y}
\begin{bchapter}
\bauthor{\bsnm{{Yuan}}, \binits{W.}},
\bauthor{\bsnm{{Zhang}}, \binits{C.}},
\bauthor{\bsnm{{Chen}}, \binits{Y.}},
\bauthor{\bsnm{{Ling}}, \binits{Z.}}:
\bctitle{{The Einstein Probe Mission}}.
In: \beditor{\bsnm{{Bambi}}, \binits{C.}},
\beditor{\bsnm{{Sangangelo}}, \binits{A.}} (eds.)
\bbtitle{Handbook of X-ray and Gamma-ray Astrophysics},
p. \bfpage{86}
(\byear{2022}).
\doiurl{10.1007/978-981-16-4544-0_151-1}
\end{bchapter}
\endbibitem

\bibitem[\protect\citeauthoryear{{Chen} and {Ruderman}}{1993}]{1993ApJ...402..264C}
\begin{barticle}
\bauthor{\bsnm{{Chen}}, \binits{K.}},
\bauthor{\bsnm{{Ruderman}}, \binits{M.}}:
\batitle{{Pulsar Death Lines and Death Valley}}.
\bjtitle{\apj}
\bvolume{402},
\bfpage{264}
(\byear{1993})
\doiurl{10.1086/172129}
\end{barticle}
\endbibitem

\bibitem[\protect\citeauthoryear{{Zhang} et~al.}{2000}]{2000ApJ...531L.135Z}
\begin{barticle}
\bauthor{\bsnm{{Zhang}}, \binits{B.}},
\bauthor{\bsnm{{Harding}}, \binits{A.K.}},
\bauthor{\bsnm{{Muslimov}}, \binits{A.G.}}:
\batitle{{Radio Pulsar Death Line Revisited: Is PSR J2144-3933 Anomalous?}}
\bjtitle{\apjl}
\bvolume{531}(\bissue{2}),
\bfpage{135}--\blpage{138}
(\byear{2000})
\doiurl{10.1086/312542}
{\href{https://arxiv.org/abs/astro-ph/0001341}{{arXiv:astro-ph/0001341}}}
{[astro-ph]}
\end{barticle}
\endbibitem

\bibitem[\protect\citeauthoryear{{Harding} and {Muslimov}}{2011}]{2011ApJ...726L..10H}
\begin{barticle}
\bauthor{\bsnm{{Harding}}, \binits{A.K.}},
\bauthor{\bsnm{{Muslimov}}, \binits{A.G.}}:
\batitle{{Pulsar Pair Cascades in a Distorted Magnetic Dipole Field}}.
\bjtitle{\apjl}
\bvolume{726}(\bissue{1}),
\bfpage{10}
(\byear{2011})
\doiurl{10.1088/2041-8205/726/1/L10}
{\href{https://arxiv.org/abs/1012.0451}{{arXiv:1012.0451}}}
{[astro-ph.HE]}
\end{barticle}
\endbibitem

\bibitem[\protect\citeauthoryear{{Becker} and {Truemper}}{1997}]{1997A&A...326..682B}
\begin{botherref}
\oauthor{\bsnm{{Becker}}, \binits{W.}},
\oauthor{\bsnm{{Truemper}}, \binits{J.}}:
{The X-ray luminosity of rotation-powered neutron stars.}
\textbf{326},
682--691
(1997)
{\href{https://arxiv.org/abs/astro-ph/9708169}{{arXiv:astro-ph/9708169}}}
{[astro-ph]}
\end{botherref}
\endbibitem

\bibitem[\protect\citeauthoryear{{Saumon} et~al.}{2022}]{2022PhR...988....1S}
\begin{barticle}
\bauthor{\bsnm{{Saumon}}, \binits{D.}},
\bauthor{\bsnm{{Blouin}}, \binits{S.}},
\bauthor{\bsnm{{Tremblay}}, \binits{P.-E.}}:
\batitle{{Current challenges in the physics of white dwarf stars}}.
\bjtitle{\physrep}
\bvolume{988},
\bfpage{1}--\blpage{63}
(\byear{2022})
\doiurl{10.1016/j.physrep.2022.09.001}
{\href{https://arxiv.org/abs/2209.02846}{{arXiv:2209.02846}}}
{[astro-ph.SR]}
\end{barticle}
\endbibitem

\bibitem[\protect\citeauthoryear{{Heise}}{1985}]{1985SSRv...40...79H}
\begin{barticle}
\bauthor{\bsnm{{Heise}}, \binits{J.}}:
\batitle{{X-Ray Emission from Isolated Hot White Dwarfs}}.
\bjtitle{\ssr}
\bvolume{40}(\bissue{1-2}),
\bfpage{79}--\blpage{90}
(\byear{1985})
\doiurl{10.1007/BF00212870}
\end{barticle}
\endbibitem

\bibitem[\protect\citeauthoryear{{Beniamini} et~al.}{2023}]{2023MNRAS.520.1872B}
\begin{barticle}
\bauthor{\bsnm{{Beniamini}}, \binits{P.}},
\bauthor{\bsnm{{Wadiasingh}}, \binits{Z.}},
\bauthor{\bsnm{{Hare}}, \binits{J.}},
\bauthor{\bsnm{{Rajwade}}, \binits{K.M.}},
\bauthor{\bsnm{{Younes}}, \binits{G.}},
\bauthor{\bsnm{{van der Horst}}, \binits{A.J.}}:
\batitle{{Evidence for an abundant old population of Galactic ultra-long period magnetars and implications for fast radio bursts}}.
\bjtitle{\mnras}
\bvolume{520}(\bissue{2}),
\bfpage{1872}--\blpage{1894}
(\byear{2023})
\doiurl{10.1093/mnras/stad208}
{\href{https://arxiv.org/abs/2210.09323}{{arXiv:2210.09323}}}
{[astro-ph.HE]}
\end{barticle}
\endbibitem

\bibitem[\protect\citeauthoryear{{Marsh} et~al.}{2016}]{2016Natur.537..374M}
\begin{barticle}
\bauthor{\bsnm{{Marsh}}, \binits{T.R.}},
\bauthor{\bsnm{{G{\"a}nsicke}}, \binits{B.T.}},
\bauthor{\bsnm{{H{\"u}mmerich}}, \binits{S.}},
\bauthor{\bsnm{{Hambsch}}, \binits{F.-J.}},
\bauthor{\bsnm{{Bernhard}}, \binits{K.}},
\bauthor{\bsnm{{Lloyd}}, \binits{C.}},
\bauthor{\bsnm{{Breedt}}, \binits{E.}},
\bauthor{\bsnm{{Stanway}}, \binits{E.R.}},
\bauthor{\bsnm{{Steeghs}}, \binits{D.T.}},
\bauthor{\bsnm{{Parsons}}, \binits{S.G.}},
\bauthor{\bsnm{{Toloza}}, \binits{O.}},
\bauthor{\bsnm{{Schreiber}}, \binits{M.R.}},
\bauthor{\bsnm{{Jonker}}, \binits{P.G.}},
\bauthor{\bsnm{{van Roestel}}, \binits{J.}},
\bauthor{\bsnm{{Kupfer}}, \binits{T.}},
\bauthor{\bsnm{{Pala}}, \binits{A.F.}},
\bauthor{\bsnm{{Dhillon}}, \binits{V.S.}},
\bauthor{\bsnm{{Hardy}}, \binits{L.K.}},
\bauthor{\bsnm{{Littlefair}}, \binits{S.P.}},
\bauthor{\bsnm{{Aungwerojwit}}, \binits{A.}},
\bauthor{\bsnm{{Arjyotha}}, \binits{S.}},
\bauthor{\bsnm{{Koester}}, \binits{D.}},
\bauthor{\bsnm{{Bochinski}}, \binits{J.J.}},
\bauthor{\bsnm{{Haswell}}, \binits{C.A.}},
\bauthor{\bsnm{{Frank}}, \binits{P.}},
\bauthor{\bsnm{{Wheatley}}, \binits{P.J.}}:
\batitle{{A radio-pulsing white dwarf binary star}}.
\bjtitle{\nat}
\bvolume{537}(\bissue{7620}),
\bfpage{374}--\blpage{377}
(\byear{2016})
\doiurl{10.1038/nature18620}
{\href{https://arxiv.org/abs/1607.08265}{{arXiv:1607.08265}}}
{[astro-ph.SR]}
\end{barticle}
\endbibitem

\bibitem[\protect\citeauthoryear{{Pelisoli} et~al.}{2023}]{2023NatAs...7..931P}
\begin{barticle}
\bauthor{\bsnm{{Pelisoli}}, \binits{I.}},
\bauthor{\bsnm{{Marsh}}, \binits{T.R.}},
\bauthor{\bsnm{{Buckley}}, \binits{D.A.H.}},
\bauthor{\bsnm{{Heywood}}, \binits{I.}},
\bauthor{\bsnm{{Potter}}, \binits{S.B.}},
\bauthor{\bsnm{{Schwope}}, \binits{A.}},
\bauthor{\bsnm{{Brink}}, \binits{J.}},
\bauthor{\bsnm{{Standke}}, \binits{A.}},
\bauthor{\bsnm{{Woudt}}, \binits{P.A.}},
\bauthor{\bsnm{{Parsons}}, \binits{S.G.}},
\bauthor{\bsnm{{Green}}, \binits{M.J.}},
\bauthor{\bsnm{{Kepler}}, \binits{S.O.}},
\bauthor{\bsnm{{Munday}}, \binits{J.}},
\bauthor{\bsnm{{Romero}}, \binits{A.D.}},
\bauthor{\bsnm{{Breedt}}, \binits{E.}},
\bauthor{\bsnm{{Brown}}, \binits{A.J.}},
\bauthor{\bsnm{{Dhillon}}, \binits{V.S.}},
\bauthor{\bsnm{{Dyer}}, \binits{M.J.}},
\bauthor{\bsnm{{Kerry}}, \binits{P.}},
\bauthor{\bsnm{{Littlefair}}, \binits{S.P.}},
\bauthor{\bsnm{{Sahman}}, \binits{D.I.}},
\bauthor{\bsnm{{Wild}}, \binits{J.F.}}:
\batitle{{A 5.3-min-period pulsing white dwarf in a binary detected from radio to X-rays}}.
\bjtitle{Nature Astronomy}
\bvolume{7},
\bfpage{931}--\blpage{942}
(\byear{2023})
\doiurl{10.1038/s41550-023-01995-x}
{\href{https://arxiv.org/abs/2306.09272}{{arXiv:2306.09272}}}
{[astro-ph.SR]}
\end{barticle}
\endbibitem

\bibitem[\protect\citeauthoryear{{Hurley-Walker} et~al.}{2024}]{2024arXiv240815757H}
\begin{botherref}
\oauthor{\bsnm{{Hurley-Walker}}, \binits{N.}},
\oauthor{\bsnm{{McSweeney}}, \binits{S.J.}},
\oauthor{\bsnm{{Bahramian}}, \binits{A.}},
\oauthor{\bsnm{{Rea}}, \binits{N.}},
\oauthor{\bsnm{{Horvath}}, \binits{C.}},
\oauthor{\bsnm{{Buchner}}, \binits{S.}},
\oauthor{\bsnm{{Williams}}, \binits{A.}},
\oauthor{\bsnm{{Meyers}}, \binits{B.W.}},
\oauthor{\bsnm{{Strader}}, \binits{J.}},
\oauthor{\bsnm{{Aydi}}, \binits{E.}},
\oauthor{\bsnm{{Urquhart}}, \binits{R.}},
\oauthor{\bsnm{{Chomiuk}}, \binits{L.}},
\oauthor{\bsnm{{Galvin}}, \binits{T.J.}},
\oauthor{\bsnm{{Coti Zelati}}, \binits{F.}},
\oauthor{\bsnm{{Bailes}}, \binits{M.}}:
{A 2.9-hour periodic radio transient with an optical counterpart}.
arXiv e-prints,
2408--15757
(2024)
\doiurl{10.48550/arXiv.2408.15757}
{\href{https://arxiv.org/abs/2408.15757}{{arXiv:2408.15757}}}
{[astro-ph.SR]}
\end{botherref}
\endbibitem

\bibitem[\protect\citeauthoryear{{Bagnulo} and {Landstreet}}{2022}]{2022Msngr.186...14B}
\begin{barticle}
\bauthor{\bsnm{{Bagnulo}}, \binits{S.}},
\bauthor{\bsnm{{Landstreet}}, \binits{J.D.}}:
\batitle{{The Isolated Magnetic White Dwarfs}}.
\bjtitle{The Messenger}
\bvolume{186},
\bfpage{14}--\blpage{18}
(\byear{2022})
\doiurl{10.18727/0722-6691/5257}
\end{barticle}
\endbibitem

\bibitem[\protect\citeauthoryear{{Kaspi} and {Beloborodov}}{2017}]{2017ARA&A..55..261K}
\begin{barticle}
\bauthor{\bsnm{{Kaspi}}, \binits{V.M.}},
\bauthor{\bsnm{{Beloborodov}}, \binits{A.M.}}:
\batitle{{Magnetars}}.
\bjtitle{\araa}
\bvolume{55}(\bissue{1}),
\bfpage{261}--\blpage{301}
(\byear{2017})
\doiurl{10.1146/annurev-astro-081915-023329}
{\href{https://arxiv.org/abs/1703.00068}{{arXiv:1703.00068}}}
{[astro-ph.HE]}
\end{barticle}
\endbibitem

\bibitem[\protect\citeauthoryear{{Esposito} et~al.}{2021}]{2021ASSL..461...97E}
\begin{bchapter}
\bauthor{\bsnm{{Esposito}}, \binits{P.}},
\bauthor{\bsnm{{Rea}}, \binits{N.}},
\bauthor{\bsnm{{Israel}}, \binits{G.L.}}:
\bctitle{{Magnetars: A Short Review and Some Sparse Considerations}}.
In: \beditor{\bsnm{{Belloni}}, \binits{T.M.}},
\beditor{\bsnm{{M{\'e}ndez}}, \binits{M.}},
\beditor{\bsnm{{Zhang}}, \binits{C.}} (eds.)
\bbtitle{Timing Neutron Stars: Pulsations, Oscillations and Explosions}.
\bsertitle{Astrophysics and Space Science Library},
vol. \bseriesno{461},
pp. \bfpage{97}--\blpage{142}
(\byear{2021}).
\doiurl{10.1007/978-3-662-62110-3_3}
\end{bchapter}
\endbibitem

\bibitem[\protect\citeauthoryear{{Beniamini} et~al.}{2020}]{2020MNRAS.496.3390B}
\begin{barticle}
\bauthor{\bsnm{{Beniamini}}, \binits{P.}},
\bauthor{\bsnm{{Wadiasingh}}, \binits{Z.}},
\bauthor{\bsnm{{Metzger}}, \binits{B.D.}}:
\batitle{{Periodicity in recurrent fast radio bursts and the origin of ultralong period magnetars}}.
\bjtitle{\mnras}
\bvolume{496}(\bissue{3}),
\bfpage{3390}--\blpage{3401}
(\byear{2020})
\doiurl{10.1093/mnras/staa1783}
{\href{https://arxiv.org/abs/2003.12509}{{arXiv:2003.12509}}}
{[astro-ph.HE]}
\end{barticle}
\endbibitem

\bibitem[\protect\citeauthoryear{{Camilo} et~al.}{2006}]{2006Natur.442..892C}
\begin{botherref}
\oauthor{\bsnm{{Camilo}}, \binits{F.}},
\oauthor{\bsnm{{Ransom}}, \binits{S.M.}},
\oauthor{\bsnm{{Halpern}}, \binits{J.P.}},
\oauthor{\bsnm{{Reynolds}}, \binits{J.}},
\oauthor{\bsnm{{Helfand}}, \binits{D.J.}},
\oauthor{\bsnm{{Zimmerman}}, \binits{N.}},
\oauthor{\bsnm{{Sarkissian}}, \binits{J.}}:
{Transient pulsed radio emission from a magnetar}
\textbf{442}(7105),
892--895
(2006)
\doiurl{10.1038/nature04986}
{\href{https://arxiv.org/abs/astro-ph/0605429}{{arXiv:astro-ph/0605429}}}
{[astro-ph]}
\end{botherref}
\endbibitem

\bibitem[\protect\citeauthoryear{{Coti Zelati} et~al.}{2018}]{2018MNRAS.474..961C}
\begin{barticle}
\bauthor{\bsnm{{Coti Zelati}}, \binits{F.}},
\bauthor{\bsnm{{Rea}}, \binits{N.}},
\bauthor{\bsnm{{Pons}}, \binits{J.A.}},
\bauthor{\bsnm{{Campana}}, \binits{S.}},
\bauthor{\bsnm{{Esposito}}, \binits{P.}}:
\batitle{{Systematic study of magnetar outbursts}}.
\bjtitle{\mnras}
\bvolume{474}(\bissue{1}),
\bfpage{961}--\blpage{1017}
(\byear{2018})
\doiurl{10.1093/mnras/stx2679}
{\href{https://arxiv.org/abs/1710.04671}{{arXiv:1710.04671}}}
{[astro-ph.HE]}
\end{barticle}
\endbibitem

\bibitem[\protect\citeauthoryear{{Vigan{\`o}} et~al.}{2013}]{2013MNRAS.434..123V}
\begin{barticle}
\bauthor{\bsnm{{Vigan{\`o}}}, \binits{D.}},
\bauthor{\bsnm{{Rea}}, \binits{N.}},
\bauthor{\bsnm{{Pons}}, \binits{J.A.}},
\bauthor{\bsnm{{Perna}}, \binits{R.}},
\bauthor{\bsnm{{Aguilera}}, \binits{D.N.}},
\bauthor{\bsnm{{Miralles}}, \binits{J.A.}}:
\batitle{{Unifying the observational diversity of isolated neutron stars via magneto-thermal evolution models}}.
\bjtitle{\mnras}
\bvolume{434}(\bissue{1}),
\bfpage{123}--\blpage{141}
(\byear{2013})
\doiurl{10.1093/mnras/stt1008}
{\href{https://arxiv.org/abs/1306.2156}{{arXiv:1306.2156}}}
{[astro-ph.SR]}
\end{barticle}
\endbibitem

\bibitem[\protect\citeauthoryear{{Dehman} et~al.}{2023}]{2023MNRAS.518.1222D}
\begin{barticle}
\bauthor{\bsnm{{Dehman}}, \binits{C.}},
\bauthor{\bsnm{{Vigan{\`o}}}, \binits{D.}},
\bauthor{\bsnm{{Pons}}, \binits{J.A.}},
\bauthor{\bsnm{{Rea}}, \binits{N.}}:
\batitle{{3D code for MAgneto-Thermal evolution in Isolated Neutron Stars, MATINS: the magnetic field formalism}}.
\bjtitle{\mnras}
\bvolume{518}(\bissue{1}),
\bfpage{1222}--\blpage{1242}
(\byear{2023})
\doiurl{10.1093/mnras/stac2761}
{\href{https://arxiv.org/abs/2209.12920}{{arXiv:2209.12920}}}
{[astro-ph.HE]}
\end{barticle}
\endbibitem

\bibitem[\protect\citeauthoryear{{Caleb} et~al.}{2022}]{2022NatAs...6..828C}
\begin{barticle}
\bauthor{\bsnm{{Caleb}}, \binits{M.}},
\bauthor{\bsnm{{Heywood}}, \binits{I.}},
\bauthor{\bsnm{{Rajwade}}, \binits{K.}},
\bauthor{\bsnm{{Malenta}}, \binits{M.}},
\bauthor{\bsnm{{Stappers}}, \binits{B.W.}},
\bauthor{\bsnm{{Barr}}, \binits{E.}},
\bauthor{\bsnm{{Chen}}, \binits{W.}},
\bauthor{\bsnm{{Morello}}, \binits{V.}},
\bauthor{\bsnm{{Sanidas}}, \binits{S.}},
\bauthor{\bsnm{{van den Eijnden}}, \binits{J.}},
\bauthor{\bsnm{{Kramer}}, \binits{M.}},
\bauthor{\bsnm{{Buckley}}, \binits{D.}},
\bauthor{\bsnm{{Brink}}, \binits{J.}},
\bauthor{\bsnm{{Motta}}, \binits{S.E.}},
\bauthor{\bsnm{{Woudt}}, \binits{P.}},
\bauthor{\bsnm{{Weltevrede}}, \binits{P.}},
\bauthor{\bsnm{{Jankowski}}, \binits{F.}},
\bauthor{\bsnm{{Surnis}}, \binits{M.}},
\bauthor{\bsnm{{Buchner}}, \binits{S.}},
\bauthor{\bsnm{{Bezuidenhout}}, \binits{M.C.}},
\bauthor{\bsnm{{Driessen}}, \binits{L.N.}},
\bauthor{\bsnm{{Fender}}, \binits{R.}}:
\batitle{{Discovery of a radio-emitting neutron star with an ultra-long spin period of 76 s}}.
\bjtitle{Nature Astronomy}
\bvolume{6},
\bfpage{828}--\blpage{836}
(\byear{2022})
\doiurl{10.1038/s41550-022-01688-x}
{\href{https://arxiv.org/abs/2206.01346}{{arXiv:2206.01346}}}
{[astro-ph.HE]}
\end{barticle}
\endbibitem

\bibitem[\protect\citeauthoryear{{Lander} et~al.}{2024}]{2024arXiv241108020L}
\begin{botherref}
\oauthor{\bsnm{{Lander}}, \binits{S.K.}},
\oauthor{\bsnm{{Gourgouliatos}}, \binits{K.N.}},
\oauthor{\bsnm{{Wadiasingh}}, \binits{Z.}},
\oauthor{\bsnm{{Antonopoulou}}, \binits{D.}}:
{Observing the Meissner effect in neutron stars}.
arXiv e-prints,
2411--08020
(2024)
{\href{https://arxiv.org/abs/2411.08020}{{arXiv:2411.08020}}}
{[astro-ph.HE]}
\end{botherref}
\endbibitem

\bibitem[\protect\citeauthoryear{{Pritchard} et~al.}{2021}]{2021MNRAS.502.5438P}
\begin{barticle}
\bauthor{\bsnm{{Pritchard}}, \binits{J.}},
\bauthor{\bsnm{{Murphy}}, \binits{T.}},
\bauthor{\bsnm{{Zic}}, \binits{A.}},
\bauthor{\bsnm{{Lynch}}, \binits{C.}},
\bauthor{\bsnm{{Heald}}, \binits{G.}},
\bauthor{\bsnm{{Kaplan}}, \binits{D.L.}},
\bauthor{\bsnm{{Anderson}}, \binits{C.}},
\bauthor{\bsnm{{Banfield}}, \binits{J.}},
\bauthor{\bsnm{{Hale}}, \binits{C.}},
\bauthor{\bsnm{{Hotan}}, \binits{A.}},
\bauthor{\bsnm{{Lenc}}, \binits{E.}},
\bauthor{\bsnm{{Leung}}, \binits{J.K.}},
\bauthor{\bsnm{{McConnell}}, \binits{D.}},
\bauthor{\bsnm{{Moss}}, \binits{V.A.}},
\bauthor{\bsnm{{Raja}}, \binits{W.}},
\bauthor{\bsnm{{Stewart}}, \binits{A.J.}},
\bauthor{\bsnm{{Whiting}}, \binits{M.}}:
\batitle{{A circular polarization survey for radio stars with the Australian SKA Pathfinder}}.
\bjtitle{\mnras}
\bvolume{502}(\bissue{4}),
\bfpage{5438}--\blpage{5454}
(\byear{2021})
\doiurl{10.1093/mnras/stab299}
{\href{https://arxiv.org/abs/2102.01801}{{arXiv:2102.01801}}}
{[astro-ph.SR]}
\end{barticle}
\endbibitem

\bibitem[\protect\citeauthoryear{{Guzman} et~al.}{2019}]{2019ascl.soft12003G}
\begin{botherref}
\oauthor{\bsnm{{Guzman}}, \binits{J.}},
\oauthor{\bsnm{{Whiting}}, \binits{M.}},
\oauthor{\bsnm{{Voronkov}}, \binits{M.}},
\oauthor{\bsnm{{Mitchell}}, \binits{D.}},
\oauthor{\bsnm{{Ord}}, \binits{S.}},
\oauthor{\bsnm{{Collins}}, \binits{D.}},
\oauthor{\bsnm{{Marquarding}}, \binits{M.}},
\oauthor{\bsnm{{Lahur}}, \binits{P.}},
\oauthor{\bsnm{{Maher}}, \binits{T.}},
\oauthor{\bsnm{{Van Diepen}}, \binits{G.}},
\oauthor{\bsnm{{Bannister}}, \binits{K.}},
\oauthor{\bsnm{{Wu}}, \binits{X.}},
\oauthor{\bsnm{{Lenc}}, \binits{E.}},
\oauthor{\bsnm{{Khoo}}, \binits{J.}},
\oauthor{\bsnm{{Bastholm}}, \binits{E.}}:
{ASKAPsoft: ASKAP Science Data Processor Software}
\end{botherref}
\endbibitem

\bibitem[\protect\citeauthoryear{{Purcell} et~al.}{2020}]{2020ascl.soft05003P}
\begin{botherref}
\oauthor{\bsnm{{Purcell}}, \binits{C.R.}},
\oauthor{\bsnm{{Van Eck}}, \binits{C.L.}},
\oauthor{\bsnm{{West}}, \binits{J.}},
\oauthor{\bsnm{{Sun}}, \binits{X.H.}},
\oauthor{\bsnm{{Gaensler}}, \binits{B.M.}}:
{RM-Tools: Rotation Measure (RM) Synthesis and Stokes QU-fitting}
\end{botherref}
\endbibitem

\bibitem[\protect\citeauthoryear{{McConnell} et~al.}{2020}]{2020PASA...37...48M}
\begin{barticle}
\bauthor{\bsnm{{McConnell}}, \binits{D.}},
\bauthor{\bsnm{{Hale}}, \binits{C.L.}},
\bauthor{\bsnm{{Lenc}}, \binits{E.}},
\bauthor{\bsnm{{Banfield}}, \binits{J.K.}},
\bauthor{\bsnm{{Heald}}, \binits{G.}},
\bauthor{\bsnm{{Hotan}}, \binits{A.W.}},
\bauthor{\bsnm{{Leung}}, \binits{J.K.}},
\bauthor{\bsnm{{Moss}}, \binits{V.A.}},
\bauthor{\bsnm{{Murphy}}, \binits{T.}},
\bauthor{\bsnm{{O'Brien}}, \binits{A.}},
\bauthor{\bsnm{{Pritchard}}, \binits{J.}},
\bauthor{\bsnm{{Raja}}, \binits{W.}},
\bauthor{\bsnm{{Sadler}}, \binits{E.M.}},
\bauthor{\bsnm{{Stewart}}, \binits{A.}},
\bauthor{\bsnm{{Thomson}}, \binits{A.J.M.}},
\bauthor{\bsnm{{Whiting}}, \binits{M.}},
\bauthor{\bsnm{{Allison}}, \binits{J.R.}},
\bauthor{\bsnm{{Amy}}, \binits{S.W.}},
\bauthor{\bsnm{{Anderson}}, \binits{C.}},
\bauthor{\bsnm{{Ball}}, \binits{L.}},
\bauthor{\bsnm{{Bannister}}, \binits{K.W.}},
\bauthor{\bsnm{{Bell}}, \binits{M.}},
\bauthor{\bsnm{{Bock}}, \binits{D.C.-J.}},
\bauthor{\bsnm{{Bolton}}, \binits{R.}},
\bauthor{\bsnm{{Bunton}}, \binits{J.D.}},
\bauthor{\bsnm{{Chippendale}}, \binits{A.P.}},
\bauthor{\bsnm{{Collier}}, \binits{J.D.}},
\bauthor{\bsnm{{Cooray}}, \binits{F.R.}},
\bauthor{\bsnm{{Cornwell}}, \binits{T.J.}},
\bauthor{\bsnm{{Diamond}}, \binits{P.J.}},
\bauthor{\bsnm{{Edwards}}, \binits{P.G.}},
\bauthor{\bsnm{{Gupta}}, \binits{N.}},
\bauthor{\bsnm{{Hayman}}, \binits{D.B.}},
\bauthor{\bsnm{{Heywood}}, \binits{I.}},
\bauthor{\bsnm{{Jackson}}, \binits{C.A.}},
\bauthor{\bsnm{{Koribalski}}, \binits{B.S.}},
\bauthor{\bsnm{{Lee-Waddell}}, \binits{K.}},
\bauthor{\bsnm{{McClure-Griffiths}}, \binits{N.M.}},
\bauthor{\bsnm{{Ng}}, \binits{A.}},
\bauthor{\bsnm{{Norris}}, \binits{R.P.}},
\bauthor{\bsnm{{Phillips}}, \binits{C.}},
\bauthor{\bsnm{{Reynolds}}, \binits{J.E.}},
\bauthor{\bsnm{{Roxby}}, \binits{D.N.}},
\bauthor{\bsnm{{Schinckel}}, \binits{A.E.T.}},
\bauthor{\bsnm{{Shields}}, \binits{M.}},
\bauthor{\bsnm{{Tremblay}}, \binits{C.}},
\bauthor{\bsnm{{Tzioumis}}, \binits{A.}},
\bauthor{\bsnm{{Voronkov}}, \binits{M.A.}},
\bauthor{\bsnm{{Westmeier}}, \binits{T.}}:
\batitle{{The Rapid ASKAP Continuum Survey I: Design and first results}}.
\bjtitle{\pasa}
\bvolume{37},
\bfpage{048}
(\byear{2020})
\doiurl{10.1017/pasa.2020.41}
{\href{https://arxiv.org/abs/2012.00747}{{arXiv:2012.00747}}}
{[astro-ph.IM]}
\end{barticle}
\endbibitem

\bibitem[\protect\citeauthoryear{{Hale} et~al.}{2021}]{2021PASA...38...58H}
\begin{barticle}
\bauthor{\bsnm{{Hale}}, \binits{C.L.}},
\bauthor{\bsnm{{McConnell}}, \binits{D.}},
\bauthor{\bsnm{{Thomson}}, \binits{A.J.M.}},
\bauthor{\bsnm{{Lenc}}, \binits{E.}},
\bauthor{\bsnm{{Heald}}, \binits{G.H.}},
\bauthor{\bsnm{{Hotan}}, \binits{A.W.}},
\bauthor{\bsnm{{Leung}}, \binits{J.K.}},
\bauthor{\bsnm{{Moss}}, \binits{V.A.}},
\bauthor{\bsnm{{Murphy}}, \binits{T.}},
\bauthor{\bsnm{{Pritchard}}, \binits{J.}},
\bauthor{\bsnm{{Sadler}}, \binits{E.M.}},
\bauthor{\bsnm{{Stewart}}, \binits{A.J.}},
\bauthor{\bsnm{{Whiting}}, \binits{M.T.}}:
\batitle{{The Rapid ASKAP Continuum Survey Paper II: First Stokes I Source Catalogue Data Release}}.
\bjtitle{\pasa}
\bvolume{38},
\bfpage{058}
(\byear{2021})
\doiurl{10.1017/pasa.2021.47}
{\href{https://arxiv.org/abs/2109.00956}{{arXiv:2109.00956}}}
{[astro-ph.GA]}
\end{barticle}
\endbibitem

\bibitem[\protect\citeauthoryear{{Sault} et~al.}{1995}]{1995ASPC...77..433S}
\begin{bchapter}
\bauthor{\bsnm{{Sault}}, \binits{R.J.}},
\bauthor{\bsnm{{Teuben}}, \binits{P.J.}},
\bauthor{\bsnm{{Wright}}, \binits{M.C.H.}}:
\bctitle{{A Retrospective View of MIRIAD}}.
In: \beditor{\bsnm{{Shaw}}, \binits{R.A.}},
\beditor{\bsnm{{Payne}}, \binits{H.E.}},
\beditor{\bsnm{{Hayes}}, \binits{J.J.E.}} (eds.)
\bbtitle{Astronomical Data Analysis Software and Systems IV}.
\bsertitle{Astronomical Society of the Pacific Conference Series},
vol. \bseriesno{77},
p. \bfpage{433}
(\byear{1995}).
\doiurl{10.48550/arXiv.astro-ph/0612759}
\end{bchapter}
\endbibitem

\bibitem[\protect\citeauthoryear{{Heywood}}{2020}]{2020ascl.soft09003H}
\begin{botherref}
\oauthor{\bsnm{{Heywood}}, \binits{I.}}:
{oxkat: Semi-automated Imaging of MeerKAT Observations}
\end{botherref}
\endbibitem

\bibitem[\protect\citeauthoryear{{McMullin} et~al.}{2007}]{2007ASPC..376..127M}
\begin{bchapter}
\bauthor{\bsnm{{McMullin}}, \binits{J.P.}},
\bauthor{\bsnm{{Waters}}, \binits{B.}},
\bauthor{\bsnm{{Schiebel}}, \binits{D.}},
\bauthor{\bsnm{{Young}}, \binits{W.}},
\bauthor{\bsnm{{Golap}}, \binits{K.}}:
\bctitle{{CASA Architecture and Applications}}.
In: \beditor{\bsnm{{Shaw}}, \binits{R.A.}},
\beditor{\bsnm{{Hill}}, \binits{F.}},
\beditor{\bsnm{{Bell}}, \binits{D.J.}} (eds.)
\bbtitle{Astronomical Data Analysis Software and Systems XVI}.
\bsertitle{Astronomical Society of the Pacific Conference Series},
vol. \bseriesno{376},
p. \bfpage{127}
(\byear{2007})
\end{bchapter}
\endbibitem

\bibitem[\protect\citeauthoryear{{Hugo} et~al.}{2022}]{2022ASPC..532..541H}
\begin{bchapter}
\bauthor{\bsnm{{Hugo}}, \binits{B.V.}},
\bauthor{\bsnm{{Perkins}}, \binits{S.}},
\bauthor{\bsnm{{Merry}}, \binits{B.}},
\bauthor{\bsnm{{Mauch}}, \binits{T.}},
\bauthor{\bsnm{{Smirnov}}, \binits{O.M.}}:
\bctitle{{Tricolour: An Optimized SumThreshold Flagger for MeerKAT}}.
In: \beditor{\bsnm{{Ruiz}}, \binits{J.E.}},
\beditor{\bsnm{{Pierfedereci}}, \binits{F.}},
\beditor{\bsnm{{Teuben}}, \binits{P.}} (eds.)
\bbtitle{Astronomical Data Analysis Software and Systems XXX}.
\bsertitle{Astronomical Society of the Pacific Conference Series},
vol. \bseriesno{532},
p. \bfpage{541}
(\byear{2022}).
\doiurl{10.48550/arXiv.2206.09179}
\end{bchapter}
\endbibitem

\bibitem[\protect\citeauthoryear{{Kenyon} et~al.}{2018}]{2018MNRAS.478.2399K}
\begin{barticle}
\bauthor{\bsnm{{Kenyon}}, \binits{J.S.}},
\bauthor{\bsnm{{Smirnov}}, \binits{O.M.}},
\bauthor{\bsnm{{Grobler}}, \binits{T.L.}},
\bauthor{\bsnm{{Perkins}}, \binits{S.J.}}:
\batitle{{CUBICAL - fast radio interferometric calibration suite exploiting complex optimization}}.
\bjtitle{\mnras}
\bvolume{478}(\bissue{2}),
\bfpage{2399}--\blpage{2415}
(\byear{2018})
\doiurl{10.1093/mnras/sty1221}
{\href{https://arxiv.org/abs/1805.03410}{{arXiv:1805.03410}}}
{[astro-ph.IM]}
\end{barticle}
\endbibitem

\bibitem[\protect\citeauthoryear{{Offringa} et~al.}{2014}]{2014MNRAS.444..606O}
\begin{barticle}
\bauthor{\bsnm{{Offringa}}, \binits{A.R.}},
\bauthor{\bsnm{{McKinley}}, \binits{B.}},
\bauthor{\bsnm{{Hurley-Walker}}, \binits{N.}},
\bauthor{\bsnm{{Briggs}}, \binits{F.H.}},
\bauthor{\bsnm{{Wayth}}, \binits{R.B.}},
\bauthor{\bsnm{{Kaplan}}, \binits{D.L.}},
\bauthor{\bsnm{{Bell}}, \binits{M.E.}},
\bauthor{\bsnm{{Feng}}, \binits{L.}},
\bauthor{\bsnm{{Neben}}, \binits{A.R.}},
\bauthor{\bsnm{{Hughes}}, \binits{J.D.}},
\bauthor{\bsnm{{Rhee}}, \binits{J.}},
\bauthor{\bsnm{{Murphy}}, \binits{T.}},
\bauthor{\bsnm{{Bhat}}, \binits{N.D.R.}},
\bauthor{\bsnm{{Bernardi}}, \binits{G.}},
\bauthor{\bsnm{{Bowman}}, \binits{J.D.}},
\bauthor{\bsnm{{Cappallo}}, \binits{R.J.}},
\bauthor{\bsnm{{Corey}}, \binits{B.E.}},
\bauthor{\bsnm{{Deshpande}}, \binits{A.A.}},
\bauthor{\bsnm{{Emrich}}, \binits{D.}},
\bauthor{\bsnm{{Ewall-Wice}}, \binits{A.}},
\bauthor{\bsnm{{Gaensler}}, \binits{B.M.}},
\bauthor{\bsnm{{Goeke}}, \binits{R.}},
\bauthor{\bsnm{{Greenhill}}, \binits{L.J.}},
\bauthor{\bsnm{{Hazelton}}, \binits{B.J.}},
\bauthor{\bsnm{{Hindson}}, \binits{L.}},
\bauthor{\bsnm{{Johnston-Hollitt}}, \binits{M.}},
\bauthor{\bsnm{{Jacobs}}, \binits{D.C.}},
\bauthor{\bsnm{{Kasper}}, \binits{J.C.}},
\bauthor{\bsnm{{Kratzenberg}}, \binits{E.}},
\bauthor{\bsnm{{Lenc}}, \binits{E.}},
\bauthor{\bsnm{{Lonsdale}}, \binits{C.J.}},
\bauthor{\bsnm{{Lynch}}, \binits{M.J.}},
\bauthor{\bsnm{{McWhirter}}, \binits{S.R.}},
\bauthor{\bsnm{{Mitchell}}, \binits{D.A.}},
\bauthor{\bsnm{{Morales}}, \binits{M.F.}},
\bauthor{\bsnm{{Morgan}}, \binits{E.}},
\bauthor{\bsnm{{Kudryavtseva}}, \binits{N.}},
\bauthor{\bsnm{{Oberoi}}, \binits{D.}},
\bauthor{\bsnm{{Ord}}, \binits{S.M.}},
\bauthor{\bsnm{{Pindor}}, \binits{B.}},
\bauthor{\bsnm{{Procopio}}, \binits{P.}},
\bauthor{\bsnm{{Prabu}}, \binits{T.}},
\bauthor{\bsnm{{Riding}}, \binits{J.}},
\bauthor{\bsnm{{Roshi}}, \binits{D.A.}},
\bauthor{\bsnm{{Shankar}}, \binits{N.U.}},
\bauthor{\bsnm{{Srivani}}, \binits{K.S.}},
\bauthor{\bsnm{{Subrahmanyan}}, \binits{R.}},
\bauthor{\bsnm{{Tingay}}, \binits{S.J.}},
\bauthor{\bsnm{{Waterson}}, \binits{M.}},
\bauthor{\bsnm{{Webster}}, \binits{R.L.}},
\bauthor{\bsnm{{Whitney}}, \binits{A.R.}},
\bauthor{\bsnm{{Williams}}, \binits{A.}},
\bauthor{\bsnm{{Williams}}, \binits{C.L.}}:
\batitle{{WSCLEAN: an implementation of a fast, generic wide-field imager for radio astronomy}}.
\bjtitle{\mnras}
\bvolume{444},
\bfpage{606}--\blpage{619}
(\byear{2014})
\doiurl{10.1093/mnras/stu1368}
{\href{https://arxiv.org/abs/1407.1943}{{arXiv:1407.1943}}}
{[astro-ph.IM]}
\end{barticle}
\endbibitem

\bibitem[\protect\citeauthoryear{{Collier} et~al.}{2021}]{2021ursi.confE...4C}
\begin{bchapter}
\bauthor{\bsnm{{Collier}}, \binits{J.D.}},
\bauthor{\bsnm{{Frank}}, \binits{B.}},
\bauthor{\bsnm{{Sekhar}}, \binits{S.}},
\bauthor{\bsnm{{Taylor}}, \binits{A.R.}}:
\bctitle{{The IDIA PROCESSMEERKAT pipeline: Fast CASA Processing on a Cloud-Based HPC Cluster}}.
In: \bbtitle{2021 XXXIVth General Assembly and Scientific Symposium of the International Union of Radio Science (URSI GASS},
p. \bfpage{4}
(\byear{2021}).
\doiurl{10.23919/URSIGASS51995.2021.9560276}
\end{bchapter}
\endbibitem

\bibitem[\protect\citeauthoryear{{Deller} et~al.}{2011}]{2011PASP..123..275D}
\begin{barticle}
\bauthor{\bsnm{{Deller}}, \binits{A.T.}},
\bauthor{\bsnm{{Brisken}}, \binits{W.F.}},
\bauthor{\bsnm{{Phillips}}, \binits{C.J.}},
\bauthor{\bsnm{{Morgan}}, \binits{J.}},
\bauthor{\bsnm{{Alef}}, \binits{W.}},
\bauthor{\bsnm{{Cappallo}}, \binits{R.}},
\bauthor{\bsnm{{Middelberg}}, \binits{E.}},
\bauthor{\bsnm{{Romney}}, \binits{J.}},
\bauthor{\bsnm{{Rottmann}}, \binits{H.}},
\bauthor{\bsnm{{Tingay}}, \binits{S.J.}},
\bauthor{\bsnm{{Wayth}}, \binits{R.}}:
\batitle{{DiFX-2: A More Flexible, Efficient, Robust, and Powerful Software Correlator}}.
\bjtitle{\pasp}
\bvolume{123}(\bissue{901}),
\bfpage{275}
(\byear{2011})
\doiurl{10.1086/658907}
{\href{https://arxiv.org/abs/1101.0885}{{arXiv:1101.0885}}}
{[astro-ph.IM]}
\end{barticle}
\endbibitem

\bibitem[\protect\citeauthoryear{{Kettenis} et~al.}{2006}]{2006ASPC..351..497K}
\begin{bchapter}
\bauthor{\bsnm{{Kettenis}}, \binits{M.}},
\bauthor{\bsnm{{van Langevelde}}, \binits{H.J.}},
\bauthor{\bsnm{{Reynolds}}, \binits{C.}},
\bauthor{\bsnm{{Cotton}}, \binits{B.}}:
\bctitle{{ParselTongue: AIPS Talking Python}}.
In: \beditor{\bsnm{{Gabriel}}, \binits{C.}},
\beditor{\bsnm{{Arviset}}, \binits{C.}},
\beditor{\bsnm{{Ponz}}, \binits{D.}},
\beditor{\bsnm{{Enrique}}, \binits{S.}} (eds.)
\bbtitle{Astronomical Data Analysis Software and Systems XV}.
\bsertitle{Astronomical Society of the Pacific Conference Series},
vol. \bseriesno{351},
p. \bfpage{497}
(\byear{2006})
\end{bchapter}
\endbibitem

\bibitem[\protect\citeauthoryear{{Ding} et~al.}{2024}]{2024ApJ...971L..13D}
\begin{barticle}
\bauthor{\bsnm{{Ding}}, \binits{H.}},
\bauthor{\bsnm{{Lower}}, \binits{M.E.}},
\bauthor{\bsnm{{Deller}}, \binits{A.T.}},
\bauthor{\bsnm{{Shannon}}, \binits{R.M.}},
\bauthor{\bsnm{{Camilo}}, \binits{F.}},
\bauthor{\bsnm{{Sarkissian}}, \binits{J.}}:
\batitle{{VLBA Astrometry of the Fastest-spinning Magnetar Swift J1818.0‑1607: A Large Trigonometric Distance and a Small Transverse Velocity}}.
\bjtitle{\apjl}
\bvolume{971}(\bissue{1}),
\bfpage{13}
(\byear{2024})
\doiurl{10.3847/2041-8213/ad5550}
{\href{https://arxiv.org/abs/2406.04674}{{arXiv:2406.04674}}}
{[astro-ph.HE]}
\end{barticle}
\endbibitem

\bibitem[\protect\citeauthoryear{{Polisensky} et~al.}{2016}]{2016ApJ...832...60P}
\begin{barticle}
\bauthor{\bsnm{{Polisensky}}, \binits{E.}},
\bauthor{\bsnm{{Lane}}, \binits{W.M.}},
\bauthor{\bsnm{{Hyman}}, \binits{S.D.}},
\bauthor{\bsnm{{Kassim}}, \binits{N.E.}},
\bauthor{\bsnm{{Giacintucci}}, \binits{S.}},
\bauthor{\bsnm{{Clarke}}, \binits{T.E.}},
\bauthor{\bsnm{{Cotton}}, \binits{W.D.}},
\bauthor{\bsnm{{Cleland}}, \binits{E.}},
\bauthor{\bsnm{{Frail}}, \binits{D.A.}}:
\batitle{{Exploring the Transient Radio Sky with VLITE: Early Results}}.
\bjtitle{\apj}
\bvolume{832}(\bissue{1}),
\bfpage{60}
(\byear{2016})
\doiurl{10.3847/0004-637X/832/1/60}
{\href{https://arxiv.org/abs/1604.00667}{{arXiv:1604.00667}}}
{[astro-ph.HE]}
\end{barticle}
\endbibitem

\bibitem[\protect\citeauthoryear{{Clarke} et~al.}{2016}]{2016SPIE.9906E..5BC}
\begin{bchapter}
\bauthor{\bsnm{{Clarke}}, \binits{T.E.}},
\bauthor{\bsnm{{Kassim}}, \binits{N.E.}},
\bauthor{\bsnm{{Brisken}}, \binits{W.}},
\bauthor{\bsnm{{Helmboldt}}, \binits{J.}},
\bauthor{\bsnm{{Peters}}, \binits{W.}},
\bauthor{\bsnm{{Ray}}, \binits{P.S.}},
\bauthor{\bsnm{{Polisensky}}, \binits{E.}},
\bauthor{\bsnm{{Giacintucci}}, \binits{S.}}:
\bctitle{{Commensal low frequency observing on the NRAO VLA: VLITE status and future plans}}.
In: \beditor{\bsnm{{Hall}}, \binits{H.J.}},
\beditor{\bsnm{{Gilmozzi}}, \binits{R.}},
\beditor{\bsnm{{Marshall}}, \binits{H.K.}} (eds.)
\bbtitle{Ground-based and Airborne Telescopes VI}.
\bsertitle{Society of Photo-Optical Instrumentation Engineers (SPIE) Conference Series},
vol. \bseriesno{9906},
p. \bfpage{99065}
(\byear{2016}).
\doiurl{10.1117/12.2233036}
\end{bchapter}
\endbibitem

\bibitem[\protect\citeauthoryear{{Cotton}}{2008}]{2008PASP..120..439C}
\begin{barticle}
\bauthor{\bsnm{{Cotton}}, \binits{W.D.}}:
\batitle{{Obit: A Development Environment for Astronomical Algorithms}}.
\bjtitle{\pasp}
\bvolume{120}(\bissue{866}),
\bfpage{439}
(\byear{2008})
\doiurl{10.1086/586754}
\end{barticle}
\endbibitem

\bibitem[\protect\citeauthoryear{{Polisensky} et~al.}{2019}]{2019ASPC..523..441P}
\begin{bchapter}
\bauthor{\bsnm{{Polisensky}}, \binits{E.}},
\bauthor{\bsnm{{Richards}}, \binits{E.}},
\bauthor{\bsnm{{Clarke}}, \binits{T.}},
\bauthor{\bsnm{{Peters}}, \binits{W.}},
\bauthor{\bsnm{{Kassim}}, \binits{N.}}:
\bctitle{{The VLITE Database Pipeline}}.
In: \beditor{\bsnm{{Teuben}}, \binits{P.J.}},
\beditor{\bsnm{{Pound}}, \binits{M.W.}},
\beditor{\bsnm{{Thomas}}, \binits{B.A.}},
\beditor{\bsnm{{Warner}}, \binits{E.M.}} (eds.)
\bbtitle{Astronomical Data Analysis Software and Systems XXVII}.
\bsertitle{Astronomical Society of the Pacific Conference Series},
vol. \bseriesno{523},
p. \bfpage{441}
(\byear{2019})
\end{bchapter}
\endbibitem

\bibitem[\protect\citeauthoryear{{Mohan} and {Rafferty}}{2015}]{2015ascl.soft02007M}
\begin{botherref}
\oauthor{\bsnm{{Mohan}}, \binits{N.}},
\oauthor{\bsnm{{Rafferty}}, \binits{D.}}:
{PyBDSF: Python Blob Detection and Source Finder}
\end{botherref}
\endbibitem

\bibitem[\protect\citeauthoryear{{Lomb}}{1976}]{1976Ap&SS..39..447L}
\begin{barticle}
\bauthor{\bsnm{{Lomb}}, \binits{N.R.}}:
\batitle{{Least-Squares Frequency Analysis of Unequally Spaced Data}}.
\bjtitle{\apss}
\bvolume{39}(\bissue{2}),
\bfpage{447}--\blpage{462}
(\byear{1976})
\doiurl{10.1007/BF00648343}
\end{barticle}
\endbibitem

\bibitem[\protect\citeauthoryear{{Balucinska-Church} and {McCammon}}{1992}]{1992ApJ...400..699B}
\begin{barticle}
\bauthor{\bsnm{{Balucinska-Church}}, \binits{M.}},
\bauthor{\bsnm{{McCammon}}, \binits{D.}}:
\batitle{{Photoelectric Absorption Cross Sections with Variable Abundances}}.
\bjtitle{\apj}
\bvolume{400},
\bfpage{699}
(\byear{1992})
\doiurl{10.1086/172032}
\end{barticle}
\endbibitem

\bibitem[\protect\citeauthoryear{{Lodders}}{2003}]{2003ApJ...591.1220L}
\begin{barticle}
\bauthor{\bsnm{{Lodders}}, \binits{K.}}:
\batitle{{Solar System Abundances and Condensation Temperatures of the Elements}}.
\bjtitle{\apj}
\bvolume{591}(\bissue{2}),
\bfpage{1220}--\blpage{1247}
(\byear{2003})
\doiurl{10.1086/375492}
\end{barticle}
\endbibitem

\bibitem[\protect\citeauthoryear{{HI4PI Collaboration} et~al.}{2016}]{2016A&A...594A.116H}
\begin{barticle}
\bauthor{\bsnm{{HI4PI Collaboration}}},
\bauthor{\bsnm{{Ben Bekhti}}, \binits{N.}},
\bauthor{\bsnm{{Fl{\"o}er}}, \binits{L.}},
\bauthor{\bsnm{{Keller}}, \binits{R.}},
\bauthor{\bsnm{{Kerp}}, \binits{J.}},
\bauthor{\bsnm{{Lenz}}, \binits{D.}},
\bauthor{\bsnm{{Winkel}}, \binits{B.}},
\bauthor{\bsnm{{Bailin}}, \binits{J.}},
\bauthor{\bsnm{{Calabretta}}, \binits{M.R.}},
\bauthor{\bsnm{{Dedes}}, \binits{L.}},
\bauthor{\bsnm{{Ford}}, \binits{H.A.}},
\bauthor{\bsnm{{Gibson}}, \binits{B.K.}},
\bauthor{\bsnm{{Haud}}, \binits{U.}},
\bauthor{\bsnm{{Janowiecki}}, \binits{S.}},
\bauthor{\bsnm{{Kalberla}}, \binits{P.M.W.}},
\bauthor{\bsnm{{Lockman}}, \binits{F.J.}},
\bauthor{\bsnm{{McClure-Griffiths}}, \binits{N.M.}},
\bauthor{\bsnm{{Murphy}}, \binits{T.}},
\bauthor{\bsnm{{Nakanishi}}, \binits{H.}},
\bauthor{\bsnm{{Pisano}}, \binits{D.J.}},
\bauthor{\bsnm{{Staveley-Smith}}, \binits{L.}}:
\batitle{{HI4PI: A full-sky H I survey based on EBHIS and GASS}}.
\bjtitle{\aap}
\bvolume{594},
\bfpage{116}
(\byear{2016})
\doiurl{10.1051/0004-6361/201629178}
{\href{https://arxiv.org/abs/1610.06175}{{arXiv:1610.06175}}}
{[astro-ph.GA]}
\end{barticle}
\endbibitem

\bibitem[\protect\citeauthoryear{{He} et~al.}{2013}]{2013ApJ...768...64H}
\begin{barticle}
\bauthor{\bsnm{{He}}, \binits{C.}},
\bauthor{\bsnm{{Ng}}, \binits{C.-Y.}},
\bauthor{\bsnm{{Kaspi}}, \binits{V.M.}}:
\batitle{{The Correlation between Dispersion Measure and X-Ray Column Density from Radio Pulsars}}.
\bjtitle{\apj}
\bvolume{768}(\bissue{1}),
\bfpage{64}
(\byear{2013})
\doiurl{10.1088/0004-637X/768/1/64}
{\href{https://arxiv.org/abs/1303.5170}{{arXiv:1303.5170}}}
{[astro-ph.HE]}
\end{barticle}
\endbibitem

\bibitem[\protect\citeauthoryear{{Marino} et~al.}{2024}]{2024NatAs...8.1020M}
\begin{barticle}
\bauthor{\bsnm{{Marino}}, \binits{A.}},
\bauthor{\bsnm{{Dehman}}, \binits{C.}},
\bauthor{\bsnm{{Kovlakas}}, \binits{K.}},
\bauthor{\bsnm{{Rea}}, \binits{N.}},
\bauthor{\bsnm{{Pons}}, \binits{J.A.}},
\bauthor{\bsnm{{Vigan{\`o}}}, \binits{D.}}:
\batitle{{Constraints on the dense matter equation of state from young and cold isolated neutron stars}}.
\bjtitle{Nature Astronomy}
\bvolume{8},
\bfpage{1020}--\blpage{1030}
(\byear{2024})
\doiurl{10.1038/s41550-024-02291-y}
{\href{https://arxiv.org/abs/2404.05371}{{arXiv:2404.05371}}}
{[astro-ph.HE]}
\end{barticle}
\endbibitem

\bibitem[\protect\citeauthoryear{{Vigan{\`o}} et~al.}{2021}]{2021CoPhC.26508001V}
\begin{barticle}
\bauthor{\bsnm{{Vigan{\`o}}}, \binits{D.}},
\bauthor{\bsnm{{Garcia-Garcia}}, \binits{A.}},
\bauthor{\bsnm{{Pons}}, \binits{J.A.}},
\bauthor{\bsnm{{Dehman}}, \binits{C.}},
\bauthor{\bsnm{{Graber}}, \binits{V.}}:
\batitle{{Magneto-thermal evolution of neutron stars with coupled Ohmic, Hall and ambipolar effects via accurate finite-volume simulations}}.
\bjtitle{Computer Physics Communications}
\bvolume{265},
\bfpage{108001}
(\byear{2021})
\doiurl{10.1016/j.cpc.2021.108001}
{\href{https://arxiv.org/abs/2104.08001}{{arXiv:2104.08001}}}
{[astro-ph.HE]}
\end{barticle}
\endbibitem

\bibitem[\protect\citeauthoryear{{Hobbs} et~al.}{2006}]{2006MNRAS.369..655H}
\begin{barticle}
\bauthor{\bsnm{{Hobbs}}, \binits{G.B.}},
\bauthor{\bsnm{{Edwards}}, \binits{R.T.}},
\bauthor{\bsnm{{Manchester}}, \binits{R.N.}}:
\batitle{{TEMPO2, a new pulsar-timing package - I. An overview}}.
\bjtitle{\mnras}
\bvolume{369}(\bissue{2}),
\bfpage{655}--\blpage{672}
(\byear{2006})
\doiurl{10.1111/j.1365-2966.2006.10302.x}
{\href{https://arxiv.org/abs/astro-ph/0603381}{{arXiv:astro-ph/0603381}}}
{[astro-ph]}
\end{barticle}
\endbibitem

\bibitem[\protect\citeauthoryear{{Luo} et~al.}{2021}]{2021ApJ...911...45L}
\begin{barticle}
\bauthor{\bsnm{{Luo}}, \binits{J.}},
\bauthor{\bsnm{{Ransom}}, \binits{S.}},
\bauthor{\bsnm{{Demorest}}, \binits{P.}},
\bauthor{\bsnm{{Ray}}, \binits{P.S.}},
\bauthor{\bsnm{{Archibald}}, \binits{A.}},
\bauthor{\bsnm{{Kerr}}, \binits{M.}},
\bauthor{\bsnm{{Jennings}}, \binits{R.J.}},
\bauthor{\bsnm{{Bachetti}}, \binits{M.}},
\bauthor{\bsnm{{van Haasteren}}, \binits{R.}},
\bauthor{\bsnm{{Champagne}}, \binits{C.A.}},
\bauthor{\bsnm{{Colen}}, \binits{J.}},
\bauthor{\bsnm{{Phillips}}, \binits{C.}},
\bauthor{\bsnm{{Zimmerman}}, \binits{J.}},
\bauthor{\bsnm{{Stovall}}, \binits{K.}},
\bauthor{\bsnm{{Lam}}, \binits{M.T.}},
\bauthor{\bsnm{{Jenet}}, \binits{F.A.}}:
\batitle{{PINT: A Modern Software Package for Pulsar Timing}}.
\bjtitle{\apj}
\bvolume{911}(\bissue{1}),
\bfpage{45}
(\byear{2021})
\doiurl{10.3847/1538-4357/abe62f}
{\href{https://arxiv.org/abs/2012.00074}{{arXiv:2012.00074}}}
{[astro-ph.IM]}
\end{barticle}
\endbibitem

\bibitem[\protect\citeauthoryear{{Foreman-Mackey} et~al.}{2013}]{2013PASP..125..306F}
\begin{barticle}
\bauthor{\bsnm{{Foreman-Mackey}}, \binits{D.}},
\bauthor{\bsnm{{Hogg}}, \binits{D.W.}},
\bauthor{\bsnm{{Lang}}, \binits{D.}},
\bauthor{\bsnm{{Goodman}}, \binits{J.}}:
\batitle{{emcee: The MCMC Hammer}}.
\bjtitle{\pasp}
\bvolume{125}(\bissue{925}),
\bfpage{306}
(\byear{2013})
\doiurl{10.1086/670067}
{\href{https://arxiv.org/abs/1202.3665}{{arXiv:1202.3665}}}
{[astro-ph.IM]}
\end{barticle}
\endbibitem

\bibitem[\protect\citeauthoryear{{Manchester} et~al.}{2005}]{2005AJ....129.1993M}
\begin{barticle}
\bauthor{\bsnm{{Manchester}}, \binits{R.N.}},
\bauthor{\bsnm{{Hobbs}}, \binits{G.B.}},
\bauthor{\bsnm{{Teoh}}, \binits{A.}},
\bauthor{\bsnm{{Hobbs}}, \binits{M.}}:
\batitle{{The Australia Telescope National Facility Pulsar Catalogue}}.
\bjtitle{\aj}
\bvolume{129}(\bissue{4}),
\bfpage{1993}--\blpage{2006}
(\byear{2005})
\doiurl{10.1086/428488}
{\href{https://arxiv.org/abs/astro-ph/0412641}{{arXiv:astro-ph/0412641}}}
{[astro-ph]}
\end{barticle}
\endbibitem

\bibitem[\protect\citeauthoryear{{De Luca} et~al.}{2006}]{2006Sci...313..814D}
\begin{barticle}
\bauthor{\bsnm{{De Luca}}, \binits{A.}},
\bauthor{\bsnm{{Caraveo}}, \binits{P.A.}},
\bauthor{\bsnm{{Mereghetti}}, \binits{S.}},
\bauthor{\bsnm{{Tiengo}}, \binits{A.}},
\bauthor{\bsnm{{Bignami}}, \binits{G.F.}}:
\batitle{{A Long-Period, Violently Variable X-ray Source in a Young Supernova Remnant}}.
\bjtitle{Science}
\bvolume{313}(\bissue{5788}),
\bfpage{814}--\blpage{817}
(\byear{2006})
\doiurl{10.1126/science.1129185}
{\href{https://arxiv.org/abs/astro-ph/0607173}{{arXiv:astro-ph/0607173}}}
{[astro-ph]}
\end{barticle}
\endbibitem

\bibitem[\protect\citeauthoryear{{Rea} et~al.}{2016}]{2016ApJ...828L..13R}
\begin{barticle}
\bauthor{\bsnm{{Rea}}, \binits{N.}},
\bauthor{\bsnm{{Borghese}}, \binits{A.}},
\bauthor{\bsnm{{Esposito}}, \binits{P.}},
\bauthor{\bsnm{{Coti Zelati}}, \binits{F.}},
\bauthor{\bsnm{{Bachetti}}, \binits{M.}},
\bauthor{\bsnm{{Israel}}, \binits{G.L.}},
\bauthor{\bsnm{{De Luca}}, \binits{A.}}:
\batitle{{Magnetar-like Activity from the Central Compact Object in the SNR RCW103}}.
\bjtitle{\apjl}
\bvolume{828}(\bissue{1}),
\bfpage{13}
(\byear{2016})
\doiurl{10.3847/2041-8205/828/1/L13}
{\href{https://arxiv.org/abs/1607.04107}{{arXiv:1607.04107}}}
{[astro-ph.HE]}
\end{barticle}
\endbibitem

\bibitem[\protect\citeauthoryear{{D'A{\`\i}} et~al.}{2016}]{2016MNRAS.463.2394D}
\begin{barticle}
\bauthor{\bsnm{{D'A{\`\i}}}, \binits{A.}},
\bauthor{\bsnm{{Evans}}, \binits{P.A.}},
\bauthor{\bsnm{{Burrows}}, \binits{D.N.}},
\bauthor{\bsnm{{Kuin}}, \binits{N.P.M.}},
\bauthor{\bsnm{{Kann}}, \binits{D.A.}},
\bauthor{\bsnm{{Campana}}, \binits{S.}},
\bauthor{\bsnm{{Maselli}}, \binits{A.}},
\bauthor{\bsnm{{Romano}}, \binits{P.}},
\bauthor{\bsnm{{Cusumano}}, \binits{G.}},
\bauthor{\bsnm{{La Parola}}, \binits{V.}},
\bauthor{\bsnm{{Barthelmy}}, \binits{S.D.}},
\bauthor{\bsnm{{Beardmore}}, \binits{A.P.}},
\bauthor{\bsnm{{Cenko}}, \binits{S.B.}},
\bauthor{\bsnm{{De Pasquale}}, \binits{M.}},
\bauthor{\bsnm{{Gehrels}}, \binits{N.}},
\bauthor{\bsnm{{Greiner}}, \binits{J.}},
\bauthor{\bsnm{{Kennea}}, \binits{J.A.}},
\bauthor{\bsnm{{Klose}}, \binits{S.}},
\bauthor{\bsnm{{Melandri}}, \binits{A.}},
\bauthor{\bsnm{{Nousek}}, \binits{J.A.}},
\bauthor{\bsnm{{Osborne}}, \binits{J.P.}},
\bauthor{\bsnm{{Palmer}}, \binits{D.M.}},
\bauthor{\bsnm{{Sbarufatti}}, \binits{B.}},
\bauthor{\bsnm{{Schady}}, \binits{P.}},
\bauthor{\bsnm{{Siegel}}, \binits{M.H.}},
\bauthor{\bsnm{{Tagliaferri}}, \binits{G.}},
\bauthor{\bsnm{{Yates}}, \binits{R.}},
\bauthor{\bsnm{{Zane}}, \binits{S.}}:
\batitle{{Evidence for the magnetar nature of 1E 161348-5055 in RCW 103}}.
\bjtitle{\mnras}
\bvolume{463}(\bissue{3}),
\bfpage{2394}--\blpage{2404}
(\byear{2016})
\doiurl{10.1093/mnras/stw2023}
{\href{https://arxiv.org/abs/1607.04264}{{arXiv:1607.04264}}}
{[astro-ph.HE]}
\end{barticle}
\endbibitem

\bibitem[\protect\citeauthoryear{{Olausen} and {Kaspi}}{2014}]{2014ApJS..212....6O}
\begin{barticle}
\bauthor{\bsnm{{Olausen}}, \binits{S.A.}},
\bauthor{\bsnm{{Kaspi}}, \binits{V.M.}}:
\batitle{{The McGill Magnetar Catalog}}.
\bjtitle{\apjs}
\bvolume{212}(\bissue{1}),
\bfpage{6}
(\byear{2014})
\doiurl{10.1088/0067-0049/212/1/6}
{\href{https://arxiv.org/abs/1309.4167}{{arXiv:1309.4167}}}
{[astro-ph.HE]}
\end{barticle}
\endbibitem

\bibitem[\protect\citeauthoryear{{Tan} et~al.}{2018}]{Tan-etal2018}
\begin{barticle}
\bauthor{\bsnm{{Tan}}, \binits{C.M.}},
\bauthor{\bsnm{{Bassa}}, \binits{C.G.}},
\bauthor{\bsnm{{Cooper}}, \binits{S.}},
\bauthor{\bsnm{{Dijkema}}, \binits{T.J.}},
\bauthor{\bsnm{{Esposito}}, \binits{P.}},
\bauthor{\bsnm{{Hessels}}, \binits{J.W.T.}},
\bauthor{\bsnm{{Kondratiev}}, \binits{V.I.}},
\bauthor{\bsnm{{Kramer}}, \binits{M.}},
\bauthor{\bsnm{{Michilli}}, \binits{D.}},
\bauthor{\bsnm{{Sanidas}}, \binits{S.}},
\bauthor{\bsnm{{Shimwell}}, \binits{T.W.}},
\bauthor{\bsnm{{Stappers}}, \binits{B.W.}},
\bauthor{\bsnm{{van Leeuwen}}, \binits{J.}},
\bauthor{\bsnm{{Cognard}}, \binits{I.}},
\bauthor{\bsnm{{Grie{\ss}meier}}, \binits{J.-M.}},
\bauthor{\bsnm{{Karastergiou}}, \binits{A.}},
\bauthor{\bsnm{{Keane}}, \binits{E.F.}},
\bauthor{\bsnm{{Sobey}}, \binits{C.}},
\bauthor{\bsnm{{Weltevrede}}, \binits{P.}}:
\batitle{{LOFAR Discovery of a 23.5 s Radio Pulsar}}.
\bjtitle{\apj}
\bvolume{866}(\bissue{1}),
\bfpage{54}
(\byear{2018})
\doiurl{10.3847/1538-4357/aade88}
{\href{https://arxiv.org/abs/1809.00965}{{arXiv:1809.00965}}}
{[astro-ph.HE]}
\end{barticle}
\endbibitem

\bibitem[\protect\citeauthoryear{{Morgan} and {Ekers}}{2021}]{2021PASA...38...13M}
\begin{barticle}
\bauthor{\bsnm{{Morgan}}, \binits{J.S.}},
\bauthor{\bsnm{{Ekers}}, \binits{R.}}:
\batitle{{A measurement of source noise at low frequency: Implications for modern interferometers}}.
\bjtitle{\pasa}
\bvolume{38},
\bfpage{013}
(\byear{2021})
\doiurl{10.1017/pasa.2021.3}
{\href{https://arxiv.org/abs/2101.11851}{{arXiv:2101.11851}}}
{[astro-ph.IM]}
\end{barticle}
\endbibitem

\bibitem[\protect\citeauthoryear{{Calabretta} et~al.}{2014}]{2014PASA...31....7C}
\begin{barticle}
\bauthor{\bsnm{{Calabretta}}, \binits{M.R.}},
\bauthor{\bsnm{{Staveley-Smith}}, \binits{L.}},
\bauthor{\bsnm{{Barnes}}, \binits{D.G.}}:
\batitle{{A New 1.4 GHz Radio Continuum Map of the Sky South of Declination +25{\textdegree}}}.
\bjtitle{\pasa}
\bvolume{31},
\bfpage{007}
(\byear{2014})
\doiurl{10.1017/pasa.2013.36}
{\href{https://arxiv.org/abs/1310.2414}{{arXiv:1310.2414}}}
{[astro-ph.IM]}
\end{barticle}
\endbibitem

\bibitem[\protect\citeauthoryear{{Reid} et~al.}{2014}]{2014ApJ...783..130R}
\begin{barticle}
\bauthor{\bsnm{{Reid}}, \binits{M.J.}},
\bauthor{\bsnm{{Menten}}, \binits{K.M.}},
\bauthor{\bsnm{{Brunthaler}}, \binits{A.}},
\bauthor{\bsnm{{Zheng}}, \binits{X.W.}},
\bauthor{\bsnm{{Dame}}, \binits{T.M.}},
\bauthor{\bsnm{{Xu}}, \binits{Y.}},
\bauthor{\bsnm{{Wu}}, \binits{Y.}},
\bauthor{\bsnm{{Zhang}}, \binits{B.}},
\bauthor{\bsnm{{Sanna}}, \binits{A.}},
\bauthor{\bsnm{{Sato}}, \binits{M.}},
\bauthor{\bsnm{{Hachisuka}}, \binits{K.}},
\bauthor{\bsnm{{Choi}}, \binits{Y.K.}},
\bauthor{\bsnm{{Immer}}, \binits{K.}},
\bauthor{\bsnm{{Moscadelli}}, \binits{L.}},
\bauthor{\bsnm{{Rygl}}, \binits{K.L.J.}},
\bauthor{\bsnm{{Bartkiewicz}}, \binits{A.}}:
\batitle{{Trigonometric Parallaxes of High Mass Star Forming Regions: The Structure and Kinematics of the Milky Way}}.
\bjtitle{\apj}
\bvolume{783}(\bissue{2}),
\bfpage{130}
(\byear{2014})
\doiurl{10.1088/0004-637X/783/2/130}
{\href{https://arxiv.org/abs/1401.5377}{{arXiv:1401.5377}}}
{[astro-ph.GA]}
\end{barticle}
\endbibitem

\bibitem[\protect\citeauthoryear{{Lorimer} and {Kramer}}{2004}]{2004hpa..book.....L}
\begin{bbook}
\bauthor{\bsnm{{Lorimer}}, \binits{D.R.}},
\bauthor{\bsnm{{Kramer}}, \binits{M.}}:
\bbtitle{{Handbook of Pulsar Astronomy}}
vol. \bseriesno{4},
(\byear{2004})
\end{bbook}
\endbibitem

\bibitem[\protect\citeauthoryear{{Ranasinghe} and {Leahy}}{2018}]{2018AJ....155..204R}
\begin{barticle}
\bauthor{\bsnm{{Ranasinghe}}, \binits{S.}},
\bauthor{\bsnm{{Leahy}}, \binits{D.A.}}:
\batitle{{Revised Distances to 21 Supernova Remnants}}.
\bjtitle{\aj}
\bvolume{155}(\bissue{5}),
\bfpage{204}
(\byear{2018})
\doiurl{10.3847/1538-3881/aab9be}
{\href{https://arxiv.org/abs/1808.09082}{{arXiv:1808.09082}}}
{[astro-ph.GA]}
\end{barticle}
\endbibitem

\bibitem[\protect\citeauthoryear{{Kassim}}{1992}]{1992AJ....103..943K}
\begin{barticle}
\bauthor{\bsnm{{Kassim}}, \binits{N.E.}}:
\batitle{{330 MHz VLA Observations of 20 Galactic Supernova Remnants}}.
\bjtitle{\aj}
\bvolume{103},
\bfpage{943}
(\byear{1992})
\doiurl{10.1086/116116}
\end{barticle}
\endbibitem

\bibitem[\protect\citeauthoryear{{Dokara} et~al.}{2021}]{2021A&A...651A..86D}
\begin{barticle}
\bauthor{\bsnm{{Dokara}}, \binits{R.}},
\bauthor{\bsnm{{Brunthaler}}, \binits{A.}},
\bauthor{\bsnm{{Menten}}, \binits{K.M.}},
\bauthor{\bsnm{{Dzib}}, \binits{S.A.}},
\bauthor{\bsnm{{Reich}}, \binits{W.}},
\bauthor{\bsnm{{Cotton}}, \binits{W.D.}},
\bauthor{\bsnm{{Anderson}}, \binits{L.D.}},
\bauthor{\bsnm{{Chen}}, \binits{C.-H.R.}},
\bauthor{\bsnm{{Gong}}, \binits{Y.}},
\bauthor{\bsnm{{Medina}}, \binits{S.-N.X.}},
\bauthor{\bsnm{{Ortiz-Le{\'o}n}}, \binits{G.N.}},
\bauthor{\bsnm{{Rugel}}, \binits{M.}},
\bauthor{\bsnm{{Urquhart}}, \binits{J.S.}},
\bauthor{\bsnm{{Wyrowski}}, \binits{F.}},
\bauthor{\bsnm{{Yang}}, \binits{A.Y.}},
\bauthor{\bsnm{{Beuther}}, \binits{H.}},
\bauthor{\bsnm{{Billington}}, \binits{S.J.}},
\bauthor{\bsnm{{Csengeri}}, \binits{T.}},
\bauthor{\bsnm{{Carrasco-Gonz{\'a}lez}}, \binits{C.}},
\bauthor{\bsnm{{Roy}}, \binits{N.}}:
\batitle{{A global view on star formation: The GLOSTAR Galactic plane survey. II. Supernova remnants in the first quadrant of the Milky Way}}.
\bjtitle{\aap}
\bvolume{651},
\bfpage{86}
(\byear{2021})
\doiurl{10.1051/0004-6361/202039873}
{\href{https://arxiv.org/abs/2103.06267}{{arXiv:2103.06267}}}
{[astro-ph.GA]}
\end{barticle}
\endbibitem

\bibitem[\protect\citeauthoryear{{Leahy} and {Williams}}{2017}]{2017AJ....153..239L}
\begin{barticle}
\bauthor{\bsnm{{Leahy}}, \binits{D.A.}},
\bauthor{\bsnm{{Williams}}, \binits{J.E.}}:
\batitle{{A Python Calculator for Supernova Remnant Evolution}}.
\bjtitle{\aj}
\bvolume{153}(\bissue{5}),
\bfpage{239}
(\byear{2017})
\doiurl{10.3847/1538-3881/aa6af6}
{\href{https://arxiv.org/abs/1701.05942}{{arXiv:1701.05942}}}
{[astro-ph.HE]}
\end{barticle}
\endbibitem

\bibitem[\protect\citeauthoryear{{Su} et~al.}{2014}]{2014ApJ...796..122S}
\begin{barticle}
\bauthor{\bsnm{{Su}}, \binits{Y.}},
\bauthor{\bsnm{{Yang}}, \binits{J.}},
\bauthor{\bsnm{{Zhou}}, \binits{X.}},
\bauthor{\bsnm{{Zhou}}, \binits{P.}},
\bauthor{\bsnm{{Chen}}, \binits{Y.}}:
\batitle{{Interaction between Supernova Remnant G22.7-0.2 and the Ambient Molecular Clouds}}.
\bjtitle{\apj}
\bvolume{796}(\bissue{2}),
\bfpage{122}
(\byear{2014})
\doiurl{10.1088/0004-637X/796/2/122}
{\href{https://arxiv.org/abs/1411.0757}{{arXiv:1411.0757}}}
{[astro-ph.SR]}
\end{barticle}
\endbibitem

\bibitem[\protect\citeauthoryear{{Arzoumanian} et~al.}{2002}]{2002ApJ...568..289A}
\begin{barticle}
\bauthor{\bsnm{{Arzoumanian}}, \binits{Z.}},
\bauthor{\bsnm{{Chernoff}}, \binits{D.F.}},
\bauthor{\bsnm{{Cordes}}, \binits{J.M.}}:
\batitle{{The Velocity Distribution of Isolated Radio Pulsars}}.
\bjtitle{\apj}
\bvolume{568}(\bissue{1}),
\bfpage{289}--\blpage{301}
(\byear{2002})
\doiurl{10.1086/338805}
{\href{https://arxiv.org/abs/astro-ph/0106159}{{arXiv:astro-ph/0106159}}}
{[astro-ph]}
\end{barticle}
\endbibitem

\bibitem[\protect\citeauthoryear{{Verbunt} et~al.}{2017}]{2017A&A...608A..57V}
\begin{barticle}
\bauthor{\bsnm{{Verbunt}}, \binits{F.}},
\bauthor{\bsnm{{Igoshev}}, \binits{A.}},
\bauthor{\bsnm{{Cator}}, \binits{E.}}:
\batitle{{The observed velocity distribution of young pulsars}}.
\bjtitle{\aap}
\bvolume{608},
\bfpage{57}
(\byear{2017})
\doiurl{10.1051/0004-6361/201731518}
{\href{https://arxiv.org/abs/1708.08281}{{arXiv:1708.08281}}}
{[astro-ph.HE]}
\end{barticle}
\endbibitem

\bibitem[\protect\citeauthoryear{{Pecaut} and {Mamajek}}{2013}]{2013ApJS..208....9P}
\begin{barticle}
\bauthor{\bsnm{{Pecaut}}, \binits{M.J.}},
\bauthor{\bsnm{{Mamajek}}, \binits{E.E.}}:
\batitle{{Intrinsic Colors, Temperatures, and Bolometric Corrections of Pre-main-sequence Stars}}.
\bjtitle{\apjs}
\bvolume{208}(\bissue{1}),
\bfpage{9}
(\byear{2013})
\doiurl{10.1088/0067-0049/208/1/9}
{\href{https://arxiv.org/abs/1307.2657}{{arXiv:1307.2657}}}
{[astro-ph.SR]}
\end{barticle}
\endbibitem

\bibitem[\protect\citeauthoryear{{Green}}{2018}]{2018JOSS....3..695G}
\begin{barticle}
\bauthor{\bsnm{{Green}}, \binits{G.M.}}:
\batitle{{dustmaps: A Python interface for maps of interstellar dust}}.
\bjtitle{The Journal of Open Source Software}
\bvolume{3}(\bissue{26}),
\bfpage{695}
(\byear{2018})
\doiurl{10.21105/joss.00695}
\end{barticle}
\endbibitem

\bibitem[\protect\citeauthoryear{{van Soelen} et~al.}{2024}]{2024MNRAS.529L.102V}
\begin{barticle}
\bauthor{\bsnm{{van Soelen}}, \binits{B.}},
\bauthor{\bsnm{{Bordas}}, \binits{P.}},
\bauthor{\bsnm{{Negueruela}}, \binits{I.}},
\bauthor{\bsnm{{de O{\~n}a Wilhelmi}}, \binits{E.}},
\bauthor{\bsnm{{Papitto}}, \binits{A.}},
\bauthor{\bsnm{{Rib{\'o}}}, \binits{M.}}:
\batitle{{NIR spectral classification of the companion in the gamma-ray binary HESS J1832-093 as an O6 V star}}.
\bjtitle{\mnras}
\bvolume{529}(\bissue{1}),
\bfpage{102}--\blpage{107}
(\byear{2024})
\doiurl{10.1093/mnrasl/slae007}
{\href{https://arxiv.org/abs/2401.05838}{{arXiv:2401.05838}}}
{[astro-ph.HE]}
\end{barticle}
\endbibitem

\bibitem[\protect\citeauthoryear{{Buisson} et~al.}{2021}]{2021MNRAS.503.5600B}
\begin{barticle}
\bauthor{\bsnm{{Buisson}}, \binits{D.J.K.}},
\bauthor{\bsnm{{Altamirano}}, \binits{D.}},
\bauthor{\bsnm{{Armas Padilla}}, \binits{M.}},
\bauthor{\bsnm{{Arzoumanian}}, \binits{Z.}},
\bauthor{\bsnm{{Bult}}, \binits{P.}},
\bauthor{\bsnm{{Castro Segura}}, \binits{N.}},
\bauthor{\bsnm{{Charles}}, \binits{P.A.}},
\bauthor{\bsnm{{Degenaar}}, \binits{N.}},
\bauthor{\bsnm{{D{\'\i}az Trigo}}, \binits{M.}},
\bauthor{\bsnm{{van den Eijnden}}, \binits{J.}},
\bauthor{\bsnm{{Fogantini}}, \binits{F.}},
\bauthor{\bsnm{{Gandhi}}, \binits{P.}},
\bauthor{\bsnm{{Gendreau}}, \binits{K.}},
\bauthor{\bsnm{{Hare}}, \binits{J.}},
\bauthor{\bsnm{{Homan}}, \binits{J.}},
\bauthor{\bsnm{{Knigge}}, \binits{C.}},
\bauthor{\bsnm{{Malacaria}}, \binits{C.}},
\bauthor{\bsnm{{Mendez}}, \binits{M.}},
\bauthor{\bsnm{{Mu{\~n}oz Darias}}, \binits{T.}},
\bauthor{\bsnm{{Ng}}, \binits{M.}},
\bauthor{\bsnm{{{\"O}zbey Arabac{\i}}}, \binits{M.}},
\bauthor{\bsnm{{Remillard}}, \binits{R.}},
\bauthor{\bsnm{{Strohmayer}}, \binits{T.E.}},
\bauthor{\bsnm{{Tombesi}}, \binits{F.}},
\bauthor{\bsnm{{Tomsick}}, \binits{J.A.}},
\bauthor{\bsnm{{Vincentelli}}, \binits{F.}},
\bauthor{\bsnm{{Walton}}, \binits{D.J.}}:
\batitle{{Dips and eclipses in the X-ray binary Swift J1858.6-0814 observed with NICER}}.
\bjtitle{\mnras}
\bvolume{503}(\bissue{4}),
\bfpage{5600}--\blpage{5610}
(\byear{2021})
\doiurl{10.1093/mnras/stab863}
{\href{https://arxiv.org/abs/2103.12787}{{arXiv:2103.12787}}}
{[astro-ph.HE]}
\end{barticle}
\endbibitem

\bibitem[\protect\citeauthoryear{{Bernardini} et~al.}{2012}]{2012A&A...542A..22B}
\begin{barticle}
\bauthor{\bsnm{{Bernardini}}, \binits{F.}},
\bauthor{\bsnm{{de Martino}}, \binits{D.}},
\bauthor{\bsnm{{Falanga}}, \binits{M.}},
\bauthor{\bsnm{{Mukai}}, \binits{K.}},
\bauthor{\bsnm{{Matt}}, \binits{G.}},
\bauthor{\bsnm{{Bonnet-Bidaud}}, \binits{J.-M.}},
\bauthor{\bsnm{{Masetti}}, \binits{N.}},
\bauthor{\bsnm{{Mouchet}}, \binits{M.}}:
\batitle{{Characterization of new hard X-ray cataclysmic variables}}.
\bjtitle{\aap}
\bvolume{542},
\bfpage{22}
(\byear{2012})
\doiurl{10.1051/0004-6361/201219233}
{\href{https://arxiv.org/abs/1204.3758}{{arXiv:1204.3758}}}
{[astro-ph.HE]}
\end{barticle}
\endbibitem

\bibitem[\protect\citeauthoryear{{Hare} et~al.}{2021}]{2021ApJ...914...85H}
\begin{barticle}
\bauthor{\bsnm{{Hare}}, \binits{J.}},
\bauthor{\bsnm{{Halpern}}, \binits{J.P.}},
\bauthor{\bsnm{{Tomsick}}, \binits{J.A.}},
\bauthor{\bsnm{{Thorstensen}}, \binits{J.R.}},
\bauthor{\bsnm{{Bodaghee}}, \binits{A.}},
\bauthor{\bsnm{{Clavel}}, \binits{M.}},
\bauthor{\bsnm{{Krivonos}}, \binits{R.}},
\bauthor{\bsnm{{Mori}}, \binits{K.}}:
\batitle{{Chandra, NuSTAR, and Optical Observations of the Cataclysmic Variables IGR J17528-2022 and IGR J20063+3641}}.
\bjtitle{\apj}
\bvolume{914}(\bissue{2}),
\bfpage{85}
(\byear{2021})
\doiurl{10.3847/1538-4357/abfa96}
{\href{https://arxiv.org/abs/2104.10503}{{arXiv:2104.10503}}}
{[astro-ph.HE]}
\end{barticle}
\endbibitem

\bibitem[\protect\citeauthoryear{{Bernardini} et~al.}{2014}]{2014MNRAS.445.1403B}
\begin{barticle}
\bauthor{\bsnm{{Bernardini}}, \binits{F.}},
\bauthor{\bsnm{{de Martino}}, \binits{D.}},
\bauthor{\bsnm{{Mukai}}, \binits{K.}},
\bauthor{\bsnm{{Falanga}}, \binits{M.}}:
\batitle{{Swift J2218.4+1925: a new hard-X-ray-selected polar observed with XMM-Newton}}.
\bjtitle{\mnras}
\bvolume{445}(\bissue{2}),
\bfpage{1403}--\blpage{1411}
(\byear{2014})
\doiurl{10.1093/mnras/stu1819}
{\href{https://arxiv.org/abs/1409.2257}{{arXiv:1409.2257}}}
{[astro-ph.HE]}
\end{barticle}
\endbibitem

\bibitem[\protect\citeauthoryear{{Stanway} et~al.}{2018}]{2018A&A...611A..66S}
\begin{barticle}
\bauthor{\bsnm{{Stanway}}, \binits{E.R.}},
\bauthor{\bsnm{{Marsh}}, \binits{T.R.}},
\bauthor{\bsnm{{Chote}}, \binits{P.}},
\bauthor{\bsnm{{G{\"a}nsicke}}, \binits{B.T.}},
\bauthor{\bsnm{{Steeghs}}, \binits{D.}},
\bauthor{\bsnm{{Wheatley}}, \binits{P.J.}}:
\batitle{{VLA radio observations of AR Scorpii}}.
\bjtitle{\aap}
\bvolume{611},
\bfpage{66}
(\byear{2018})
\doiurl{10.1051/0004-6361/201732380}
{\href{https://arxiv.org/abs/1801.07258}{{arXiv:1801.07258}}}
{[astro-ph.SR]}
\end{barticle}
\endbibitem

\bibitem[\protect\citeauthoryear{{Pelisoli} et~al.}{2023}]{2023NatAs.tmp..120P}
\begin{barticle}
\bauthor{\bsnm{{Pelisoli}}, \binits{I.}},
\bauthor{\bsnm{{Marsh}}, \binits{T.R.}},
\bauthor{\bsnm{{Buckley}}, \binits{D.A.H.}},
\bauthor{\bsnm{{Heywood}}, \binits{I.}},
\bauthor{\bsnm{{Potter}}, \binits{S.B.}},
\bauthor{\bsnm{{Schwope}}, \binits{A.}},
\bauthor{\bsnm{{Brink}}, \binits{J.}},
\bauthor{\bsnm{{Standke}}, \binits{A.}},
\bauthor{\bsnm{{Woudt}}, \binits{P.A.}},
\bauthor{\bsnm{{Parsons}}, \binits{S.G.}},
\bauthor{\bsnm{{Green}}, \binits{M.J.}},
\bauthor{\bsnm{{Kepler}}, \binits{S.O.}},
\bauthor{\bsnm{{Munday}}, \binits{J.}},
\bauthor{\bsnm{{Romero}}, \binits{A.D.}},
\bauthor{\bsnm{{Breedt}}, \binits{E.}},
\bauthor{\bsnm{{Brown}}, \binits{A.J.}},
\bauthor{\bsnm{{Dhillon}}, \binits{V.S.}},
\bauthor{\bsnm{{Dyer}}, \binits{M.J.}},
\bauthor{\bsnm{{Kerry}}, \binits{P.}},
\bauthor{\bsnm{{Littlefair}}, \binits{S.P.}},
\bauthor{\bsnm{{Sahman}}, \binits{D.I.}},
\bauthor{\bsnm{{Wild}}, \binits{J.F.}}:
\batitle{{A 5.3-min-period pulsing white dwarf in a binary detected from radio to X-rays}}.
\bjtitle{Nature Astronomy}
(\byear{2023})
\doiurl{10.1038/s41550-023-01995-x}
{\href{https://arxiv.org/abs/2306.09272}{{arXiv:2306.09272}}}
{[astro-ph.SR]}
\end{barticle}
\endbibitem

\bibitem[\protect\citeauthoryear{{Barrett} et~al.}{2020}]{2020AdSpR..66.1226B}
\begin{barticle}
\bauthor{\bsnm{{Barrett}}, \binits{P.}},
\bauthor{\bsnm{{Dieck}}, \binits{C.}},
\bauthor{\bsnm{{Beasley}}, \binits{A.J.}},
\bauthor{\bsnm{{Mason}}, \binits{P.A.}},
\bauthor{\bsnm{{Singh}}, \binits{K.P.}}:
\batitle{{Radio observations of magnetic cataclysmic variables}}.
\bjtitle{Advances in Space Research}
\bvolume{66}(\bissue{5}),
\bfpage{1226}--\blpage{1234}
(\byear{2020})
\doiurl{10.1016/j.asr.2020.04.007}
{\href{https://arxiv.org/abs/2004.11418}{{arXiv:2004.11418}}}
{[astro-ph.SR]}
\end{barticle}
\endbibitem

\bibitem[\protect\citeauthoryear{{Guedel} and {Benz}}{1993}]{1993ApJ...405L..63G}
\begin{barticle}
\bauthor{\bsnm{{Guedel}}, \binits{M.}},
\bauthor{\bsnm{{Benz}}, \binits{A.O.}}:
\batitle{{X-Ray/Microwave Relation of Different Types of Active Stars}}.
\bjtitle{\apjl}
\bvolume{405},
\bfpage{63}
(\byear{1993})
\doiurl{10.1086/186766}
\end{barticle}
\endbibitem

\bibitem[\protect\citeauthoryear{{Benz} and {Guedel}}{1994}]{1994A&A...285..621B}
\begin{barticle}
\bauthor{\bsnm{{Benz}}, \binits{A.O.}},
\bauthor{\bsnm{{Guedel}}, \binits{M.}}:
\batitle{{X-ray/microwave ratio of flares and coronae}}.
\bjtitle{\aap}
\bvolume{285},
\bfpage{621}--\blpage{630}
(\byear{1994})
\end{barticle}
\endbibitem

\bibitem[\protect\citeauthoryear{{Callingham} et~al.}{2024}]{2024arXiv240915507C}
\begin{botherref}
\oauthor{\bsnm{{Callingham}}, \binits{J.R.}},
\oauthor{\bsnm{{Pope}}, \binits{B.J.S.}},
\oauthor{\bsnm{{Kavanagh}}, \binits{R.D.}},
\oauthor{\bsnm{{Bellotti}}, \binits{S.}},
\oauthor{\bsnm{{Daley-Yates}}, \binits{S.}},
\oauthor{\bsnm{{Damasso}}, \binits{M.}},
\oauthor{\bsnm{{Grie{\ss}meier}}, \binits{J.-M.}},
\oauthor{\bsnm{{G{\"u}del}}, \binits{M.}},
\oauthor{\bsnm{{G{\"u}nther}}, \binits{M.}},
\oauthor{\bsnm{{Kao}}, \binits{M.M.}},
\oauthor{\bsnm{{Klein}}, \binits{B.}},
\oauthor{\bsnm{{Mahadevan}}, \binits{S.}},
\oauthor{\bsnm{{Morin}}, \binits{J.}},
\oauthor{\bsnm{{Nichols}}, \binits{J.D.}},
\oauthor{\bsnm{{Osten}}, \binits{R.A.}},
\oauthor{\bsnm{{P{\'e}rez-Torres}}, \binits{M.}},
\oauthor{\bsnm{{Pineda}}, \binits{J.S.}},
\oauthor{\bsnm{{Rigney}}, \binits{J.}},
\oauthor{\bsnm{{Saur}}, \binits{J.}},
\oauthor{\bsnm{{Stef{\'a}nsson}}, \binits{G.}},
\oauthor{\bsnm{{Turner}}, \binits{J.D.}},
\oauthor{\bsnm{{Vedantham}}, \binits{H.}},
\oauthor{\bsnm{{Vidotto}}, \binits{A.A.}},
\oauthor{\bsnm{{Villadsen}}, \binits{J.}},
\oauthor{\bsnm{{Zarka}}, \binits{P.}}:
{Radio Signatures of Star-Planet Interactions, Exoplanets, and Space Weather}.
arXiv e-prints,
2409--15507
(2024)
\doiurl{10.48550/arXiv.2409.15507}
{\href{https://arxiv.org/abs/2409.15507}{{arXiv:2409.15507}}}
{[astro-ph.EP]}
\end{botherref}
\endbibitem

\bibitem[\protect\citeauthoryear{{Hu} et~al.}{2022}]{2022ApJ...928...82H}
\begin{barticle}
\bauthor{\bsnm{{Hu}}, \binits{K.}},
\bauthor{\bsnm{{Baring}}, \binits{M.G.}},
\bauthor{\bsnm{{Barchas}}, \binits{J.A.}},
\bauthor{\bsnm{{Younes}}, \binits{G.}}:
\batitle{{Intensity and Polarization Characteristics of Extended Neutron Star Surface Regions}}.
\bjtitle{\apj}
\bvolume{928}(\bissue{1}),
\bfpage{82}
(\byear{2022})
\doiurl{10.3847/1538-4357/ac4ae8}
{\href{https://arxiv.org/abs/2201.06537}{{arXiv:2201.06537}}}
{[astro-ph.HE]}
\end{barticle}
\endbibitem

\bibitem[\protect\citeauthoryear{{Rajwade} et~al.}{2022}]{2022MNRAS.512.1687R}
\begin{barticle}
\bauthor{\bsnm{{Rajwade}}, \binits{K.M.}},
\bauthor{\bsnm{{Stappers}}, \binits{B.W.}},
\bauthor{\bsnm{{Lyne}}, \binits{A.G.}},
\bauthor{\bsnm{{Shaw}}, \binits{B.}},
\bauthor{\bsnm{{Mickaliger}}, \binits{M.B.}},
\bauthor{\bsnm{{Liu}}, \binits{K.}},
\bauthor{\bsnm{{Kramer}}, \binits{M.}},
\bauthor{\bsnm{{Desvignes}}, \binits{G.}},
\bauthor{\bsnm{{Karuppusamy}}, \binits{R.}},
\bauthor{\bsnm{{Enoto}}, \binits{T.}},
\bauthor{\bsnm{{G{\"u}ver}}, \binits{T.}},
\bauthor{\bsnm{{Hu}}, \binits{C.-P.}},
\bauthor{\bsnm{{Surnis}}, \binits{M.P.}}:
\batitle{{Long term radio and X-ray evolution of the magnetar Swift J1818.0-1607}}.
\bjtitle{\mnras}
\bvolume{512}(\bissue{2}),
\bfpage{1687}--\blpage{1695}
(\byear{2022})
\doiurl{10.1093/mnras/stac446}
{\href{https://arxiv.org/abs/2202.07548}{{arXiv:2202.07548}}}
{[astro-ph.HE]}
\end{barticle}
\endbibitem

\bibitem[\protect\citeauthoryear{{Ibrahim} et~al.}{2023}]{2023ApJ...943...20I}
\begin{barticle}
\bauthor{\bsnm{{Ibrahim}}, \binits{A.Y.}},
\bauthor{\bsnm{{Borghese}}, \binits{A.}},
\bauthor{\bsnm{{Rea}}, \binits{N.}},
\bauthor{\bsnm{{Coti Zelati}}, \binits{F.}},
\bauthor{\bsnm{{Parent}}, \binits{E.}},
\bauthor{\bsnm{{Russell}}, \binits{T.D.}},
\bauthor{\bsnm{{Ascenzi}}, \binits{S.}},
\bauthor{\bsnm{{Sathyaprakash}}, \binits{R.}},
\bauthor{\bsnm{{G{\"o}tz}}, \binits{D.}},
\bauthor{\bsnm{{Mereghetti}}, \binits{S.}},
\bauthor{\bsnm{{Topinka}}, \binits{M.}},
\bauthor{\bsnm{{Rigoselli}}, \binits{M.}},
\bauthor{\bsnm{{Savchenko}}, \binits{V.}},
\bauthor{\bsnm{{Campana}}, \binits{S.}},
\bauthor{\bsnm{{Israel}}, \binits{G.L.}},
\bauthor{\bsnm{{Tiengo}}, \binits{A.}},
\bauthor{\bsnm{{Perna}}, \binits{R.}},
\bauthor{\bsnm{{Turolla}}, \binits{R.}},
\bauthor{\bsnm{{Zane}}, \binits{S.}},
\bauthor{\bsnm{{Esposito}}, \binits{P.}},
\bauthor{\bsnm{{Rodr{\'\i}guez Castillo}}, \binits{G.A.}},
\bauthor{\bsnm{{Graber}}, \binits{V.}},
\bauthor{\bsnm{{Possenti}}, \binits{A.}},
\bauthor{\bsnm{{Dehman}}, \binits{C.}},
\bauthor{\bsnm{{Ronchi}}, \binits{M.}},
\bauthor{\bsnm{{Loru}}, \binits{S.}}:
\batitle{{Deep X-Ray and Radio Observations of the First Outburst of the Young Magnetar Swift J1818.0-1607}}.
\bjtitle{\apj}
\bvolume{943}(\bissue{1}),
\bfpage{20}
(\byear{2023})
\doiurl{10.3847/1538-4357/aca528}
{\href{https://arxiv.org/abs/2211.12391}{{arXiv:2211.12391}}}
{[astro-ph.HE]}
\end{barticle}
\endbibitem

\bibitem[\protect\citeauthoryear{{Lander}}{2024}]{2024MNRAS.tmp.2386L}
\begin{barticle}
\bauthor{\bsnm{{Lander}}, \binits{S.K.}}:
\batitle{{The Meissner effect in neutron stars}}.
\bjtitle{\mnras}
(\byear{2024})
\doiurl{10.1093/mnras/stae2453}
\end{barticle}
\endbibitem

\bibitem[\protect\citeauthoryear{{De Luca} et~al.}{2006}]{DeLuca2006}
\begin{barticle}
\bauthor{\bsnm{{De Luca}}, \binits{A.}},
\bauthor{\bsnm{{Caraveo}}, \binits{P.A.}},
\bauthor{\bsnm{{Mereghetti}}, \binits{S.}},
\bauthor{\bsnm{{Tiengo}}, \binits{A.}},
\bauthor{\bsnm{{Bignami}}, \binits{G.F.}}:
\batitle{{A Long-Period, Violently Variable X-ray Source in a Young Supernova Remnant}}.
\bjtitle{Science}
\bvolume{313}(\bissue{5788}),
\bfpage{814}--\blpage{817}
(\byear{2006})
\doiurl{10.1126/science.1129185}
{\href{https://arxiv.org/abs/astro-ph/0607173}{{arXiv:astro-ph/0607173}}}
{[astro-ph]}
\end{barticle}
\endbibitem

\bibitem[\protect\citeauthoryear{{Ho} and {Andersson}}{2017}]{2017MNRAS.464L..65H}
\begin{barticle}
\bauthor{\bsnm{{Ho}}, \binits{W.C.G.}},
\bauthor{\bsnm{{Andersson}}, \binits{N.}}:
\batitle{{Ejector and propeller spin-down: how might a superluminous supernova millisecond magnetar become the 6.67 h pulsar in RCW 103}}.
\bjtitle{\mnras}
\bvolume{464}(\bissue{1}),
\bfpage{65}--\blpage{69}
(\byear{2017})
\doiurl{10.1093/mnrasl/slw186}
{\href{https://arxiv.org/abs/1608.03149}{{arXiv:1608.03149}}}
{[astro-ph.SR]}
\end{barticle}
\endbibitem

\bibitem[\protect\citeauthoryear{{Borghese} et~al.}{2018}]{2018MNRAS.478..741B}
\begin{barticle}
\bauthor{\bsnm{{Borghese}}, \binits{A.}},
\bauthor{\bsnm{{Coti Zelati}}, \binits{F.}},
\bauthor{\bsnm{{Esposito}}, \binits{P.}},
\bauthor{\bsnm{{Rea}}, \binits{N.}},
\bauthor{\bsnm{{De Luca}}, \binits{A.}},
\bauthor{\bsnm{{Bachetti}}, \binits{M.}},
\bauthor{\bsnm{{Israel}}, \binits{G.L.}},
\bauthor{\bsnm{{Perna}}, \binits{R.}},
\bauthor{\bsnm{{Pons}}, \binits{J.A.}}:
\batitle{{Gazing at the ultraslow magnetar in RCW 103 with NuSTAR and Swift}}.
\bjtitle{\mnras}
\bvolume{478}(\bissue{1}),
\bfpage{741}--\blpage{748}
(\byear{2018})
\doiurl{10.1093/mnras/sty1119}
{\href{https://arxiv.org/abs/1805.00496}{{arXiv:1805.00496}}}
{[astro-ph.HE]}
\end{barticle}
\endbibitem

\bibitem[\protect\citeauthoryear{{Koljonen} and {Linares}}{2023}]{2023MNRAS.525.3963K}
\begin{barticle}
\bauthor{\bsnm{{Koljonen}}, \binits{K.I.I.}},
\bauthor{\bsnm{{Linares}}, \binits{M.}}:
\batitle{{A Gaia view of the optical and X-ray luminosities of compact binary millisecond pulsars}}.
\bjtitle{\mnras}
\bvolume{525}(\bissue{3}),
\bfpage{3963}--\blpage{3985}
(\byear{2023})
\doiurl{10.1093/mnras/stad2485}
{\href{https://arxiv.org/abs/2308.07377}{{arXiv:2308.07377}}}
{[astro-ph.HE]}
\end{barticle}
\endbibitem

\bibitem[\protect\citeauthoryear{{Papitto} and {de Martino}}{2022}]{Papitto2022}
\begin{bchapter}
\bauthor{\bsnm{{Papitto}}, \binits{A.}},
\bauthor{\bsnm{{de Martino}}, \binits{D.}}:
\bctitle{{Transitional Millisecond Pulsars}}.
In: \beditor{\bsnm{{Bhattacharyya}}, \binits{S.}},
\beditor{\bsnm{{Papitto}}, \binits{A.}},
\beditor{\bsnm{{Bhattacharya}}, \binits{D.}} (eds.)
\bbtitle{Astrophysics and Space Science Library}.
\bsertitle{Astrophysics and Space Science Library},
vol. \bseriesno{465},
pp. \bfpage{157}--\blpage{200}
(\byear{2022}).
\doiurl{10.1007/978-3-030-85198-9_6}
\end{bchapter}
\endbibitem

\bibitem[\protect\citeauthoryear{{Schreiber} et~al.}{2021}]{2021NatAs...5..648S}
\begin{barticle}
\bauthor{\bsnm{{Schreiber}}, \binits{M.R.}},
\bauthor{\bsnm{{Belloni}}, \binits{D.}},
\bauthor{\bsnm{{G{\"a}nsicke}}, \binits{B.T.}},
\bauthor{\bsnm{{Parsons}}, \binits{S.G.}},
\bauthor{\bsnm{{Zorotovic}}, \binits{M.}}:
\batitle{{The origin and evolution of magnetic white dwarfs in close binary stars}}.
\bjtitle{Nature Astronomy}
\bvolume{5},
\bfpage{648}--\blpage{654}
(\byear{2021})
\doiurl{10.1038/s41550-021-01346-8}
{\href{https://arxiv.org/abs/2104.14607}{{arXiv:2104.14607}}}
{[astro-ph.SR]}
\end{barticle}
\endbibitem

\bibitem[\protect\citeauthoryear{{Lamb} et~al.}{1983}]{1983ApJ...274L..71L}
\begin{barticle}
\bauthor{\bsnm{{Lamb}}, \binits{F.K.}},
\bauthor{\bsnm{{Aly}}, \binits{J.-J.}},
\bauthor{\bsnm{{Cook}}, \binits{M.C.}},
\bauthor{\bsnm{{Lamb}}, \binits{D.Q.}}:
\batitle{{Synchronization of magnetic stars in binary systems.}}
\bjtitle{\apjl}
\bvolume{274},
\bfpage{71}--\blpage{75}
(\byear{1983})
\doiurl{10.1086/184153}
\end{barticle}
\endbibitem

\bibitem[\protect\citeauthoryear{{Gangadhara}}{1997}]{1997A&A...327..155G}
\begin{barticle}
\bauthor{\bsnm{{Gangadhara}}, \binits{R.T.}}:
\batitle{{Orthogonal polarization mode phenomenon in pulsars.}}
\bjtitle{\aap}
\bvolume{327},
\bfpage{155}--\blpage{166}
(\byear{1997})
\doiurl{10.48550/arXiv.astro-ph/9707168}
{\href{https://arxiv.org/abs/astro-ph/9707168}{{arXiv:astro-ph/9707168}}}
{[astro-ph]}
\end{barticle}
\endbibitem

\bibitem[\protect\citeauthoryear{{Lower} et~al.}{2024}]{2024NatAs...8..606L}
\begin{barticle}
\bauthor{\bsnm{{Lower}}, \binits{M.E.}},
\bauthor{\bsnm{{Johnston}}, \binits{S.}},
\bauthor{\bsnm{{Lyutikov}}, \binits{M.}},
\bauthor{\bsnm{{Melrose}}, \binits{D.B.}},
\bauthor{\bsnm{{Shannon}}, \binits{R.M.}},
\bauthor{\bsnm{{Weltevrede}}, \binits{P.}},
\bauthor{\bsnm{{Caleb}}, \binits{M.}},
\bauthor{\bsnm{{Camilo}}, \binits{F.}},
\bauthor{\bsnm{{Cameron}}, \binits{A.D.}},
\bauthor{\bsnm{{Dai}}, \binits{S.}},
\bauthor{\bsnm{{Hobbs}}, \binits{G.}},
\bauthor{\bsnm{{Li}}, \binits{D.}},
\bauthor{\bsnm{{Rajwade}}, \binits{K.M.}},
\bauthor{\bsnm{{Reynolds}}, \binits{J.E.}},
\bauthor{\bsnm{{Sarkissian}}, \binits{J.M.}},
\bauthor{\bsnm{{Stappers}}, \binits{B.W.}}:
\batitle{{Linear to circular conversion in the polarized radio emission of a magnetar}}.
\bjtitle{Nature Astronomy}
\bvolume{8},
\bfpage{606}--\blpage{616}
(\byear{2024})
\doiurl{10.1038/s41550-024-02225-8}
{\href{https://arxiv.org/abs/2311.04195}{{arXiv:2311.04195}}}
{[astro-ph.HE]}
\end{barticle}
\endbibitem

\bibitem[\protect\citeauthoryear{{du Plessis} et~al.}{2019}]{2019ApJ...887...44D}
\begin{barticle}
\bauthor{\bsnm{{du Plessis}}, \binits{L.}},
\bauthor{\bsnm{{Wadiasingh}}, \binits{Z.}},
\bauthor{\bsnm{{Venter}}, \binits{C.}},
\bauthor{\bsnm{{Harding}}, \binits{A.K.}}:
\batitle{{Constraining the Emission Geometry and Mass of the White Dwarf Pulsar AR Sco Using the Rotating Vector Model}}.
\bjtitle{\apj}
\bvolume{887}(\bissue{1}),
\bfpage{44}
(\byear{2019})
\doiurl{10.3847/1538-4357/ab4e19}
{\href{https://arxiv.org/abs/1910.07401}{{arXiv:1910.07401}}}
{[astro-ph.HE]}
\end{barticle}
\endbibitem

\bibitem[\protect\citeauthoryear{{Ferrario} et~al.}{2015}]{2015SSRv..191..111F}
\begin{barticle}
\bauthor{\bsnm{{Ferrario}}, \binits{L.}},
\bauthor{\bsnm{{de Martino}}, \binits{D.}},
\bauthor{\bsnm{{G{\"a}nsicke}}, \binits{B.T.}}:
\batitle{{Magnetic White Dwarfs}}.
\bjtitle{\ssr}
\bvolume{191}(\bissue{1-4}),
\bfpage{111}--\blpage{169}
(\byear{2015})
\doiurl{10.1007/s11214-015-0152-0}
{\href{https://arxiv.org/abs/1504.08072}{{arXiv:1504.08072}}}
{[astro-ph.SR]}
\end{barticle}
\endbibitem

\bibitem[\protect\citeauthoryear{{Hyman} et~al.}{2005}]{2005Natur.434...50H}
\begin{botherref}
\oauthor{\bsnm{{Hyman}}, \binits{S.D.}},
\oauthor{\bsnm{{Lazio}}, \binits{T.J.W.}},
\oauthor{\bsnm{{Kassim}}, \binits{N.E.}},
\oauthor{\bsnm{{Ray}}, \binits{P.S.}},
\oauthor{\bsnm{{Markwardt}}, \binits{C.B.}},
\oauthor{\bsnm{{Yusef-Zadeh}}, \binits{F.}}:
{A powerful bursting radio source towards the Galactic Centre}
\textbf{434}(7029),
50--52
(2005)
\doiurl{10.1038/nature03400}
{\href{https://arxiv.org/abs/astro-ph/0503052}{{arXiv:astro-ph/0503052}}}
{[astro-ph]}
\end{botherref}
\endbibitem

\bibitem[\protect\citeauthoryear{{Keane}}{2018}]{2018NatAs...2..865K}
\begin{botherref}
\oauthor{\bsnm{{Keane}}, \binits{E.F.}}:
{The future of fast radio burst science}
\textbf{2},
865--872
(2018)
\doiurl{10.1038/s41550-018-0603-0}
{\href{https://arxiv.org/abs/1811.00899}{{arXiv:1811.00899}}}
{[astro-ph.HE]}
\end{botherref}
\endbibitem

\end{thebibliography}

\end{document}